\documentclass[aap,MSNbibl,nameyear,seceqn,dvips]{arximspdf}
\usepackage{graphicx}
\doi{10.1214/13-AAP936} 
\volume{24}
\issue{2}
\pubyear{2014}
\firstpage{811}
\lastpage{856}

\makeatletter

\newcommand{\rrvert}{\vert}
\newcommand{\llvert}{\vert}
\def\cal{\mathcal}
\newcommand{\eqref}[1]{(\ref{#1})}

\newtheorem{proposition}{Proposition}[section]
\newtheorem{corollary}{Corollary}[section]

\newproclaim{definition}{Definition}[section]
\newproclaim{assumption}{Assumption}[section]
\newtheorem{theorem}{Theorem}[section]
\newproclaim{remark}{Remark}[section]
\newproclaim{vg}{Example}[section]
\newcommand{\calZ}{{\cal Z}}
\newcommand{\calT}{{\cal T}}
\newcommand{\calS}{{\cal S}}
\newcommand{\calQ}{{\cal Q}}
\newcommand{\calP}{{\cal P}}
\newcommand{\calL}{{\cal L}}
\newcommand{\calH}{{\cal H}}
\newcommand{\calG}{{\cal G}}
\newcommand{\calF}{{\cal F}}
\newcommand{\calE}{{\cal E}}
\newcommand{\calA}{{\cal A}}
\newcommand{\bbP}{\mathbb{P}}
\newcommand{\bbE}{\mathbb{E}}
\makeatother

\begin{document}
\begin{frontmatter}

\title{Time-changed CIR default intensities with two-sided mean-reverting
jumps}
\runtitle{Time-changed CIR default intensities}

\begin{aug}
\author[a]{\fnms{Rafael} \snm{Mendoza-Arriaga}\ead[label=e1]{rafael.mendoza-arriaga@mccombs.utexas.edu}}
\and
\author[b]{\fnms{Vadim} \snm{Linetsky}\corref{}\thanksref{t1}\ead[label=e2]{linetsky@iems.northwestern.edu}}
\affiliation{University of Texas at Austin and Northwestern University}
\thankstext{t1}{Supported by NSF Grants DMS-08-02720 and
CMMI-1030486.}
\runauthor{R. Mendoza-Arriaga and V. Linetsky}
\address[a]{IROM\\ McCombs School of Business\\
University of Texas at Austin\\
CBA 5.202, B6500\\
Austin, Texas 78712\\
USA\\
\printead{e1}}

\address[b]{IEMS\\
McCormick School of Engineering\\
\quad and Applied Sciences\\
Northwestern University\\
2145 Sheridan Road\\
Evanston, Illinois 60208\\
USA\\
\printead{e2}}
\end{aug}

\received{\smonth{9} \syear{2012}}
\revised{\smonth{3} \syear{2013}}

%
\begin{abstract}
The present paper introduces a jump-diffusion extension of the
classical diffusion default intensity model by means of subordination
in the sense of Bochner. We start from the bi-variate process $(X,D)$
of a diffusion state variable $X$ driving default intensity and a
default indicator process $D$ and time change it with a L\'evy
subordinator $\calT$. We characterize the
time-changed process $(X^\phi_t,D^\phi_t)=(X(\calT_t),D(\calT_t))$
as a Markovian--It\^{o} semimartingale and show from the Doob--Meyer
decomposition of $D^\phi$ that the default time in the time-changed model
has a jump-diffusion or a pure jump intensity. When $X$ is a CIR
diffusion with mean-reverting drift, the default intensity of the
subordinate model (\textit{SubCIR}) is a jump-diffusion or a pure jump
process with mean-reverting jumps in both directions that stays
nonnegative. The SubCIR default intensity model is analytically
tractable by means of explicitly computed eigenfunction expansions of
relevant semigroups, yielding closed-form pricing of credit-sensitive
securities.
\end{abstract}

%

\pdfkeywords{91G40, 60J75, 91G30, 60G55, Default, default intensity, credit spread,
corporate bond, credit derivative, CIR process, time change, subordinator, Bochner subordination,
jump-diffusion process, state dependent Levy measure, spectral expansion}

\end{frontmatter}

\section{Introduction}\label{intro}

The classical \citet{cox-ingersoll-ross-1} (CIR)/ \citet{feller-1}
square-root diffusion has been a workhorse in the stochastic intensity
approach to the modeling of default risk in financial markets since the
seminal work of \citet{jarrow-lando-turnbull-1} and \citet{duffie-singleton-1} on reduced-form default modeling; see monographs
\citet{bielecki-rutkowski-1}, \citet{duffie-singleton-2} and \citet{jeanblanc-yor-chesney-1} for surveys. In this framework, the default
time can be thought of as the first jump time of a doubly stochastic
Poisson process (Cox process) with stochastic intensity following a
diffusion process. The attractiveness of the CIR diffusion as the model
for intensity stems from, on one hand, its dynamics and, on the other
hand, its analytical tractability.
If the coefficient of the linear term in the drift is negative, and the
constant term is positive, then CIR diffusion is mean-reverting, which
is an important empirical feature observed in credit markets. At the
same time, the process stays nonnegative due to vanishing volatility
and positive drift near the origin. Its analytical tractability stems,
on one hand, from its close connection with Bessel processes [\citet{pitman-yor-1}, \citet{going-jaeschke-yor-1}, Chapter~6 of \citet{jeanblanc-yor-chesney-1}, \citet{revuz-yor}] and, on the other hand,
from its membership in the class of affine processes [\citet{duffie-kan-2}, \citet{duffie-pan-singleton-1}, \citet{duffie-filipovic-schachermayer-1},
\citet{keller-ressel-schachermayer-teichmann-1}]. The former connection
yields explicit expressions for the CIR transition density and the
associated Feynman--Kac semigroup, while the later connection yields an
explicit expression for Laplace transform of the time integral of the
CIR process, giving rise to a closed-form solution for the survival
probability in the CIR default intensity model that essentially
coincides with the expression for the bond price in CIR interest rate model.
These properties lead to analytical pricing for a wide range of
credit-sensitive instruments in CIR-based models [e.g., \citet{brigo-alfonsi-1}, \citet{bielecki-jeanblanc-rutkowski-1}].

A limitation of the CIR default intensity model is its inability to
capture jumps in credit spreads and prices of credit-sensitive
securities (other than the default event itself).
This led a number of authors to introduce jumps into the CIR model
[\citet{duffie-garleanu-1}, \citet{filipovic-1}, \citeauthor{brigo-el-bachir-2} (\citeyear{brigo-el-bachir-2,brigo-el-bachir-1})]. To preserve analytical
tractability, all of the models considered in the literature so far
have been in the affine class. The most general extensions of the
one-dimensional CIR diffusion with jumps that remain in the affine
class are continuous state branching processes with immigration (CBI)
of \citet{kawazu-watanabe-1}; see \citet{filipovic-1} for a detailed
treatment in the context of applications to interest rate term
structure modeling. Roughly speaking, CBI-processes are nonnegative
Feller processes with CIR-type diffusion components and one-sided,
positive jumps with the compensator measure of the form $m(dy)+x\mu
(dy)$, where $m$ is the L\'{e}vy measure of a subordinator, and $\mu
(dy)$ is the L\'{e}vy measure of a spectrally positive L\'{e}vy
process; see Theorem 4.3 in \citet{filipovic-1} for the explicit
expression of their infinitesimal generator and the summary of their properties.

A limitation of CBI-processes is the one-sided nature of their jumps.
From the standpoint of financial applications, their sample path
behavior is somewhat unnatural. CBI processes can only jump up, and can
never jump down. Assuming the drift of the CBI-process is
mean-reverting, if the process experiences a large jump up bringing it
far away from its long-run mean, the only mechanism for it to return
back to its long-run mean is via its continuous mean-reverting drift,
with no possibility to jump back down. Moreover, jumps of CBI-processes
are either state independent (governed by $m$ if $\mu=0$), or depend
linearly of the current state via $x$ multiplying $\mu$. The one-sided
nature of jumps and their affine dependence on the state are common to
general affine processes, for example, \citet{cuchiero-filipovic-mayerhofer-teichmann-1} and \citet{Cuchiero-Keller-Ressel-Mayerhofer-Teichmann-1}. 
However, this is in contrast to the behavior often observed in
financial markets where a jump in one direction may be followed by a
jump in the opposite direction. This behavior is often observed in\vadjust{\goodbreak}
energy markets, where mean-reverting models are commonly used to
capture the spike-like behavior of the spot price of electricity [e.g.,
\citet{Barlow-1}, \citet{Geman-Roncoroni-1} and \citet{Meyer-Brandis-Tankov-1}]. This is also relevant in credit markets,
where the succession of good and bad news about the financial health of
an obligor, such as a firm or a sovereign viewed by the markets to be
in distress, can result in sharp changes in its market credit spreads
over relatively short periods of time (witness the recent behavior of
some European credit spreads sea-sawing under the influence of the
rapidly changing flow of economic and political news).
Recent empirical literature studying positive and negative jumps in
credit spreads includes \citet{zhang-zhou-zhu-1}, \citet{Elkamhi-Jacobs-Langlois-Ornthanalai-1} and \citet{Kita-1}.

This paper proposes a new approach to introducing more realistic
two-sided jump behavior into diffusion intensity models via Bochner's
subordination. We start with a nonnegative diffusion intensity model
and time change it with a subordinator, that is, a nonnegative L\'
{e}vy process with positive jumps and nonnegative drift. We show that
this results in a jump-diffusion (when the subordinator has a positive
drift) or a pure-jump (when the subordinator is driftless) intensity
model with two-sided jumps that stays nonnegative. In particular, when
the diffusion is CIR, the time-changed model possesses a nonnegative
intensity process with two-sided, mean-reverting jumps. The compensator
measure of this intensity process is state-dependent, and the
state-dependence is such that it automatically prevents the process
from going negative. While the process can experience downward jumps,
the magnitude of negative jumps depends on the pre-jump state of the
process to keep the process nonnegative. While the structure of the
process is highly state-dependent (and obviously nonaffine),
remarkably, the model remains fully analytically tractable by means of
eigenfunction expansions.

The rest of the paper is organized as follows. In Section~\ref{Biv} we
review diffusion intensity models in the particular setting convenient
for our purposes. Namely, we consider a bi-variate process
$(X,D)$, where $X$ is the state variable following a nonnegative
diffusion process (pre-intensity), and $D$ is an
event indicator process $D$. The bi-variate process is a Markov process
on ${\mathbb R}_+\cup\{0,1\}$ and a semimartingale.
This section contains a detailed discussion of the bi-variate process
$(X,D)$ both from the Markovian and from the semimartingale points
of view. While the diffusion intensity model is very well known, this
detailed presentation in the bi-variate form is provided for the
reader's convenience to set up notation in preparation for our
treatment of the time-changed (subordinated) model by both Markovian
and semimartingale methods.
We note that this bi-variate point of view of diffusion default
intensity models is also followed in some interesting recent papers by
\citet{bielecki-crepey-jeanblanc-rutkowski-2},
\citet{bielecki-crepey-jeanblanc-zargari-1} and
\citet{bielecki-cousin-crepey-herbertsson-1} in the context of pricing
multi-name credit derivatives.
In Section~\ref{sub.sec} we time change the bi-variate process $(X,D)$
with a L\'{e}vy subordinator $\calT$ with Laplace exponent $\phi$. The
resulting time-changed process $(X^\phi,D^\phi)$ is a Markov process
with its infinitesimal generator given by the Phillips theorem. We
explicitly compute the generator from the Phillips theorem and obtain
its representation as an integro-differential operator. Being a time
change of a semimartingale, the bi-variate process is also a
semimartingale. We then identify its predictable characteristics from
the generator and obtain L\'{e}vy--It\^{o} decomposition of $X^\phi$,
Doob--Meyer decomposition of $D^\phi$ and It\^{o} formula for functions
$f(t,X^\phi,D^\phi)$. We then identify the process $D^\phi$ with the
default indicator in our model, so that the default time is the jump
time of $D^\phi$, and from its Doob--Meyer decomposition identify
explicitly the default intensity as $\lambda^\phi_t=(1-D^\phi_{t})
k^\phi(X_t^\phi)$, where $k^\phi(x)$ is an explicitly determined
positive function. In Section~\ref{sectpricing} we apply the results to
the pricing of credit-sensitive securities.
In Section~\ref{sub.spec.dec} we detail the eigenfunction expansion
approach to calculate the semigroup associated to the bi-variate
process $(X^\phi,D^\phi)$ and, in particular, to calculate the survival
probability and prices of defaultable securities. In Section~\ref{subcirspec} we specialize $X$ to be the CIR diffusion and thus obtain
the subordinate CIR (SubCIR) default intensity model. This section
contains explicit expressions of all the quantities relevant to the
SubCIR model and, in particular, explicit eigenfunction expansions for
the SubCIR semigroups.
These explicit solutions are then applied to give numerical
illustrations of the SubCIR default intensity model.

\section{The diffusion default intensity model} \label{Biv}


We start with a complete probability space $(\Omega,\calF,\bbP)$ on
which a one-dimensional standard Brownian motion $\{B_t,t\geq0\}$ is
defined. Let ${\mathbb F}^B=({\cal F}^B_t)_{t\geq0}$ denote its
completed natural filtration.
We model the state variable
as the unique strong solution of the
stochastic differential equation (SDE)
%
%
\begin{equation}
X_t=x+\int_0^t b(X_u)\,du+
\int_0^t\sigma(X_u)\,dB_u,\qquad
t\geq0. \label{SDE1}
\end{equation}
We assume that the drift and diffusion coefficients $b(x)$ and $\sigma
(x)$ are continuous on $(0,\infty)$, $\sigma(x)>0$ on $(0,\infty)$, and
are such that for each positive initial condition $X_0=x>0$ this SDE
admits a unique strong solution that stays nonnegative for all $t>0$.
Thus the state variable $(X_t)_{t\geq0}$ is a one-dimensional
diffusion, as well as a nonnegative continuous semimartingale.

Under our assumptions, the boundary at zero is either \emph{natural} (in
which case the process cannot reach zero when started from a positive
value $x>0$, and cannot be started at zero), \emph{entrance} [in which
case the process cannot reach zero when started from a positive value
$x>0$, but can be started at zero, in which case it instantaneously
enters the interval $(0,\infty)$ and never comes back to zero], \emph{instantaneously reflecting}, or \emph{absorbing} (in which case $X_t=0$
for all $t\geq T_0$, where $T_0$ is the first hitting time of zero). We
refer the reader to \citet{ethier-kurtz-1}, pages 366--367, and \citet{borodin-salminen}, Chapter II, for detailed expositions of Feller's
classification of boundaries of one-dimensional diffusions.
In this paper we exclude absorption, assuming that zero is either
unattainable (natural or entrance), or an instantaneously reflecting boundary.
Since we assume that the SDE~\eqref{SDE1} has a unique strong solution,
the process does not explode to infinity when started from any $x>0$;
that is, infinity is an unattainable boundary. The state space of the
diffusion $X$ will be denoted by $I$, and $I=(0,\infty)$ if $0$ is
unattainable (natural or entrance) or $I=[0,\infty)$ if zero is
instantaneously reflecting.

\begin{vg}[(The CIR SDE)]\label{EG.1}
The key example of interest to us in this paper is the CIR SDE with
\[
\sigma(x)=\sigma\sqrt{x} \qquad\mbox{with } \sigma>0,\qquad b(x)=\kappa(\theta-x)\qquad \mbox{with }
\kappa\theta>0. 
\]
Surveys of CIR processes and their relationship with Bessel processes
can be found in \citet{going-jaeschke-yor-1} and \citet{jeanblanc-yor-chesney-1}, Section~6.3.
The drift coefficient $b(x)$ is Lipschitz, and the diffusion
coefficient $\sigma(x)$ satisfies the Yamada--Watanabe condition [cf.
\citet{revuz-yor}, Theorem~IX.1.7], so the SDE has a unique strong
solution for any $x\geq0$.
Since for $\theta=0$ and $x=0$ the solution is $X_t=0$, by the
comparison theorem for one-dimensional SDEs [cf. \citet{revuz-yor},
Theorem~IX.3.7], the solutions for $\kappa\theta>0$ and $x\geq0$ stay
nonnegative, $X_t\geq0$ for all $t\geq0$.
Furthermore, when the Feller condition is satisfied,
$2\kappa\theta\geq\sigma^2$,
the process stays strictly positive when started from any $x>0$, that
is, ${\mathbb P}(T_0=\infty)=1$, where $T_0$ is the first hitting time
of zero. It can also be started from $x=0$, in which case it
immediately enters the interval $(0,\infty)$ and stays strictly
positive for all $t>0$. In this case the boundary at zero is an
entrance boundary. When the Feller condition is not satisfied, $0<
2\kappa\theta< \sigma^2$, the process can reach zero when started from
$x>0$, and zero is an instantaneously reflecting boundary.\looseness=1
\end{vg}

Let $C([0,\infty])$ denote the Banach space
of functions continuous on $(0,\infty)$ and such that the limits $\lim_{x\rightarrow0}f(x)$ and $\lim_{x\rightarrow\infty}f(x)$ exist and
are finite and endowed with the usual supremum norm.
As shown in, for example, \citet{ethier-kurtz-1}, page 366, the
transition function $P^0_t(x,dy)={\mathbb P}(X_t^x\in dy)$ of the
diffusion process\vadjust{\goodbreak} $X^x$ started at $x$ defines a Feller semigroup
$({\cal P}_t^0)_{t\geq0}$ acting on $C([0,\infty])$
by
%
%
\begin{equation}
{\cal P}_t^0f(x)=\bbE^x \bigl[f(X_t)
\bigr]=\int_I f(y) P^0_t(x,dy),
\label{smgn.1}
\end{equation}
where ${\mathbb E}^{x}$ denotes the expectation corresponding to the
law ${\mathbb P}^{x}$ of $(X_t^x)_{t\geq0}$ started at $x$.
The infinitesimal generator of ${\cal P}^0$ is a second-order
differential operator of the form
\[
{\cal A}^0f(x)= \tfrac{1}{2}\sigma^2(x)f^{\prime\prime
}(x)+b(x)f^\prime(x)
\label{infg.1}
\]
with the domain $D({\cal A}^0)=\{f\in C([0,\infty])\cap C^2((0,\infty
))\dvtx {\cal A}^0f\in C([0,\infty])\}$ if zero is an unattainable boundary.
If zero is an instantaneously reflecting boundary, the Neumann-type
boundary condition is additionally imposed at zero [\citet{ethier-kurtz-1}, page 367, equation (1.11) with $q_0=0$].
We also note that when zero and infinity are both natural boundaries,
the semigroup leaves the space $C_0((0,\infty))\subset C([0,\infty])$
of functions continuous on $(0,\infty)$ and having zero limits $\lim_{x\rightarrow0}f(x)=0$ and $\lim_{x\rightarrow0}f(x)=0$ invariant
and is a Feller semigroup on it. If zero is not a natural boundary,
while infinity is, then the semigroup is Feller on $C_0([0,\infty))$.

We next assume that our probability space supports a unit-mean
exponential random variable $\calE\sim\operatorname{Exp}(1)$ independent of the
Brownian motion $B$ (and hence, of $X$). Define a random time $\zeta$
\[
\zeta:=\inf \biggl\{t\geq0\dvtx \int_0^t
k(X_u)\,du\geq\calE \biggr\}, \label{ffte}
\]
where $k(x)\geq0$ is a given function assumed continuous on $(0,\infty
)$. If zero is an instantaneously reflecting boundary, we assume that
there is a finite limit $\lim_{x\rightarrow0}k(x)<\infty$. If zero is
unattainable, we do not make any assumptions about the behavior of
$k(x)$ as $x\rightarrow0$.
Under these assumptions, $\int_0^t k(X_u)\,du<\infty$ a.s. for any
initial condition $X_0=x>0$ [from our assumptions that $X$ does not
explode to infinity, and that $k(x)$ is continuous on $(0,\infty)$ and
has a finite limit at zero if zero is an attainable boundary for $X$].
Key examples of functions $k(x)$ of interest to us for credit risk
applications are given in Examples~\ref{EG.2}--\ref{EG.6} at the end of
this section.


We denote
by $({\cal P}_t^\beta)_{t\geq0}$ the Feynman--Kac semigroup associated
with the positive continuous additive functional $\int_0^t\beta
k(X_u)\,du$ with $\beta>0$,
%
%
\begin{equation}
{\cal P}^\beta_tf(x)=\bbE^x
\bigl[e^{-\beta\int_0^t
k(X_u)\,du}f(X_t) \bigr]. \label{fk.1}
\end{equation}
Under our assumptions, it is a sub-Markovian--Feller semigroup on
$C([0,\infty])$
with the generator
\[
{\cal A}^\beta f(x)={\cal A}^0f(x)-\beta k(x)f(x)
\label{infg.1.b}
\]
with the domain $D({\cal A}^\beta)\subseteq D({\cal A}^0)$. More
precisely, $D({\cal A}^\beta)=\{f\in C([0,\infty])\cap C^2((0,\infty
))\dvtx {\cal A}^\beta f\in C([0,\infty])\}$ if zero is an
unattainable\vadjust{\goodbreak}
boundary for a \emph{diffusion with killing at the rate} $\beta k(x)$;
see \citet{borodin-salminen}, pages 16--17, for Feller's boundary
classification of one-dimensional diffusions with killing. If zero is
instantaneously reflecting for the diffusion with killing at the rate
$\beta k(x)$, the Neumann-type boundary condition is imposed at zero
[\citet{ethier-kurtz-1}, page 367, equation (1.11) with $q_0=0$].


We next associate to the random time $\zeta$ the event indicator
process (one-point process) $(D_t)_{t\geq0}$ defined by
\[
D_t:=\mathbf{1}_{\{\zeta\leq t\}}, \qquad t\geq0,\label{indic.1}
\]
denote by ${\mathbb D}=({\cal D}_t)_{t\geq0}$ its (completed) natural
filtration
and define an enlarged filtration ${\mathbb G}=(\calG_t)_{t\geq0}$ with
$\calG_t={\cal F}^B_t\vee{\cal D}_t$.
This filtration is the smallest one which contains ${\mathbb F}^B$ and
such that the random time $\zeta$ is a stopping time; cf. \citet{jeanblanc-yor-chesney-1}, Section~7.3.3.
From \citet{jeanblanc-yor-chesney-1}, Proposition 5.9.1.1 and Remark 7.5.1.2, we
observe that the filtrations ${\mathbb F}^B$ and ${\mathbb G}$,
${\mathbb F}^B\subset{\mathbb G}$, satisfy the \textbf{H}-\emph{Hypothesis}.
As a result, any ${\mathbb F}^B$-local martingale is also
a ${\mathbb G}$-local martingale.

We will now study the bi-variate process $(X_t,D_t)_{t\geq0}$ of the
state variable $X$ and the event indicator $D$.
Given our assumptions, for any initial conditions $X_0=x>0$ and
$D_0=d\in\{0,1\}$, $(X,D)$ is
a Markovian semimartingale taking values in ${\mathbb R}_+\times\{0,1\}
\subset{\mathbb R}^2$ ($D_0=1$ corresponds to $\zeta=0$, and hence
$D_t=1$ for all $t>0$ when $D_0=1$).
We first characterize its Markovian nature.
To this end, observe that any function $f(x,d)\in C([0,\infty]\times\{
0,1\})$ can be written in the form
%
%
\begin{equation}
f(x,d)=f_1(x)+(1-d) \bigl(f_0(x)-f_1(x)
\bigr), \label{eqr.2}
\end{equation}
where $f_0(x):=f(x,0)\in C([0,\infty])$ and $f_1(x):=f(x,1)\in
C([0,\infty])$.
%
\begin{theorem}[{[Markovian characterization
of $(X,D)$]}]\label{th.mkv.0}
\textup{(i)} The bi-variate process $(X,D)$ is a Feller process whose Feller
semigroup $({\cal P}_t)_{t\geq0}$ acts on $f\in C([0,\infty]\times\{
0,1\})$ according to
%
%
\begin{equation}
\calP_t f(x,d)=\calP^0_tf_1(x)+(1-d)
\calP^1_t(f_0-f_1)
(x),\label{semi.bi.1}
\end{equation}
where $f_0(x)=f(x,0)\in C([0,\infty])$, $f_1(x)=f(x,1)\in C([0,\infty
])$, $({\cal P}_t^0)_{t\geq0}$ is the transition semigroup~\eqref
{smgn.1} of $X$ on $C([0,\infty])$ and $({\cal P}_t^1)_{t\geq0}$ is
the Feynman--Kac semigroup~\eqref{fk.1}
on $C([0,\infty])$.

\textup{(ii)} The infinitesimal generator of the Feller semigroup $({\cal
P}_t)_{t\geq0}$ is given by
\begin{eqnarray*}
\label{eqm.3} \calA f(x,d)&=&{\cal A}^0f_1(x)+(1-d){\cal
A}^1(f_0-f_1) (x)
\\
&=&{\cal A}^0f(x,d) +(1-d)k(x) \bigl(f(x,1)-f(x,0)\bigr),
\end{eqnarray*}
where
${\cal A}^0$ and ${\cal A}^1$ are the generators of $({\cal
P}_t^0)_{t\geq0}$
and $({\cal P}_t^1)_{t\geq0}$, respectively.

\textup{(iii)} If $f(x,d)\in D({\cal A})$ [i.e., $f$ is of the form~\eqref
{eqr.2} with $f_0,f_1\in D({\cal A}^1)$] and $(X,D)$ starts from
$X_0=x>0$ and $D_0=d\in\{0,1\}$, then the process
\[
M^f_t:=f(X_t,D_t)-f(x,d)-\int
_0^t {\cal A}f(X_s,D_s)\,ds
\label{eqfmis.1}
\]
is a ${\mathbb G}$-martingale.
\end{theorem}
\begin{pf}
(i)
For all $0\leq s<t$, we have
\begin{eqnarray*}
\label{eqm.6} \bbE \bigl[f(X_t,D_t)|
\calG_s \bigr] &=&\bbE\bigl[(1-D_t) (f_0-f_1)
(X_t)|\calG_s\bigr]+\bbE\bigl[f_1(X_t)|
\calG_s\bigr]
\\
&=&(1-D_s)\bbE\bigl[e^{-\int_s^tk(X_u)\,du}(f_0-f_1)
(X_t)|\calF^B_s\bigr]+\bbE
\bigl[f_1(X_t)|\calF^B_s\bigr]
\\
&=&(1-D_s){\cal P}^1_{t-s}(f_0-f_1)
(X_s)+{\cal P}^0_{t-s}f_1(X_s)
\\
&=&{\cal P}_{t-s}f(X_s,D_s).
\end{eqnarray*}
The first equality follows from the representation~\eqref{eqr.2}, the
second equality is a standard result in intensity modeling in credit
risk [e.g., \citet{jeanblanc-yor-chesney-1}, Corollary 7.3.4.2, or
\citet{bielecki-rutkowski-1}, Corollary~5.1.1],
the third equality follows from the Markov property and time
homogeneity of $X$.
Since the operators $({\cal P}_{t}^0)_{t\geq0}$ and $({\cal
P}_{t}^1)_{t\geq0}$ form Feller semigroups on $C([0,\infty])$, it is
then immediate that the operators $({\cal P}_{t})_{t\geq0}$ form a
Feller semigroup on $C([0,\infty]\times\{0,1\})$.
Thus, the bi-variate process $(X,D)$ is a Feller process whose
semigroup action on $C([0,\infty]\times\{0,1\})$ is given by
equation~\eqref{semi.bi.1}.
(ii)~The expression for the generator ${\cal A}$ follows from
equation~\eqref{semi.bi.1}, given ${\cal A}^0$ and ${\cal A}^1$ are the
generators of ${\cal P}^0$ and ${\cal P}^1$.
Part (iii) follows from \citet{ethier-kurtz-1}, Proposition 1.7,
page 162.
\end{pf}

Since $X$ is a continuous semimartingale, and $D$ is a one-point point
process, the bi-variate process $(X,D)$ is a special semimartingale.
We can write
It\^{o} formula for functions of time and the bi-variate process in the
useful form that separates the process $f(s,X_s,D_s)$ into a
predictable finite variation process, a continuous local martingale
that is the stochastic integral with respect to Brownian motion, and a
discontinuous martingale that is the integral with respect to the
compensated one-point process.
%
\begin{theorem}[{[It\^o formula for $(X,D)$]}]\label{th.xD.0}
\textup{(i)} The one-point point process $D$ has the following Doob--Meyer
decomposition:
\[
D_t=A_t+M_t, \qquad A_t=\int
_0^t (1-D_{s})k(X_s)\,ds,\qquad
M_t=D_t-A_t, \label{teq.1}
\]
where $A$ is the predictable ${\mathbb G}$-compensator of $D$ [so that
$\lambda_t^{\mathbb G}:=(1-D_{t})k(X_t)$ is its ${\mathbb
G}$-intensity] and $M$ is a ${\mathbb G}$-martingale.

\textup{(ii)} Suppose the semimartingale $(X,D)$ starts from $X_0=x>0$ and
$D_0=d\in\{0,1\}$.
For any function $f(t,x,d)=f_1(t,x)+(1-d)(f_0(t,x)-f_1(t,x))$ with
$f_i(t,x)\in C^{1,2}({\mathbb R}_+ \times(0,\infty))$ if zero is an
unattainable boundary for the process $X$ or $f_i(t,x)\in
C^{1,2}({\mathbb R}_+ \times[0,\infty))$ if zero is attainable for
$X$, the process $f(t,X_t,D_t)$ is a special ${\mathbb
G}$-semimartingale with the following canonical decomposition into a
predictable finite variation process, a continuous local martingale,
and a purely discontinuous martingale,
\begin{eqnarray*}
\label{eqm.8} f(t,X_t,D_t)&=&f(0,x,d)+\int
_0^t (\partial_s+{\cal A}
)f(s,X_s,D_s)\,ds
\\
&&{}+\int_0^t
\sigma(X_s)\partial_x f(s,X_s,D_s)\,dB_s
\\
&&{}+\int_0^t (1-D_{s-})
\bigl(f(s,X_s,1)-f(s,X_s,0)\bigr)\,dM_s.
\end{eqnarray*}
\end{theorem}
\begin{pf}
(i) This is a standard result; cf. Lemma 7.3.4.3(ii) in \citet{jeanblanc-yor-chesney-1}, page 421.

(ii)
Since $X$ is a nonnegative semimartingale, the functions $f_i(t,x)$
only need to be defined for $x\geq0$. In order for all the terms in
It\^{o}'s formula to be well defined, when $X$ is strictly positive,
the functions $f_i$ need only be $C^{1,2}({\mathbb R}_+ \times
(0,\infty
))$, while in the case when $X$ can hit zero $f_i$ and their first and
second derivatives in $x$ and first derivatives in $t$ need to have
finite limits as $x\rightarrow0$, so that $f_i\in C^{1,2}({\mathbb
R}_+ \times[0,\infty))$.
With these observations, this form of
It\^{o}'s formula immediately follows from the form of It\^{o}'s formula
for special semimartingales in \citet{jacod-1}, Theorem~3.89, page 109.
\end{pf}

\begin{vg}[(CIR intensity model, Example~\ref{EG.1} continued)]\label{EG.2}
Assuming that $\kappa,\theta,\sigma>0$, the CIR diffusion has the gamma
stationary density
%
%
\begin{equation}
\pi(x)=\frac{a^b x^{b-1}}{\Gamma(b)}e^{-a x},\qquad  b:=\frac{2\kappa
\theta
}{\sigma^2}, \qquad a:=
\frac{2\kappa}{\sigma^2}. \label{eqr.4}
\end{equation}
%
That is, for all $x\in I$ and $a,b>0$, $\lim_{t\rightarrow\infty
}P^0_t(x,dy)=\pi(y)\,dy$. With this choice of parameters,
$\lim_{t\rightarrow\infty}\bbE^x[X_t]=\int_I y \pi(y)\,dy=\theta$,
and $\theta$ is referred to as the long-run mean and $\kappa$ as the
rate of mean reversion of the CIR state variable.

Let $k(x)=x$
in the CIR intensity model. Then, the ${\mathbb G}$-intensity of the
stopping time $\zeta$ is $\lambda^{{\mathbb G}}_t=(1-D_t)X_t$, and the
indicator process $D_t$ has a ${\mathbb G}$-compensator $A_t=\int_0^t
(1-D_s)X_s\,ds$.
If $D$ is interpreted as the default indicator, then (under the
assumption of zero recovery) the instantaneous credit spread is equal
to the ${\mathbb G}$-intensity $\lambda^{{\mathbb G}}_t$.
The corresponding default intensity model goes back to \citet{duffie-singleton-1}.
Since zero is either an entrance or an instantaneously reflecting
boundary and infinity is a natural boundary,
the CIR Feynman--Kac semigroup $({\cal P}^1_t)_{t\geq0}$ is a
sub-Markovian--Feller semigroup on $C_0([0,\infty))$ [and also on
$C_0((0,\infty))$, when the Feller condition is satisfied and zero is a
natural boundary]. It coincides with the pricing semigroup in the CIR
interest rate model. Explicit expressions for the densities of the CIR
transition semigroup $({\cal P}^0_t)_{t\geq0}$ and the CIR
Feynman--Kac semigroup $({\cal P}^\beta_t)_{t\geq0}$ with $\beta>0$
are given in Section~\ref{subcirspec}.
\end{vg}

\begin{vg}[(Reciprocal CIR intensity model)]\label{EG.3}
Let $X$ follow the
CIR process as in Example~\ref{EG.1} and assume that the Feller
condition is satisfied, but take $k(x)=1/x$ instead of $k(x)=x$. This
choice leads to the \emph{reciprocal CIR intensity model}. It was
applied in credit modeling by \citet{andreasen-1}. Applying It\^{o}'s
formula to the process $Y_t=1/X_t$ (justified when the Feller condition
is satisfied, since in that case the process stays strictly positive),
we obtain the SDE for $Y$,
\[
Y_t=y+\int_0^t \tilde{\kappa}(
\tilde{\theta}-Y_s)Y_s \,ds - \int_0^t
\sigma Y^{3/2}_s\,dB_s, \label{eqm.10}
\]
where $Y_0=y=1/x$ and $\tilde{\kappa}=\kappa/(\kappa\theta-\sigma^2)$
and $\tilde{\theta}=\kappa\theta-\sigma^2$. This SDE has quadratic
drift and the so-called $3/2$ volatility.
When $\tilde{\theta}>0$ and $\tilde{\kappa}>0$, which requires
$\kappa
\theta> \sigma^2$ for the CIR process $X$,
this SDE also appeared as the model for the instantaneous inflation
rate in \citet{cox-ingersoll-ross-1} and as the model for the
instantaneous nominal interest rate in \citet{lewis-4} and \citet{ahn-gao-1} (the so-called $3/2$ model).
In this case the process has a stationary density
%
%
\[
\pi(y)= \frac{\alpha^\beta}{\Gamma(\beta)} y^{-\beta-1} e^{-\alpha/y}\qquad \mbox{where }
\alpha:=\frac{2\tilde\kappa\tilde\theta}{\sigma
^2},\qquad \beta:=\frac{2(\sigma^2+\tilde\kappa)}{\sigma^2}. \label{eqr.5}
\]
The ${\mathbb G}$-intensity in this model is $\lambda^{{\mathbb
G}}_t=(1-D_t)Y_t=(1-D_t)/X_t$, where $Y$ is the $3/2$-diffusion, or
equivalently $X$ is the CIR diffusion. The semigroup $({\cal
P}^1)_{t\geq0}$ can be obtained explicitly in this case and coincides
with the pricing semigroup in the $3/2$ interest rate model.
\end{vg}

\begin{vg}[{[Quadratic Ornstein--Uhlenbeck (OU) model]}]\label{EG.5}
Consider
an SDE \eqref{SDE1} with
%
%
\begin{equation}
\sigma(x)=2\sigma\sqrt{x},\qquad  b(x)=2\kappa(a+\theta\sqrt{x}-x) \label{eqr.7}
\end{equation}
with $\sigma>0$, $\kappa>0$, $\theta\geq0$, and $a=\sigma
^2/(2\kappa
)$. This SDE is similar to the CIR SDE, but has an extra term
with $\sqrt{x}$ in the drift. Let $Y_t$ be the OU process solving the
SDE $Y_t = y+\int_0^t \kappa(\theta-Y_u)\,du+\sigma B_t$. Applying It\^
{o}'s formula to the square of the OU process, $X_t=Y_t^2$, we verify
that $X$ satisfies the SDE with the coefficients~\eqref{eqr.7}. The
Feynman--Kac semigroup $({\cal P}_t^1)_{t\geq0}$ of the quadratic OU
model coincides with the pricing semigroup in the quadratic OU interest
rate model studied in \citet{beaglehole-tenney-1} and \citet{jamshidian-3}.
\end{vg}

\begin{vg}[{[\citet{carr-linetsky-1} JDCEV credit-equity model]}]\label{EG.4}
A~jump-to-default extended constant elasticity of variance (JDCEV)
diffusion of \citet{carr-linetsky-1} models the pre-default stock price
of a firm as the diffusion with
\begin{eqnarray*}\label{eqr.6}
\sigma(x)&=&ax^{\beta+1},\qquad b(x)=\bigl(r-q +k(x)\bigr)x,\\
 k(x)&=&b+c
\sigma^2(x)=b + c a^2 x^{2\beta},
\end{eqnarray*}
where $a>0$ fixes the volatility scale, the constant elasticity of
variance $\beta<0$ is assumed negative to capture the leverage effect
(volatility of the stock price increases as the stock price falls),
$r\geq0$ is the risk-free rate, $q\geq0$ is the dividend yield, and
$k(x)$ is the function defining the default intensity in the JDCEV
model, where $b\geq0$ is the constant part and $c\geq0$ is the
sensitivity of the default intensity to the instantaneous variance of
the stock price. $k(x)x$ is added to the drift to compensate for a jump
to default to ensure that, under the risk-neutral measure, the
discounted stock price with dividends reinvested and subject to default
is a martingale. Thus, in the JDCEV model the stock price of a firm
subject to default risk is $S_t=(1-D_t)X_t$, where $D_t$ is the default
indicator (stock price drops to zero when the firm defaults on its
debt). The ${\mathbb G}$-intensity is $\lambda_t^{\mathbb
G}=(1-D_t)(b+ca^2X_{t}^{2\beta})$.

For any $a>0$, $\mu:=r-q+b\in{\mathbb R}$, $\beta<0$ and $c\in
[(1/2+\beta)^+,\infty)$, the JDCEV SDE can be reduced to the CIR SDE as
follows. Let $(Y_t)_{t\geq0}$ be the unique strong solution of the CIR
SDE with $Y_0=y>0$ and parameters satisfying $\kappa\theta>0$ and
$\sigma>0$. For all $t\geq0$ define a new process $X_t = (Y_t)^{{1}/{(2|\beta|)}}$
with the initial condition $X_0=x=y^{{1}/{(2|\beta|)}}>0$. Then by It\^o's formula (the application is justified since
$|\beta|>0$), the process $X_t$ solves the JDCEV SDE with $a=\frac
{\sigma}{2|\beta|}$, $\mu=-\frac{\kappa}{2|\beta|}$, and
$c=1/2+|\beta
|(\frac{2\kappa\theta}{\sigma^2}-1)$. Since we are only interested in
nonnegative default intensities, we impose the condition $1/2+|\beta
|(\frac{2\kappa\theta}{\sigma^2}-1)\geq0$.
When the CIR process $Y$ satisfies Feller's condition, $2\kappa\theta
\geq\sigma^2$, the JDCEV parameter satisfies $c\in[1/2,\infty)$. In
this case, the boundary at zero is entrance for both the CIR and
JDCEV diffusions. When $2\kappa\theta\in(0,\sigma^2)$, the resulting
JDCEV parameter satisfies $c\in((1/2+\beta)^+,1/2)$. In this case, the
boundary at zero is instantaneously reflecting for both CIR and JDCEV.
In both cases, the killing rate $k$ reduces to $k(x)=b+ca^2/y$ in terms
of the CIR variable $y$, and hence the JDCEV FK semigroup $({\cal
P}^1_t)_{t\geq0}$ reduces to the Feynman--Kac semigroup in the
reciprocal CIR model of Example \ref{EG.3} (with the constant $b$ added).
Finally, we remark that while the JDCEV diffusion can also be defined
when $-1/2<\beta<0$ and $c\in[0,(1/2+\beta)^+)$ (in this case zero is
an exit boundary), it \emph{cannot} be reduced to the CIR diffusion for
this set of parameters. Since in this paper we do not consider exit
boundaries, we are not concerned with this case in the present paper.
\end{vg}

\begin{vg}[{[\citet{linetsky-1} credit-equity model]}]\label{EG.6}
Also in the context
of credit-equity models, \citet{linetsky-1} studies an extension of\vadjust{\goodbreak} the
Black--Scholes--Merton (BSM) model with bankruptcy where the
killing
rate is a negative power of the state variable. The pre-default
dynamics of the stock price are determined by
\[
\sigma(x)=\sigma x,\qquad b(x)=\bigl(r-q +k(x)\bigr)x, \qquad k(x)=\alpha x^{-p},
\label{lin.1a}
\]
where $\sigma>$ is the constant volatility, $r\geq0$ is the risk-free
rate, $q\geq0$ is the dividend yield and $k(x)$ is the killing rate
specified to be a negative power of the stock price with $\alpha>0$ and
$p>0$. As in Example \ref{EG.4}, the killing rate $k(x)$ is added into
the drift to compensate for the jump to default that makes the stock
price worthless in default. By specifying $k(x)$ to be the negative
power of the stock price, this model is able to exhibit implied
volatility skews in stock option prices, with the parameters $\alpha$
and $p$ of the killing rate specification controlling the slope of the
skew, thus establishing a link between implied volatility skews and
credit spreads (as the stock price drops, the implied volatility and
the probability of default increase).
In this case, the stock price is $S_t=(1-D_t)X_t$ and the ${\mathbb
G}$-intensity of default in this model is $\lambda_t^{\mathbb
G}=(1-D_t)\alpha X_t^p$.
\end{vg}

\section{The subordinated diffusion default intensity model}\label{sub.sec}

We next assume that our probability space $(\Omega,{\cal F},{\mathbb
P})$ also supports a L\'{e}vy subordinator $({\cal T}_t)_{t\geq0}$
independent of both
the Brownian motion $B$ and the exponential random variable~${\cal E}$
and thus is independent of the bi-variate process $(X,D)$. Recall that
a L\'{e}vy subordinator is a nondecreasing L\'{e}vy process, that is,
a L\'{e}vy process with one-sided positive jumps and nonnegative drift
and no diffusion component.
The Laplace transform of a L\'{e}vy subordinator $({\cal T}_t)_{t\geq
0}$ is given by
the L\'{e}vy--Khintchine formula
%
%
\begin{eqnarray}\label{Subord.1}
\bbE\bigl[e^{-\lambda\calT_t}\bigr]=\int_{[0,\infty
)}e^{-\lambda s}
\pi _t(ds)=e^{-t \phi(\lambda)}\nonumber \\
\eqntext{\displaystyle\mbox{with } \phi(\lambda)=\gamma \lambda
+\int_{(0,\infty)}\bigl(1-e^{-\lambda s}\bigr)
\nu(ds).}
\end{eqnarray}
Here $\pi_t(ds)$ is the transition kernel, $\phi(\lambda)$ is a L\'
{e}vy exponent, $\gamma\geq0$ is the nonnegative drift and $\nu(ds)$
is a L\'{e}vy measure of the subordinator that satisfies the
integrability condition $\int_{(0,\infty)}(s\wedge1)\nu(ds)<\infty$
[standard references on subordinators are \citeauthor{bertoin-1} (\citeyear{bertoin-1,bertoin-2}),
\citet{sato} and \citet{schilling-song-vondraceck-1}].

\begin{vg}[(Tempered stable and related subordinators)]\label{TempStab}
A family
of subordinators important in financial applications is defined by the
following three-parameter family of L\'{e}vy measures:
%
%
\begin{equation}
\nu(ds)=Cs^{-\alpha-1}e^{-\eta s}\,ds \label{Subord.7}
\end{equation}
with $C>0$, $\eta>0$, and $\alpha<1$. For $\alpha\in(0,1)$ these are
the so-called \emph{tempered stable subordinators}
[exponentially\vadjust{\goodbreak} dampened counterparts of the $\alpha$-stable
subordinators with $\nu(ds)=Cs^{-\alpha-1}\,ds$].
The special case $\alpha=1/2$ is
the \emph{inverse Gaussian process} [\citet{barndorff-nielsen-1}]. The
limiting case $\alpha=0$ is the \emph{gamma
process} [\citet{madan-carr-chang-1}]. Subordinators with $\alpha\in
[0,1)$ are infinite activity processes. Subordinators with $\alpha<0$ are
compound Poisson processes with gamma distributed jump sizes. The
compound Poisson process with the L\'{e}vy measure $\nu(ds)=\omega
\eta
e^{-\eta s}\,ds$ with
exponential jumps is a special case with $\alpha=-1$ and $C=\omega
\eta
$, where $\omega$ is the jump arrival rate, and $1/\eta$ is the mean of
the exponential jump size distribution. The Laplace exponent is given
by
\[
\phi(\lambda)=\cases{ %
\gamma\lambda-C\Gamma(-
\alpha)\bigl[(\lambda+\eta )^\alpha -\eta^\alpha\bigr],& \quad$\alpha
\neq0,$
\vspace*{2pt}\cr
\gamma\lambda+ C\ln(1+\lambda /\eta),& \quad$\alpha=0,$}
 \label{Subord.8}
\]
where $\Gamma(x)$ is the gamma function.
\end{vg}
We now time change the bi-variate process $(X,D)$ of the previous
section with a subordinator ${\cal T}$. That is, we define a new
bi-variate process $(X^\phi_t,D^\phi_t)_{t\geq0}$ by
%
%
\begin{equation}
X^\phi_t:=X\bigl({\cal T}(t)\bigr), \qquad D^\phi_t:=D
\bigl({\cal T}(t)\bigr) \label{eqr.8}
\end{equation}
and assume that $(D^\phi_t)_{t\geq0}$ is the default indicator process
(i.e., the default time is the first time $D^\phi$ is equal to one),
and $X^\phi$ is the state variable that models the credit health of
the obligor.
We also define the time changed filtration as follows.
Define an inverse subordinator process as the right inverse $(L_t:=\inf
\{s\geq0\dvtx {\cal T}_s>t\})_{t\geq0}$. Since ${\cal T}$ is c\'{a}dl\'
{a}g, so is $L$. Let ${\mathbb L}=(\calL_t)_{t\geq0}$ be its completed
natural filtration. Let ${\mathbb H}=({\cal H}_t)_{t\geq0}$ denote the
enlarged filtration with ${\cal H}_t={\cal G}_t\vee{\cal L}_t$, where
$\calG_t$ refers to the filtration ${\mathbb G}=(\calG_t)_{t\geq0}$ of
Section~\ref{Biv}. Then $({\cal T}_t)_{t\geq0}$ is an increasing
family of ${\mathbb
H}$-stopping times, and we can define the time changed filtration
${\mathbb H}^\phi=({\cal H}^\phi_t)_{t\geq0}$ by ${\cal H}^\phi
_t=\calH
_{\calT
_t}$ for all $t\geq0$. The time changed bi-variate process $(X^\phi
_t,D^\phi_t)_{t\geq0}$ is obviously ${\mathbb H}^\phi$-adapted and
c\'
{a}dl\'{a}g.
%
\begin{proposition}\label{prop.xD.0}
The process $(X^\phi_t,D^\phi_t)_{t\geq0}$
is an ${\mathbb H}^\phi$-semimartingale.
\end{proposition}

\begin{pf}
Since $(X^\phi_t,D^\phi_t)_{t\geq0}$ is a time change of a
semimartingale $(X,D)$, it is a semimartingale by Corollary 10.12 in
\citet{jacod-1}, page 315.
\end{pf}

We will now investigate its properties. In particular, we show that
$(X^\phi_t,\break D^\phi_t)_{t\geq0}$ is a Feller process with the Feller
semigroup on $C([0,\infty]\times\{0,1\})$, explicitly compute its
infinitesimal generator, obtain its predictable semimartingale
characteristics and give It\^{o}'s formula.

We first recall some key results about the subordination in the sense
of Bochner of semigroups of operators in Banach spaces. The procedure
of subordination goes back to \citet{bochner-1}. The expression for the
generator constitutes the Phillips theorem [\citet{phillips-1}]. The
formulation below is reproduced from \citet{sato}, Theorem 32.1.
%
\begin{theorem}[(Subordination in the sense
of Bochner; Phillips theorem)]\label{phillips.theo}Let $({\cal T}_t)_{t\geq0}$ be a
subordinator with L\'
{e}vy measure $\nu$, drift $\gamma$, Laplace exponent $\phi(\lambda)$
and transition function $\pi_t(ds)$. Let $(\calP_t)_{t\geq0}$ be a strongly
continuous contraction semigroup of linear operators on a Banach space
${\mathfrak B}$ with infinitesimal generator~$\calA$.

\begin{longlist}[(ii)]
\item[(i)] Define
%
%
\begin{equation}
\calP^\phi_tf(x)=\int_{[0,\infty)}
\calP_s f(x)\pi _t(ds),\qquad t\geq0, f\in {\mathfrak
B}.\label{subsem.1}
\end{equation}
Then $(\calP_t^\phi)_{t\geq0}$ is a strongly continuous contraction
semigroup of linear
operators on ${\mathfrak B}$ called {subordinate semigroup of $(\calP
_t)_{t\geq0}$
with respect to the subordinator} $({\cal T}_t)_{t\geq0}$.

\item[(ii)] Denote the infinitesimal generator of $({\cal P}_t^\phi)_{t\geq
0}$ by ${\cal A}^\phi$. Then the domain of ${\cal A}$ is a core of
${\cal A}^\phi$ and
%
%
\begin{equation}
{\cal A}^\phi f = \gamma\calA f + \int_{(0,\infty)} (
\calP_s f-f)\nu(ds),\qquad f\in\operatorname{Dom}(\calA). \label{subgen.1}
\end{equation}
\end{longlist}
\end{theorem}
We will need the following corollary.
%
\begin{corollary}\label{subFellerpospres} If $(\calP_t)_{t\geq0}$ is
a Feller
semigroup on $C([0,\infty])$, then the subordinate semigroup $(\calP
_t^\phi)_{t\geq0}$
is also a Feller semigroup on $C([0,\infty])$.
\end{corollary}

\begin{pf}
The space $C([0,\infty])$ consists of continuous functions
on $[0,\infty]$ or, equivalently, continuous functions on $(0,\infty)$
with finite limits at $0$ and $\infty$.
A~strongly continuous contraction semigroup on $C([0,\infty])$ is
Feller if it is positivity preserving. Suppose $(\calP_t)_{t\geq0}$
is Feller on
$C([0,\infty])$.
Then $(\calP_t^\phi)_{t\geq0}$ is a strongly continuous contraction
semigroup on
$C([0,\infty])$ by Theorem~\ref{phillips.theo}(i) with ${\mathfrak
B}=C([0,\infty])$. Since Bochner's integral in equation~\eqref
{subsem.1} is positivity preserving, for all $u\in C([0,\infty])$ such
that $0\leq u\leq1$, we have $0\leq{\cal P}_t^\phi u\leq1$. Thus
$(\calP_t^\phi)_{t\geq0}$ is positivity preserving and, hence,
Feller on $C([0,\infty
])$.
\end{pf}

Recall that $I = (0,\infty)$ if $0$ is unattainable and $I = [0,\infty
)$ if $0$ is reflecting. Under our assumptions, the transition kernels
of the semigroups $({\cal P}_t^\beta)_{t\geq0}$ have densities with
respect to the Lebesgue measure, $P^\beta_t(x,dy)=p^\beta(t,x,y)\,dy$,
where $p^\beta(t,x,y)$ are jointly continuous in $t,x,y$. This follows
from the fact that any one-dimensional diffusion has a density with
respect to the speed measure that is jointly continuous in $t,x,y$; cf.
\citet{mckean-1} or \citet{borodin-salminen}, page 13. Under our
assumptions, the speed measure is absolutely continuous with respect to
the Lebesgue measure [cf. \citet{borodin-salminen}, page 17], and hence
the semigroups have densities with respect to the Lebesgue measure.
For $\beta=0$ the density is the proper probability density on $I$,
$P_t^0(x,I)=\int_{I}p^1(t,x,y)\,dy=1$ for each $x\in I$. For $\beta>0$,
the density is generally defective, $P_t^\beta(x,I)=\int_{I}p^\beta
(t,x,y)\,dy\leq1$.
For notational convenience we extend the densities from $I$ to
${\mathbb R}$ by setting $p^\beta(t,x,y)\equiv0$ for $y<0$ for all
$x\in I$ and $t>0$.
We are now ready to give the Markovian characterization of the
time-changed process $(X^\phi,D^\phi)$ defined by~\eqref{eqr.8} based
on Phillips Theorem~\ref{phillips.theo} and Corollary~\ref{subFellerpospres}.

\begin{theorem}[{[Markovian
characterization of $(X^\phi,D^\phi)$]}]\label{mkvian.sub.biv}
\textup{(i)} The bi-variate process $(X^\phi,D^\phi)$ is a Feller process with
the Feller semigroup $({\cal P}^\phi_t)_{t\geq0}$ acting on $f\in
C([0,\infty]\times\{0,1\})$ by
%
%
\begin{equation}
\calP_t^\phi f(x,d)={\cal P}^{0,\phi}_tf_1(x)+(1-d){
\cal P}^{1,\phi
}_t(f_0-f_1) (x),
\label{semi.sub.bi.1}
\end{equation}
where $f_0(x)=f(x,0)\in C([0,\infty])$, $f_1(x)=f(x,1)\in C([0,\infty
])$ and $({\cal P}_t^{0,\phi})_{t\geq0}$ and
$({\cal P}_t^{1,\phi})_{t\geq0}$ are Feller semigroups obtained by
subordination in the sense of Bochner from Feller semigroups
$({\cal P}_t^{0})_{t\geq0}$ and
$({\cal P}_t^{1})_{t\geq0}$.

\textup{(ii)} The infinitesimal generator ${\cal A}^\phi$ of the Feller semigroup
$({\cal P}_t^\phi)_{t\geq0}$ has the following representation:
%
%
\begin{eqnarray}\label{subdeltsemi.3}
{\cal A}^\phi f(x,d)={\cal A}^{0,\phi
}f_1(x)+(1-d){
\cal A}^{1,\phi}(f_0-f_1) (x),
\nonumber
\\[-8pt]
\\[-8pt]
 \eqntext{f_0,f_1
\in \operatorname{Dom}\bigl({\cal A}^1\bigr),}
\end{eqnarray}
where ${\cal A}^{\beta,\phi}$, $\beta\in\{0,1\}$, are generators of
$({\cal P}_t^{\beta,\phi})_{t\geq0}$.

\textup{(iii)}
The generator ${\cal A}^{\beta,\phi}$ has the following L\'
{e}vy--Khintchine-type representations with state-dependent coefficients
%
%
\begin{eqnarray}
\label{subgen.a.1} {\cal A}^{\beta,\phi}f(x)&=&\frac{1}{2}\gamma
\sigma^2(x)f^{\prime
\prime
}(x)+b^{\beta
,\phi}(x)f^\prime(x)-
k^\phi(x)f(x)
\nonumber
\\[-8pt]
\\[-8pt]
\nonumber
&&{}+\int_{{\mathbb R}} \bigl(f(x+y)-f(x)-\mathbf{1}_{\{|y|\leq1\}}y
f^\prime (x) \bigr)\pi^{\beta,\phi}(x,y)\,dy
\end{eqnarray}
for all $ {f\in D({\cal A}^\beta)}$, where
the state-dependent L\'{e}vy density $\pi^{\beta,\phi}(x,y)$ is defined
for all $y\neq 0$ by
%
%
\begin{equation}
\pi^{\beta,\phi}(x,y)=\int_{(0,\infty)}p^\beta(s,x,x+y)
\nu(ds), \label{sublevmeas.1}
\end{equation}
and satisfies the integrability condition $\int_{{\mathbb
R}}(|y|^2\wedge1)\pi^{\beta,\phi}(x,y)\,dy<\infty$ for each $x\in I$
[recall that we extended $p(t,x,y)$ to ${\mathbb R}$ by setting
$p(t,x,y)=0$ for $y<0$],
the drift with respect to the truncation function $x\mathbf{1}_{\{|x|\leq
1\}}$ is given by
%
%
\begin{equation}
b^{\beta,\phi}(x)=\gamma b(x) + \int_{(0,\infty
)} \biggl(\int
_{\{
|y|\leq1\}} y p^\beta(s,x,x+y)\,dy \biggr)
\nu(ds),\label{subdrift.1}
\end{equation}
and the killing rate is given by
%
%
\begin{equation}
k^\phi(x)=\gamma\beta k(x) + \int_{(0,\infty)}
\bigl(1-P^\beta _s(x,I) \bigr)\nu(ds),\label{subkill.1}
\end{equation}
where $P^\beta_s(x,I)=\int_{I} p^1(s,x,y)\,dy$.

\textup{(iv)} If $f(x,d)\in D({\cal A})$ [i.e., $f$ is of the form \eqref
{eqr.2} with $f_0,f_1\in D({\cal A}^1)$] and $(X^\phi,D^\phi)$ starts
with $X^\phi_0=x>0$ and $D^\phi_0=d\in\{0,1\}$, then the process
%
%
\begin{equation}
M^f_t:=f\bigl(X^\phi_t,D^\phi_t
\bigr)-f(x,d)-\int_0^t {\cal A}^\phi
f\bigl(X^\phi _s,D^\phi_s
\bigr)\,ds \label{eqr.9}
\end{equation}
is an ${\mathbb H}^\phi$-martingale.
\end{theorem}
\begin{pf}
(i) The semigroup $({\cal P}^\phi_t)_{t\geq0}$ of
$(X^\phi,D^\phi)$ is Feller by Corollary~\ref{subFellerpospres}. The
explicit representation~\eqref{semi.sub.bi.1} for the semigroup results
from combining equation~\eqref{semi.bi.1} with~\eqref{subsem.1} of
Theorem~\ref{phillips.theo}.

(ii) Representation~\eqref{subdeltsemi.3} for the generator ${\cal
A}^\phi$ in terms of generators of subordinate semigroups $({\cal
P}^{\beta,\phi}_t)_{t\geq0}$ follows from~\eqref{subgen.1}.

(iii) Explicit representation~\eqref{subgen.a.1} for the generator
${\cal A}^{\beta,\phi}$
is shown as follows.
We start by observing that by Theorem 4.5 of \citet{mckean-1} for each
$x\in I$
the density $p^\beta(t,x,y)$ satisfies the following estimates:
%
%
\begin{eqnarray}
\int_{\{|y-x|>1\}}p^\beta(t,x,y)\,dy&\leq &C_1 t,
\label{est.1}
\\
\int_{\{|y-x|\leq1\}} (y-x)^2 p^\beta(t,x,y)\,dy
&\leq & C_2 t, \label{eqr.10}
\\
\biggl\llvert \int_{\{|y-x|\leq1\}} (y-x) p^\beta(t,x,y)\,dy
\biggr\rrvert &\leq& C_3 t, \label{eqr.11}
\\
1-\int_{I} p^\beta(t,x,y)\,dy&\leq&
C_4 t. \label{est.2}
\end{eqnarray}
For each $x\in I$ write
\begin{eqnarray*}
\label{teq.3} {\cal P}_s^\beta f(x)-f(x)&=&\int
_{{\mathbb R}} \bigl(f(x+y)-f(x)-\mathbf{ 1}_{\{|y|\leq1\}}y
f^\prime(x) \bigr)p^\beta(s,x,x+y)\,dy
\\
&&{}+ \biggl( \int_{\{|y|\leq1\}}y p^\beta(s,x,x+y)\,dy \biggr)
f^\prime(x) -\bigl(1-P_s^\beta(x,I)\bigr)f(x).
\end{eqnarray*}
Substitute the result into the Phillips
representation~\eqref{subgen.1} of the generator of the subordinate
semigroup, integrate term-by-term against the L\'{e}vy measure $\nu
(ds)$ of the subordinator and interchange the integration in $y$ and in
$s$ in the first of the three integrals. The result yields the
representation~\eqref{subgen.a.1}--\eqref{subkill.1}.
These operations are justified and the three resulting integrals are
well defined due to estimates~\eqref{est.1}--\eqref{est.2} and the
integrability of the L\'{e}vy measure of the subordinator, $\int_{(0,\infty)}(1\wedge{ s})\nu(d{ s})<\infty$.
Specifically, estimates~\eqref{est.1} and~\eqref{eqr.10} ensure that
the application of Fubini's theorem to interchange the integrations in
$s$ and $y$ is justified, and the resulting integral in~\eqref
{subgen.a.1} is well defined
for each $f\in D(\calA^\beta)$, as they ensure that the measure $\pi
^{\beta,\phi}(x,y)\,dy$ with the density~\eqref{sublevmeas.1} is a L\'
{e}vy measure for each $x\in I$
[it is similar to \citet{sato}, pages 200--201, proof that (30.8) is
the L\'{e}vy measure of the subordinate L\'{e}vy process].
Estimate~\eqref{eqr.11} ensures that the integral in~\eqref{subdrift.1}
is well defined [it is similar to \citet{sato}, proof that the drift
(30.9) of the subordinate L\'{e}vy process is well defined]. Finally,
estimate~\eqref{est.2} ensures that the integral in~\eqref{subkill.1}
is well defined due to the integrand tending to zero at the rate $s$ as
$s\rightarrow0$.

Part (iv) follows from \citet{ethier-kurtz-1}, Proposition 1.7, page 162.
\end{pf}

To obtain predictable characteristics of the semimartingale $(X^\phi
,D^\phi)$ from the explicit form of the Feller generator ${\cal
A}^\phi
$, it is convenient to first re-write the generator in the following
equivalent form.
%
\begin{corollary}[(Alternative representation
of the generator ${\cal A}^\phi$)]\label{altgen.1}
The generator ${\cal A}^\phi$ admits the following alternative representation:
%
%
\begin{eqnarray}
\label{eqr.12} &&{\cal A}^\phi f(x,d)\nonumber\\
&&\qquad=\frac{1}{2}\gamma
\sigma^2(x)\partial_x^2 f(x,d)+b^{0,\phi}(x)
\partial_x f(x,d) +(1-d)k^\phi(x)\partial_d
f(x,d)
\nonumber
\\[-8pt]
\\[-8pt]
\nonumber
&&\qquad\quad{}+\int_{{\mathbb R}^2}\bigl(f(x+y,d+z)-f(x,d)-y\mathbf{1}_{\{|y|\leq1\}
}
\partial_x f(x,d)-z\partial_d f(x,d)\bigr)\\
&&\hspace*{61pt}{}\times\Pi^\phi(x,d;dy\,dz),\nonumber
\end{eqnarray}
where
%
%
\begin{eqnarray}
\label{eqr.13}&& \Pi^\phi(x,d;dy \,dz)\nonumber\\
&&\qquad=(1-d)\gamma k(x)
\delta_0(dy)\delta_1(dz)
\nonumber
\\[-8pt]
\\[-8pt]
\nonumber
&&\qquad\quad{}+ \bigl[\pi^{0,\phi}(x,y)-(1-d) \bigl(\pi^{0,\phi}(x,y)-
\pi^{1,\phi
}(x,y)\bigr) \bigr]\,dy\, \delta_0(dz)
\\
&&\qquad\quad{}+(1-d)
\bigl(\pi^{0,\phi}(x,y)-\pi^{1,\phi}(x,y)\bigr)\,dy\,
\delta_1(dz),\nonumber
\end{eqnarray}
where $\pi^{\beta,\phi}(x,y)$ are the L\'{e}vy densities defined in
equation~\eqref{sublevmeas.1} with $\beta=0,1$, and $\delta_a$ is the
Dirac measure charging $a$.
\end{corollary}
\begin{pf}
Denote the operator defined by equations~\eqref
{subdeltsemi.3}--\eqref{subkill.1} by $\hat{{\cal A}}^\phi$. We need to
show that $\hat{{\cal A}}^\phi f(x,d)={\cal A}^\phi f(x,d)$ for all
$x\in I$ and $d\in\{0,1\}$, where ${\cal A}^\phi$ is the operator in
equation~\eqref{eqr.12}. The case with $d=1$ is immediate, $\hat
{{\cal
A}}^\phi f(x,1)={\cal A}^\phi f(x,1)={\cal A}^{0,\phi} f(x,1)$. Next
consider the case $d=0$. From~\eqref{subdeltsemi.3} we have
%
%
\begin{eqnarray}
\label{reqs.3}
\hat{\cal A}^\phi f(x,0)&=& \frac{1}{2}\gamma
\sigma^2(x) \partial_x^2 f(x,0)
\nonumber\\
&&{}+ b^{1,\phi}(x)\partial_x f(x,0)+\bigl(b^{0,\phi}(x)-b^{1,\phi
}(x)
\bigr)\partial _x f(x,1)\nonumber\\
&&{}+k^\phi(x) \bigl(f(x,1)-f(x,0)
\bigr)
\nonumber
\\[-8pt]
\\[-8pt]
\nonumber
&&{}+\int_{{\mathbb R}} \bigl(f(x+y,0)-f(x,0)-\mathbf{1}_{\{|y|\leq1\}
}y
\partial _x f(x,0) \bigr)\pi^{1,\phi}(x,y)\,dy
\\
&&{}+\int_{{\mathbb R}} \bigl(f(x+y,1)-f(x,1)-\mathbf{1}_{\{|y|\leq1\}
}y
\partial _x f(x,1) \bigr)\nonumber\\
&&\hspace*{25pt}{}\times \bigl(\pi^{0,\phi}(x,y)-
\pi^{1,\phi}(x,y)\bigr)\,dy.\nonumber
\end{eqnarray}
The last integral can be written as
%
%
\begin{eqnarray}
\label{reqs.7}
&&\int_{{\mathbb R}} \bigl(f(x+y,1)-f(x,1)-\mathbf{1}_{\{|y|\leq1\}
}y
\partial _x f(x,1) \bigr) \bigl(\pi^{0,\phi}(x,y)-
\pi^{1,\phi}(x,y)\bigr)\,dy
\nonumber\\
&&\qquad=\int_{{\mathbb R}} \bigl(f(x+y,1)-f(x,0)+f(x,0)-f(x,1)
\nonumber\\
&&\hspace*{11pt}\qquad\quad{}-\mathbf{1}_{\{|y|\leq1\}}y\bigl(\partial_x f(x,1)-
\partial_x f(x,0)+\partial_x f(x,0)\bigr) \bigr)\nonumber\\
&&\hspace*{11pt}\qquad\quad{}\times \bigl(
\pi^{0,\phi}(x,y)-\pi^{1,\phi}(x,y)\bigr)\,dy
\\
&&\qquad=\int_{{\mathbb R}} \bigl(f(x+y,1)-f(x,0)-{
\mathbf{1}}_{\{|y|\leq1\}
}y\partial _x f(x,0)-\partial_d
f(x,0) \bigr)
\nonumber\\
&&\hspace*{11pt}\qquad\quad{}\times\bigl(\pi^{0,\phi}(x,y)-\pi^{1,\phi}(x,y)\bigr)\,dy
\nonumber\\
&&\hspace*{11pt}\qquad\quad{}-\bigl(\partial_x f(x,1)-\partial_x f(x,0)\bigr)\int
_{{\mathbb R}}\mathbf{1}_{\{
|y|\leq1\}}y\bigl(\pi^{0,\phi}(x,y)-
\pi^{1,\phi}(x,y)\bigr)\,dy.\nonumber
\end{eqnarray}
From equation~\eqref{subdrift.1} observe that
\[
\int_{{\mathbb R}}\mathbf{1}_{\{|y|\leq1\}}y\bigl(\pi^{0,\phi}(x,y)-
\pi ^{1,\phi
}(x,y)\bigr)\,dy=b^{0,\phi}(x)-b^{1,\phi}(x).
\label{reqs.8}
\]
%
%
Substituting this result into \eqref{reqs.7} and substituting the
result into \eqref{reqs.3} and comparing with \eqref{eqr.12}--\eqref
{eqr.13}, we establish that $\hat{{\cal A}}^\phi f(x,0)={\cal A}^\phi
f(x,0)$.
\end{pf} 

Next we are ready to give semimartingale characterization of $(X^\phi
,D^\phi)$. For the definition of predictable characteristics of a
semimartingale, see \citet{jacod-shiryaev-1}, page 76.
%
\begin{theorem}[{[Semimartingale
characterization of $(X^\phi,D^\phi)$]}]\label{semchar.t.1}
\textup{(i)} The bi-variate ${\mathbb H}^\phi$-semimartingale $(X^\phi
,D^\phi
)$ has the following predictable characteristics.
The predictable quadratic variation of the continuous local martingale
component $X_t^{\phi,c}$ is
\[
C^{X^\phi X^\phi}_t=\int_0^t \gamma
\sigma^2\bigl(X_s^\phi\bigr)\,ds \label{eqr.16}
\]
($C^{D^\phi D^\phi}_t=0$ and $C^{X^\phi D^\phi}_t=0$ since $D^\phi$ is
purely discontinuous).
The predictable process of finite variation associated with the
truncation function $(h^{X^\phi}(x,d)=x\mathbf{1}_{\{|x|\leq1\}
},h^{D^\phi
}(x,d)=d)$ is
%
%
\begin{equation}
B_t^{X^\phi}=\int_0^t
b^{0,\phi}\bigl(X_s^\phi\bigr)\,ds,\qquad
B_t^{D^\phi}=\int_0^t
\bigl(1-D_s^\phi\bigr)k^\phi
\bigl(X_s^\phi\bigr)\,ds, \label{eqr.17}
\end{equation}
where the function $b^{0,\phi}(x)$ is defined in equation~\eqref
{subdrift.1}, and $k^\phi(X_s^\phi)$ is defined in equation~\eqref
{subkill.1}.
The compensator of the random measure $\mu(\omega;dt, dy \,dz)$
associated to the jumps of $(X^\phi,D^\phi)$ is a predictable random
measure on ${\mathbb R}_+ \times({\mathbb R}^2\backslash\{(0,0)\})$,
%
%
\begin{equation}
\nu(\omega;dt, dy \,dz) = \Pi^\phi\bigl(X^\phi_{t-},D^\phi_{t-};\,dy
\,dz\bigr)\,dt \label{eqr.18}
\end{equation}
with the measure $\Pi^\phi(x,d;dy \,dz)$ given by equation~\eqref{eqr.13}.

\textup{(ii)} The L\'{e}vy--It\^{o} canonical representation of $X^\phi$ with
respect to the truncation function $x \mathbf{1}_{\{|x|\leq1\}}$ is
\begin{eqnarray*}
\label{eqr.21} X^\phi_t&=&x+B_t^{X^\phi}+X_t^{\phi,c}
\\
&&{}+\int_0^t \int_{{\mathbb R}} y {
\mathbf{1}}_{\{|y|\leq1\}} \bigl(\mu ^{X^\phi
}(ds,dy)-\nu^{X^\phi}(ds,dy)
\bigr) \\
&&{}+\int_0^t \int_{{\mathbb R}}
y \mathbf{1}_{\{|y|>1 \}} \mu^{X^\phi}(ds,dy),
\end{eqnarray*}
where the compensator of the random measure $\mu^{X^\phi}(\omega;dt,
dy)$ associated to the jumps of $X^\phi$ is a predictable random
measure on ${\mathbb R}_+ \times({\mathbb R}\backslash\{0\})$,
%
%
\begin{equation}
\nu^{X^\phi}(\omega;dt,dy)=\pi^{0,\phi}\bigl(X^\phi_{t-},y
\bigr)\,dy \,dt, \label{eqr.22}
\end{equation}
where $\pi^{0,\phi}(x,y)$ is defined in equation~\eqref{sublevmeas.1}.

\textup{(iii)} The Doob--Meyer decomposition of $D^\phi_t$ is
\[
D^\phi_t=A_t^{D^\phi} +
M^\phi_t \label{eqr.23}
\]
with the martingale $M^\phi_t=D_t^\phi-A_t^{D^\phi}$ and the
predictable compensator $A_t^{D^\phi}= B_t^{D^\phi}$ given in
equation~\eqref{eqr.17}, so that the ${\mathbb H}^\phi$-intensity is
$\lambda_t^{\mathbb H^\phi}=(1-D_t^\phi)k^\phi(X_t^\phi)$.
\end{theorem}
\begin{pf}
(i) By Theorem 2.42 of \citet{jacod-shiryaev-1}, page 86, the
following two statements are equivalent: (i) the $n$-dimensional
semimartingale $Z$ admits characteristics $(B,C,\nu)$ with respect to
the truncation function $h$, and (ii) for each bounded function $f$
of class $C^2$ the process [using notation of equation (2.43) in \citet{jacod-shiryaev-1}, page
86]
%
%
\begin{eqnarray}
\label{mart}& &f(Z)-f(Z_0)-\sum_{i\leq n}
\partial_i f(Z_-)\bullet B^i -\frac
{1}{2}\sum
_{i,j\leq n}\partial_i\partial_jf(Z_-)
\bullet C^{ij}
\nonumber
\\[-8pt]
\\[-8pt]
\nonumber
&&\qquad{} - \biggl(f(Z_-+z)-f(Z_-)-\sum_{i\leq n}h^i(z)
\partial_i f(Z_-) \biggr)\star\nu
\end{eqnarray}
is a local martingale. In our case $Z=(X^\phi,D^\phi)$ is a
two-dimensional semimartingale such that
for any $f\in D({\cal A})$ the process~\eqref{mart} is a martingale.
Substituting expression~\eqref{eqr.12} for the generator into~\eqref
{eqr.9}, we immediately identify the characteristics of $(X^\phi
,D^\phi
)$ since the characteristics are unique (up to a null set).

(ii) The result is shown by observing that $X^\phi$ is itself
one-dimensional Markov with the generator ${\cal A}^{0,\phi}$ given
by~\eqref{subgen.a.1} with $\beta=0$, and identifying its predictable
characteristics $(B^{X^\phi},C^{X^\phi X^\phi}, \nu^{X^\phi})$ with
$\nu
^{X^\phi}$ given by~\eqref{eqr.22} from the generator ${\cal
A}^{0,\phi
}$, as we did in (i) for the bi-variate process. The canonical
representation of $X^\phi$ is then immediate by Theorem 2.34 of \citet{jacod-shiryaev-1}, page 84.

(iii) Immediate by Theorem 3.15 of \citet{jacod-shiryaev-1} (the
one-point point process $D^\phi$\vadjust{\goodbreak} is a class $D$ submartingale) and the
fact that $D^\phi=B^{D^\phi}+M^\phi$ is the canonical decomposition of
the special semimartingale $D^\phi$ [Proposition~2.29(a) of \citet{jacod-shiryaev-1}].
\end{pf} 


From Theorem~\ref{semchar.t.1}, we see that $(X^\phi,D^\phi)$ is a
Markovian It\^{o} semimartingale or It\^{o} process in the terminology
of \citet{cinlar-jacod-protter-sharpe-1}, page 165. In particular, when
$\gamma>0$, $X^\phi$ is an It\^{o} jump-diffusion with the continuous
local martingale component with quadratic variation $\gamma\int_0^t
\sigma^2(X_s)\,ds$ and with jumps with the predictable
compensator~\eqref
{eqr.22}. When $\gamma=0$, $X^\phi$ is a pure-jump process.
Recall that every It\^{o} semimartingale can be represented as a
solution of a stochastic differential equation driven by a standard
Brownian motion, Lebesgue measure and a Poisson random measure,
generally defined on an extended probability space [\citeauthor{cinlar-jacod-1} (\citeyear{cinlar-jacod-2,cinlar-jacod-1}),
\citet{jacod-protter-1}, Section~2.1.4]. If $\gamma>0$, we can thus represent the continuous local
martingale component as $X_t^{\phi,c}= \int_0^t \sqrt{\gamma}\sigma
(X^{\phi}_s)\,d\tilde{B}_s$, where $\tilde{B}$ is a standard Brownian
motion (possibly defined on an extended probability space). The jump
measure can be expressed in terms of a Poisson random measure. Such
explicit representation is useful in applications for Monte Carlo
simulation of It\^{o} semimartingales as solutions of SDEs [\citet{jacod-protter-1}]. Since our model arises as the time change, an
alternative way to simulate it is by simulating the ``background''
process $(X,D)$ and the independent subordinator~${\cal T}$.

We now formulate It\^{o}'s formula for functions of the bi-variate
process in the form convenient for our application. We first formulate
It\^{o}'s formula for functions of $X^\phi$ only.
%
\begin{theorem}[(It\^{o}'s formula for $X^\phi$)]\label{th.x.1}
Suppose $X^\phi$ starts from $X^\phi_0=x>0$.
For any function $f(t,x)\in C^{1,2}({\mathbb R}_+ \times(0,\infty))$
if zero is an unattainable boundary for the process $X$ or $f(t,x)\in
C^{1,2}({\mathbb R}_+ \times[0,\infty))$ if zero is an attainable
boundary for $X$,
It\^{o}'s formula can be written in the following form:
\begin{eqnarray*}
\label{eqr.20} f\bigl(t,X^\phi_t\bigr)&=&f(0,x)+\int
_0^t \biggl(\partial_s+
\frac{1}{2}\gamma \sigma ^2\bigl(X_s^\phi
\bigr)\partial_x^2+b^{0,\phi}\bigl(X_t^\phi
\bigr)\partial_x \biggr)f\bigl(s,X^\phi_s
\bigr)\,ds
\\
&&{}+\int_0^t \int_{{\mathbb R}}
\bigl(f\bigl(s,X^\phi_{s-}+y\bigr)-f\bigl(s,X^\phi
_{s-}\bigr)-y\partial_x f\bigl(s,X^\phi_{s-}
\bigr) \bigr)\\
&&\hspace*{38pt}{}\times\mathbf{1}_{\{|y|\leq1\}} \nu^{X^\phi}(ds, dy)
\\
&&{}+\int_0^t \int_{{\mathbb R}}
\bigl(f\bigl(s,X^\phi_{s-}+y\bigr)-f\bigl(s,X^\phi
_{s-}\bigr) \bigr)\mathbf{1}_{\{|y|> 1\}}\mu^{X^\phi}(ds,dy)
\\
&&{}+\int_0^t \int_{{\mathbb R}}
\bigl(f\bigl(s,X^\phi_{s-}+y\bigr)-f\bigl(s,X^\phi
_{s-}\bigr) \bigr)\\
&&\hspace*{38pt}{}\times\mathbf{1}_{\{|y|\leq1\}}\bigl(\mu^{X^\phi}(ds,
dy)-\nu^{X^\phi
}(ds, dy)\bigr)
\\
&&{}+\int_0^t\partial_x f
\bigl(s,X^\phi_s\bigr)\,dX^{\phi,c}_s,
\end{eqnarray*}
where $\mu^{X^\phi}$ is the random measure associated to jumps of
$X^\phi$, and $\nu^{X^\phi}$ is its compensator measure~\eqref{eqr.22}.
\end{theorem}
\begin{pf}
This form of It\^{o}'s formula based on characteristics
can be found in \citet{jacod-protter-1}, equation (2.1.20), page 32.
\end{pf}

This useful form of It\^{o}'s formula gives the canonical
representation of the semimartingale $f(t,X^\phi_t)$ in terms of the
predictable process of finite variation (``drift''), optional process of
finite variation (``large jumps''), continuous local martingale
component that is the stochastic integral with respect to $X^{\phi,c}$,
and the purely discontinuous local martingale that is the stochastic
integral with respect to the martingale random measure $\mu^{X^\phi
}-\nu
^{X^\phi}$ of compensated jumps of $X^\phi$ (``compensated small jumps'').
We are now ready to present It\^{o}'s formula for the bi-variate
process. Due to the decomposition
\[
f\bigl(t,X^\phi_t,D^\phi_t
\bigr)=f_1\bigl(t,X^\phi_t\bigr)+
\bigl(1-D^\phi_t\bigr) \bigl(f_0
\bigl(t,X^\phi _t\bigr)-f_1
\bigl(t,X^\phi_t\bigr)\bigr) \label{teq.9}
\]
and Theorem~\ref{th.x.1},
it is sufficient to give It\^{o}'s formula for the product $(1-D^\phi
_t)f(t,X^\phi_t)$.
%
\begin{theorem}[{[It\^{o}'s formula for
$(X^\phi
,D^\phi)$]}]\label{th.xD.1}
Suppose $(X^\phi,D^\phi)$ starts from $X^\phi_0=x>0$ and $D_0^\phi
=d\in
\{0,1\}$.
For any function $f(t,x)\in C^{1,2}({\mathbb R}_+ \times(0,\infty))$
if zero is an unattainable boundary for the diffusion process $X$ or
$f(t,x)\in C^{1,2}({\mathbb R}_+ \times[0,\infty))$ if zero is an
attainable boundary for $X$,
we have
%
%
\begin{eqnarray}
\label{itodecomp}&& \hspace*{-4pt}\bigl(1-D_t^\phi\bigr)f
\bigl(t,X^\phi_t\bigr)\nonumber\\
&&\hspace*{-4pt}\qquad=(1-d)f(0,x)
\nonumber\\
&&\hspace*{-9pt}\qquad\quad{}+\int_0^t \bigl(1-D_{s-}^\phi
\bigr) \biggl(\partial_s+\frac{1}{2}\gamma \sigma
^2\bigl(X_s^\phi\bigr)\partial_x^2+b^{1,\phi}
\bigl(X_t^\phi\bigr)\partial_x-k^\phi
\bigl(X_{s}^\phi\bigr) \biggr)f\bigl(s,X^\phi_s
\bigr)\,ds
\nonumber\\
&&\qquad\quad{}+\int_0^t \int_{{\mathbb R}}
\bigl(1-D_{s-}^\phi\bigr) \bigl(f\bigl(s,X^\phi
_{s-}+y\bigr)-f\bigl(s,X^\phi_{s-}\bigr)-y
\partial_x f\bigl(s,X^\phi_{s-}\bigr) \bigr)\nonumber\\
&&\hspace*{39pt}\qquad\quad{}\times\mathbf{
1}_{\{
|y|\leq1\}} \hat{\nu}(ds, dy)
\\
&&\hspace*{-9pt}\qquad\quad{}+\int_0^t \int_{{\mathbb R}}
\bigl(1-D_{s-}^\phi\bigr) \bigl(f\bigl(s,X^\phi
_{s-}+y\bigr)-f\bigl(s,X^\phi_{s-}\bigr) \bigr){
\mathbf{1}}_{\{|y|> 1\}}\hat{\mu}(ds,dy)
\nonumber\\
&&\hspace*{-9pt}\qquad\quad{}+\int_0^t\bigl(1-D_{s-}^\phi
\bigr)\partial_x f\bigl(s,X^\phi_s
\bigr)\,dX^{\phi
,c}_s-\int_0^t
\bigl(1-D_{s-}^\phi\bigr)f\bigl(s,X_{s-}^\phi
\bigr)\,dM_s^\phi
\nonumber\\
&&\hspace*{-9pt}\qquad\quad{}+\int_0^t \int_{{\mathbb R}}
\bigl(1-D_{s-}^\phi\bigr) \bigl(f\bigl(s,X^\phi
_{s-}+y\bigr)-f\bigl(s,X^\phi_{s-}\bigr) \bigr)\nonumber\\
&&\hspace*{-9pt}\hspace*{38pt}\qquad\quad{}\times{
\mathbf{1}}_{\{|y|\leq1\}} \bigl(\hat{\mu}(ds, dy)-\hat{\nu}(ds, dy)
\bigr),\nonumber
\end{eqnarray}
where we introduced a random measure associated to those jumps of
$X^\phi$ that do \emph{not} coincide with jumps of $D^\phi$,
\[
\hat{\mu}(\omega;ds,dy)=\sum_u {
\mathbf{1}}_{\{\Delta X_u^\phi(\omega
)\neq0\}} \mathbf{1}_{\{\Delta D_u^\phi(\omega)=0\}}\delta_{(u,\Delta
X_u^\phi
(\omega))}(ds,dy),
\label{teq.14}
\]
and its compensator measure
%
%
\begin{eqnarray}\label{teq.15}
\qquad&&\hat{\nu}(\omega;ds,dy)
\nonumber
\\[-8pt]
\\[-8pt]
\nonumber
&&\qquad= \bigl[\pi^{0,\phi}\bigl(X_{s-}^\phi
,y\bigr)-\bigl(1-D^\phi _{s-}\bigr) \bigl(
\pi^{0,\phi}\bigl(X_{s-}^\phi,y\bigr)-
\pi^{1,\phi}\bigl(X_{s-}^\phi,y\bigr)\bigr) \bigr]\,dy
\,ds.
\end{eqnarray}
\end{theorem}
\begin{pf}
We start with It\^{o}'s product rule,
\begin{eqnarray*}
\label{teq.16} \bigl(1-D_t^\phi\bigr)f
\bigl(t,X^\phi_t\bigr)&=&(1-d)f(0,x)+\int
_0^t \bigl(1-D_{s-}^\phi
\bigr)\,df\bigl(s,X_s^\phi\bigr)
\\
&&{}-\int_0^t f\bigl(s,X_{s-}^\phi
\bigr)\,dD_s^\phi-\sum_{s\leq t}
\Delta D^\phi_s \bigl(f\bigl(s,X_s^\phi
\bigr)-f\bigl(s,X_{s-}^\phi\bigr)\bigr).
\end{eqnarray*}
Due to Theorem~\ref{th.x.1}, the second term is
\begin{eqnarray*}
\label{teq.21}&& \int_0^t \bigl(1-D_{s-}^\phi
\bigr)\,df\bigl(X_s^\phi\bigr)\\
&&\qquad= \int_0^t
\bigl(1-D_{s-}^\phi\bigr) \biggl(\partial_s+
\frac{1}{2}\gamma\sigma ^2\bigl(X_s^\phi
\bigr)\partial_x^2+b^{0,\phi}\bigl(X_t^\phi
\bigr)\partial_x \biggr)f\bigl(s,X^\phi_s
\bigr)\,ds
\\
&&\qquad\quad{}+\int_0^t \int_{{\mathbb R}}
\bigl(1-D_{s-}^\phi\bigr) \bigl(f\bigl(s,X^\phi
_{s-}+y\bigr)-f\bigl(s,X^\phi_{s-}\bigr)-y
\partial_x f\bigl(s,X^\phi_{s-}\bigr) \bigr)\\
&&\qquad\quad{}\hspace*{38pt}\times\mathbf{
1}_{\{
|y|\leq1\}} \nu^{X^\phi}(ds, dy)
\\
&&\qquad\quad{}+\int_0^t \int_{{\mathbb R}}
\bigl(1-D_{s-}^\phi\bigr) \bigl(f\bigl(s,X^\phi
_{s-}+y\bigr)-f\bigl(s,X^\phi_{s-}\bigr) \bigr){
\mathbf{1}}_{\{|y|> 1\}}\mu^{X^\phi}(ds,dy)
\\
&&\qquad\quad{}+\int_0^t \int_{{\mathbb R}}
\bigl(1-D_{s-}^\phi\bigr) \bigl(f\bigl(s,X^\phi
_{s-}+y\bigr)-f\bigl(s,X^\phi_{s-}\bigr) \bigr)\\
&&\qquad\quad{}\hspace*{38pt}\times{
\mathbf{1}}_{\{|y|\leq1\}} \bigl(\mu ^{X^\phi
}(ds, dy)-\nu^{X^\phi}(ds, dy)
\bigr)
\\
&&\qquad\quad{}+\int_0^t\bigl(1-D_{s-}^\phi
\bigr)\partial_x f\bigl(s,X^\phi_s
\bigr)\,dX^{\phi,c}_s.
\end{eqnarray*}

The third term is
\begin{eqnarray*}
\label{teq.23}&& \int_0^t f\bigl(s,X_{s-}^\phi
\bigr)\,dD_s^\phi\\
&&\qquad=\int_0^t
\bigl(1-D_{s-}^\phi \bigr)f\bigl(s,X_{s-}^\phi
\bigr)\,dD_s^\phi
\\
&&\qquad=\int_0^t \bigl(1-D_{s-}^\phi
\bigr)f\bigl(s,X_{s-}^\phi\bigr)\,dM_s^\phi+
\int_0^t \bigl(1-D_{s-}^\phi
\bigr)f\bigl(s,X_{s-}^\phi\bigr)k^\phi
\bigl(X_{s-}^\phi\bigr)\,ds.
\end{eqnarray*}
The first equality is due to the fact that
\[
\int_0^t D_{s-}^\phi f
\bigl(X_{s-}^\phi\bigr)\,dD_s^\phi=
\sum_{s\leq t} D_{s-}^\phi f
\bigl(X_{s-}^\phi\bigr) \Delta D_s^\phi=0,
\label{teq.24}
\]
since $D_{s-}^\phi\Delta D_s^\phi=0$ (if $\Delta D_s^\phi=1$, then
$D_{s-}^\phi=0$). In the second equality we used the Doob--Meyer
decomposition of $D^\phi$.

The fourth term is (in the first equality we again use $D_{s-}^\phi
\Delta D_s^\phi=0$)
\begin{eqnarray*}
\label{teq.29} &&\sum_{s\leq t} \Delta
D^\phi_s \bigl(f\bigl(X_s^\phi
\bigr)-f\bigl(X_{s-}^\phi\bigr)\bigr)\\
&&\qquad=\sum
_{s\leq t} \bigl(1-D_{s-}^\phi\bigr)\Delta
D^\phi_s \bigl(f\bigl(X_s^\phi
\bigr)-f\bigl(X_{s-}^\phi\bigr)\bigr)
\\
&&\qquad=\int_0^t\int_{\mathbb R}
\bigl(1-D_{s-}^\phi\bigr) \bigl(f\bigl(X^\phi
_{s-}+y\bigr)-f\bigl(X^\phi _{s-}\bigr)\bigr)\\
&&\qquad\hspace*{38pt}{}\times{
\mathbf{1}}_{\{|y|\leq1\}}\bigl(\tilde{\mu}(ds,dy)-\tilde{\nu}(ds,dy)\bigr)
\\
&&\qquad\quad{}+\int_0^t\int_{\mathbb R}
\bigl(1-D_{s-}^\phi\bigr) \bigl(f\bigl(X^\phi
_{s-}+y\bigr)-f\bigl(X^\phi _{s-}\bigr)\bigr){
\mathbf{1}}_{\{|y|> 1\}}\tilde{\mu}(ds,dy)
\\
&&\qquad\quad{}+\int_0^t\int_{\mathbb R}
\bigl(1-D_{s-}^\phi\bigr) \bigl(f\bigl(X^\phi
_{s-}+y\bigr)-f\bigl(X^\phi _{s-}\bigr)-y
\partial_x f\bigl(s,X_{s-}^\phi\bigr)\bigr)\\
&&\qquad\quad\hspace*{38pt}{}\times{
\mathbf{1}}_{\{|y|\leq1\}}\tilde {\nu}(ds,dy)
\\
&&\qquad\quad{}+\int_0^t\bigl(1-D_{s-}^\phi
\bigr)\partial_x f\bigl(s,X_{s-}^\phi\bigr)\int
_{\mathbb R} y\mathbf{1}_{\{|y|\leq1\}}\tilde{\nu}(ds,dy),
\end{eqnarray*}
where we introduced a random measure associated to those jumps of
$X^\phi$ that occur contemporaneously with jumps of $D^\phi$,
\[
\tilde{\mu}(\omega;ds,dy)=\sum_u {
\mathbf{1}}_{\{\Delta X_u^\phi(\omega
)\neq0\}} \mathbf{1}_{\{\Delta D_u^\phi(\omega)=1\}}\delta_{(u,\Delta
X_u^\phi
(\omega))}(ds,dy),
\label{teq.30}
\]
and its compensator measure
\[
\tilde{\nu}(\omega;ds,dy)=\bigl(1-D^\phi_{s-}\bigr) \bigl(
\pi^{0,\phi
}\bigl(X_{s-}^\phi,y\bigr)-
\pi^{1,\phi}\bigl(X_{s-}^\phi,y\bigr)\bigr)\,dy \,ds.
\label{teq.31}
\]
To prove that this is the compensator of $\tilde{\mu}$, we note that
for any Borel set ${\cal B}\in{\mathbb R}\backslash\{0\}$ the process
\[
\tilde{\mu}_t({\cal B}) (\omega):=\tilde{\mu}\bigl(\omega;(0,t]\times {
\cal B}\bigr) \label{teq.32}
\]
is a one-point point process equal to one at time $t$ if $D^\phi
_t-D^\phi_0=1$ (i.e., a~jump of $D^\phi$ (default) occurs during the
time interval $(0,t]$) \emph{and} the process $X^\phi$ experiences a
jump at the time of default $\tau$ with size in ${\cal B}$, $\Delta
X_\tau^\phi\in{\cal B}$. The compensator of this process is readily
computed from the compensator $\nu$~\eqref{eqr.18} of the measure
$\mu$
associated to the jumps of $(X^\phi,D^\phi)$
\begin{eqnarray*}
\label{teq.33} \tilde{\nu}_t({\cal B})&=&\tilde{\nu}\bigl((0,t]\times{\cal
B}\bigr)
\\
&=&\int_0^t \int_{{\cal B}\times{\mathbb R}}yz
\nu(ds,dy\,dz)
\\
&=&\int_0^t \int_{{\cal B}}
\bigl(1-D_{s-}^\phi\bigr) \bigl(\pi^{0,\phi
}
\bigl(X_{s-}^\phi,y\bigr)-\pi^{1,\phi}
\bigl(X_{s-}^\phi,y\bigr)\bigr)\,dy,
\end{eqnarray*}
where the last equality follows by substituting equation~\eqref{eqr.13}
into equation~\eqref{eqr.18} and doing the integration.
Then $(\tilde{\mu}_t({\cal B})-\tilde{\nu}_t({\cal B}))+\tilde{\nu
}_t({\cal B})$ is the Doob--Meyer decomposition of $\tilde{\mu
}_t({\cal B})$.

We now put the pieces together and use the following identities to
combine similar terms and arrive at the final result~\eqref{itodecomp}.
First we observe that
\[
\mu^{X^\phi}=\hat{\mu}+\tilde{\mu}. \label{teq.34}
\]
This immediately follows from
\begin{eqnarray*}
\label{teq.36} &&\sum_u \mathbf{1}_{\{\Delta X_u^\phi(\omega)\neq0\}}
\delta _{(u,\Delta
X_u^\phi(\omega))}(ds,dy)
\\
&&\qquad=\sum_u \mathbf{1}_{\{\Delta X_u^\phi(\omega)\neq0\}} {
\mathbf{1}}_{\{
\Delta D_u^\phi(\omega)=0\}}\delta_{(u,\Delta X_u^\phi(\omega))}(ds,dy)
\\
&&\qquad\quad{}+\sum_u \mathbf{1}_{\{\Delta X_u^\phi(\omega)\neq0\}} {
\mathbf{1}}_{\{
\Delta D_u^\phi(\omega)=1\}}\delta_{(u,\Delta X_u^\phi(\omega))}(ds,dy)
\end{eqnarray*}
and accordingly for the compensators
\[
\nu^{X^\phi}=\hat{\nu}+\tilde{\nu}. \label{teq.37}
\]
These identities allow us to combine integrals with the same integrands
with respect to the random measures $\mu^{X^\phi}$ and $-\tilde{\mu}$
and $\nu^{X^\phi}$ and $-\tilde{\nu}$ into the ones with respect to
$\hat{\mu}$ and $\hat{\nu}$.

Finally, we use the identity
\begin{eqnarray*}
\label{teq.40} &&\int_{\mathbb R}y\mathbf{1}_{\{|y|\leq1\}}\tilde{\nu
}(ds,dy)\\
&&\qquad=\bigl(1-D_{s-}^\phi\bigr)\int_{\mathbb R}y{
\mathbf{1}}_{\{|y|\leq1\}}\bigl(\pi ^{0,\phi}\bigl(X^\phi_{s-},y
\bigr)-\pi^{1,\phi}\bigl(X^\phi_{s-},y\bigr)\bigr)\,dy
\\
&&\qquad=\bigl(1-D_{s-}^\phi\bigr)\int_{(0,\infty)}
\int_{\mathbb R}y\mathbf{1}_{\{
|y|\leq
1\}}\bigl(p^{0}
\bigl(u,X^\phi_{s-},X^\phi_{s-}+y
\bigr)\\
&&\hspace*{165pt}{}-p^1\bigl(u,X^\phi_{s-},X^\phi
_{s-}+y\bigr)\bigr)\,dy\nu(du)
\\
&&\qquad=\bigl(1-D_{s-}^\phi\bigr) \bigl(b^{0,\phi}
\bigl(X^\phi_{s-}\bigr)-b^{1,\phi}
\bigl(X^\phi _{s-}\bigr) \bigr)\,ds
\end{eqnarray*}
to simplify the drift. The interchange of integrations in $u$ and $y$
is allowed due to the estimate~\eqref{eqr.11}
and the integrability properties of the L\'{e}vy measure, $\int_{(0,\infty)}(1\wedge u)\nu(du)<\infty$.
\end{pf}

It\^{o}'s formula simplifies when the process $f(t,X_t^\phi,D_t^\phi)$
is a special semimartingale.

\begin{corollary}[{[It\^{o}'s formula for
$(X^\phi,D^\phi)$---Special semimartingale version]}]\label{cor.xD.1}
Suppose $(X^\phi,D^\phi)$ starts from $X^\phi_0=x>0$ and $D_0^\phi
=d\in
\{0,1\}$.
For any function $f(t,x,d)=f_1(t,x)+(1-d)(f_0(t,x)+f_1(t,x))$\vadjust{\goodbreak} with
$f_i(t,x)\in C^{1,2}({\mathbb R}_+ \times(0,\infty))$ if zero is an
unattainable boundary for the diffusion process $X$ or $f_i(t,x)\in
C^{1,2}({\mathbb R}_+ \times[0,\infty))$ if zero is an attainable
boundary for $X$, if $f(t,X_t^\phi,D_t^\phi)$ is a special semimartingale
[it suffices that \emph{either} $X^\phi$ is a special semimartingale
[i.e., $\int_{\mathbb R} (|y|^2\wedge|y|)\pi^{0,\phi}(x,y)\,dy<\infty$
for each $x\in I$ by Proposition 2.29 of \citet{jacod-shiryaev-1}, page
82] \emph{or} the functions $f_i(t,x)$ are bounded],
It\^{o}'s formula can be written in the following form:
\begin{eqnarray*}
\label{eqr.19.f2.b}&& f\bigl(t,X^\phi_t,D^\phi_t
\bigr)\\
&&\qquad=f(0,x,d)-\int_0^t \bigl(1-D^\phi
_{s-}\bigr) (f_0-f_1) \bigl(s,X^\phi_{s-}
\bigr)\,dM^\phi_s
\\
&&\qquad\quad{}+\int_0^t \bigl(\partial_s+{\cal
A}^{\phi} \bigr)f\bigl(s,X^\phi _s,D^\phi_t
\bigr)\,ds +\int_0^t\partial_x f
\bigl(s,X^\phi_s,D^\phi_t
\bigr)\,dX^{\phi,c}_s
\\
&&\qquad\quad{}+\int_0^t \int_{{\mathbb R}}
\bigl(f_1\bigl(s,X^\phi_{s-}+y
\bigr)-f_1\bigl(s,X^\phi _{s-}\bigr)\bigr)
\bigl(\mu^{X^\phi}(ds,dy)-\nu^{X^\phi}(ds,dy)\bigr)
\\
&&\qquad\quad{}+\int_0^t \int_{{\mathbb R}}
\bigl((f_0-f_1) \bigl(s,X^\phi
_{s-}+y\bigr)-(f_0-f_1) \bigl(s,X^\phi_{s-}
\bigr) \bigr) \bigl(1-D^\phi_{s-}\bigr)
\\
&&\hspace*{38pt}\qquad\quad{}\times\bigl(\hat{\mu}(ds,dy)-\hat{\nu}(ds,dy)\bigr),
\end{eqnarray*}
where $\mu^{X^\phi}$ is the random measure associated to jumps of
$X^\phi$, and $\hat{\mu}$ is the random measure associated to those
jumps of $X^\phi$ that do \emph{not} coincide with jumps of $D^\phi$,
and $\nu^{X^\phi}$ and $\hat{\nu}$ are their respective compensator
measures~\eqref{eqr.22} and~\eqref{teq.15}. The generator ${\cal
A}^\phi$ is given by equation~\eqref{subdeltsemi.3}.
\end{corollary}
\begin{pf}
The results follows immediately from Theorems~\ref{th.x.1}
and~\ref{th.xD.1}, expression for the generator~\eqref{subdeltsemi.3},
and the canonical decomposition for the special semimartingale; cf.
Proposition 2.29 of \citet{jacod-shiryaev-1}, page 82.
\end{pf}

This useful version of It\^{o}'s formula gives a canonical
decomposition of the special semimartingale $f(t,X_t^\phi,D_t^\phi)$
into the predictable process of finite variation (explicitly given in
terms of the generator ${\cal A}^\phi$ in the Markovian case considered
here), a continuous local martingale part and a purely discontinuous
local martingale part. The general form of it can be found in
Theorem 3.89 of \citet{jacod-1}, page 109.

Next we show the following useful sufficient condition for the
``specialness'' of the subordinate diffusion $X^\phi$.
%
\begin{theorem}[(Condition for specialness of
$X^\phi$)]\label{spec.sem}
If the diffusion $X$ has a stationary density
\[
\lim_{t\rightarrow\infty}p^0(t,x,y):=\pi(y) \label{limeq.1a}
\]
with the finite first moment $\int_{I}y\pi(y)\,dy<\infty$, then the
subordinate diffusion $X^\phi_t$ is a special semimartingale.
\end{theorem}
\begin{pf}
By Proposition 2.29 of \citet{jacod-shiryaev-1}, page 82, and given our
previous results,
it suffices to show that $\int_{\{|y|>1\}} |y|\pi^{0,\phi
}(x,y)\,dy<\infty
$ for each $x\in I$.
From \citet{mckean-1} [see also \citet{borodin-salminen}, page 13],
under our assumptions the transition density $p^0(t,x,y)$ can be
written in the form
$p^0(t,x,y)=m(y)p_{m}^0(t,x,y)$,
where $m$ is the speed density of the diffusion $X$ given by
%
%
\begin{equation}
m(x)=\frac{2}{\sigma^2(x)s(x)},\qquad s(x)=\exp \biggl\{-\int_{x_0}^x
\frac{2b(y)}{\sigma^2(y)}\,dy \biggr\}, \label{eq40}
\end{equation}
where $x_0>0$ in the definition of the scale density $s(x)$ is an
arbitrary point [see \citet{borodin-salminen} for the definitions of
the scale function and the speed measure of a one-dimensional
diffusion; under our assumptions the scale function and the speed
measure of $X$ are absolutely continuous with respect to the Lebesgue
measure with the densities given by equation~\eqref{eq40}],
and $p_{m}^0(t,x,y)=p_{m}^0(t,y,x)$ is symmetric and jointly continuous
in $t,x,y$. The diffusion $X$ admits a stationary density if and only
if the speed density is integrable on $I$ and, in this case, $\pi
(x)=m(x)/\int_I m(y)\,dy$ [cf. \citet{borodin-salminen}, page 20]. In
this case we can write the L\'{e}vy density of $X^\phi$ as
%
%
\begin{equation}
\pi^{0,\phi}(x,y)=\pi(x+y)\int_{(0,\infty)}p^0_m(s,x,x+y)
\nu(ds) \label{limeq.2.a}
\end{equation}
for all $y\neq 0$, where we chose $x_0$ in the definition of speed
density so that $\int_I m(y)\,dy=1$ and $\pi(x)=m(x)$. Since the function
$\int_{(0,\infty)}p^0_m(s,x,x+y)\nu(ds)$ is bounded on the set $\{
|y|>1\}$,
$\int_{\{|y|>1\}} |y|\pi^{0,\phi}(x,y)\,dy<\infty$ follows immediately
from the assumption $\int_{I}y\pi(y)\,dy<\infty$.
\end{pf}

We note that many diffusions $X$ used in default intensity models, such
as the CIR, the $3/2$, and the quadratic models given in the examples in
Section~\ref{Biv}, have stationary densities, so that the resulting
time changed processes $X^\phi$ turn out to be special semimartingales.
The canonical decomposition of the special semimartingale $X^\phi$ can
then be written in the following form:
\[
X^\phi_t=x+A_t^{X^\phi}+X_t^{\phi,c}+
\int_0^t \int_{{\mathbb R}} y \bigl(
\mu^{X^\phi}(ds,dy)-\pi^{0,\phi}\bigl(X^\phi_{t-},y
\bigr)\,dy \,ds \bigr) \label{specialcd}
\]
with the predictable finite variation part
\[
A_t^{X^\phi}=\int_0^t
\biggl(\gamma b\bigl(X_t^\phi\bigr)+\int
_{(0,\infty
)} \biggl(\int_{\mathbb R}yp^{0}
\bigl(u,X_t^\phi,X_t^\phi+y\bigr)\,dy
\biggr)\nu (du) \biggr)\,ds \label{limeq.2.b}
\]
with respect to the truncation function $h^{X^\phi}(x)=x$
[note that it differs from~\eqref{eqr.17} with respect to the
truncation function $h^{X^\phi}(x)=x\mathbf{1}_{\{|x|\leq1\}}$], the
continuous local martingale part that can be represented as $X_t^{\phi
,c}=\int_0^t \sqrt{\gamma} \sigma(X_s^\phi)\,d\tilde{B}_s$ and the purely
discontinuous local martingale with jumps with the compensator $\pi
^{0,\phi}(X^\phi_{t-},y)\,dy \,ds$. From Example~\ref{EG.2}, we observe
that the CIR diffusion satisfies the conditions of Theorem~\ref{spec.sem} for all $\kappa,\theta,\sigma>0$ without any further
conditions on the coefficients.

%

\section{Pricing credit-sensitive securities}\label{sectpricing}
We now discuss applications to the pricing of credit-sensitive
securities. We make the usual assumptions of frictionless
arbitrage-free markets, assume that the probability measure we are
working with is an equivalent martingale measure chosen by the market,
and that, under this probability measure, the default time $\tau$ of
the obligor is modeled by the jump time of the process $D^\phi$, that
is, $\tau=\inf\{t\geq0\dvtx D_t^\phi=1\}$ (the case of $D_0^\phi=1$ and,
hence, $\tau=0$, corresponds to the case when the obligor is already in
default at time zero).
Thus, the bivariate process $(X^\phi,D^\phi)$ under the EMM describes
all the financial information in our model relevant for the
risk-neutral pricing of credit-sensitive securities.
We remark that our model falls into the general framework of default
times of \citet{janson-mbay-protter-1}, with the underlying information
flow affecting default generated by a Markovian It\^{o} semimartingale
and with the compensator of the default indicator $D^\phi$ absolutely
continuous with respect to Lebesgue measure with intensity $(1-D_s^\phi
)k^\phi(X_t^\phi)$, which, in our case, is explicitly computed via the
application of the Phillips Theorem~\ref{phillips.theo}.

Consider a security with a promised payment $f_0(X^\phi_T)$ at maturity
$T>0$ if default does not occur by time $T$ and a ``recovery'' payment
$f_1(X^\phi_T)$ at maturity if default occurs. We generally allow the
promised payment to depend on the state variable at maturity. This is
the case when pricing options on credit spreads, where the credit
spread at option's maturity is the function of the credit state
variable at that time. This is also the case when pricing equity
options in unified credit-equity models, where the state variable also
drives the stock price observable up to the time of default.
Depending on the context of the model, the recovery payment at maturity
can be either taken constant, $f_1(x)=R$, if we do not assume that the
state variable $X^\phi_T$ is observable to the investor \emph{after}
default, or taken to be a function of the state variable at maturity if
the context of the model allows the investor to observe the state
variable after default.
In some applications, where the state variable drives the credit spread
prior to default or in the credit-equity modeling framework, where the
state variable drives the stock price prior to default, the recovery at
maturity is taken to be constant. On the other hand, if one considers
the framework where the firm defaults at time $\tau$ on its liabilities
but continues to operate through the reorganization process (such as
Chapter~11), and the final recovery settlement of the claims is made
based on the outcome of restructuring, then in such applications it may
make sense to model recovery as a function of the state variable at the
time of payment. Our mathematical framework can accommodate both types
of applications.

Thus, securities we consider are defined by payoff functions $f(x,d)$
with decomposition~\eqref{eqr.2}, where $f_0(x)$ is interpreted as the
promised payment if no default occurs by maturity and $f_1(x)$ as the
recovery paid at maturity if default occurs.
The \emph{defaultable zero-coupon bond} with unit face value is the
simplest such security with $f_0=1$ and constant recovery $f_1=R\in[0,1]$.
The security pricing in this model follows from the general results of
the previous section.
The payoff we consider is
%
%
\begin{equation}
f\bigl(X^\phi_T,D^\phi_T
\bigr)=f_1\bigl(X^\phi_T\bigr)-
\bigl(1-D^\phi_T\bigr) \bigl(f_1
\bigl(X^\phi _T\bigr)-f_0\bigl(X^\phi_T
\bigr)\bigr), \label{teq.41}
\end{equation}
at time $T$. The price process of the security with this payoff is
%
%
\begin{eqnarray}
\label{teq.42}&& f\bigl(t,X^\phi_t,D^\phi_t
\bigr)\nonumber\\
&&\qquad=e^{-r(T-t)}\bbE \bigl[f\bigl(X^\phi_T,D^\phi
_T\bigr)|{\cal H}^\phi _t \bigr]
\\
&&\qquad=e^{-r (T-t)}{\cal P}^{0,\phi}_{T-t}f_1
\bigl(X_t^\phi\bigr)+\bigl(1-D_t^\phi
\bigr)e^{-r
(T-t)}\calP^{1,\phi}_{T-t}(f_0-f_1)
\bigl(X_t^\phi\bigr),\nonumber
\end{eqnarray}
where $r\geq0$ is the risk-free interest rate assumed constant (but see
Remark~\ref{rfir} at the end of this section).
In particular, the price process of the \emph{defaultable zero-coupon
bond} with unit face value $f_0=1$ and zero recovery $f_1=0$ in the
event of default is
\[
\calZ\bigl(t,X^\phi_t,D^\phi_t;T
\bigr)=e^{-r(T-t)}\calQ \bigl(t,X^\phi_t,D^\phi_t;T
\bigr),\label{zerocupon.1a}
\]
where ${\cal Q}(t,X_t^\phi,D_t^\phi;T)$ is the \emph{survival
probability} to survive up to time $T$, given the state at time $t$,
%
%
\begin{equation}
\calQ\bigl(t,X^\phi_t,D^\phi_t;T
\bigr)=\bbE\bigl[\bigl(1-D^\phi _T\bigr)|{\cal
H}^\phi_t\bigr]=\bigl(1-D_t^\phi
\bigr)P_{T-t}^{1,\phi}\bigl(X_t^\phi,I
\bigr), \label{surv.p.1b}
\end{equation}
where $P_{t}^{1,\phi}(x,I)={\cal P}_t^{1,\phi}1(x)$.
The term structure of credit spreads for defaultable bonds of all
maturities as observed at time $t$, given the state $X_t^\phi$ and
$D_t^\phi=0$, is
%
%
\begin{equation}
{\cal S}\bigl(t,X_t^\phi;T\bigr)=-\frac{1}{(T-t)} \ln
\calP^{1,\phi
}_{T-t}1\bigl(X_t^\phi\bigr).
\label{spread.1a}
\end{equation}

In those applications where the recovery at maturity is assumed
constant, $f_1=R$, the pricing formula simplifies to
%
%
\begin{eqnarray}\label{teq.43}
f\bigl(t,X^\phi_t,D^\phi_t\bigr)
&=&e^{-r (T-t)}\bigl(1-D_t^\phi\bigr)
\calP^{1,\phi}_{T-t}f_0\bigl(X_t^\phi
\bigr)
\nonumber
\\[-8pt]
\\[-8pt]
\nonumber
&&{}+e^{-r
(T-t)}R \bigl(1 -\calQ\bigl(t,X^\phi_t,D^\phi_t;T
\bigr)\bigr),
\end{eqnarray}
and the investor who observes the price processes of traded securities
in this market can determine whether or not default has occurred, as
well as can filter out the state variable $X^\phi$ prior to default
$\tau$ from the prices of traded securities.
In this case, when the recovery payment is not allowed to depend on
$X^\phi_T$ (assumed unobservable in such applications), the investor's
filtration is smaller than the filtration generated by $(X_t^\phi
,D_t^\phi)$ since $X_t^\phi$ is only observed by the investor prior to
default time $\tau$. In fact, in such applications the investor's
filtration can be identified with the filtration generated by $(Y^\phi
_t,D_t^\phi)$, where the process
$Y_t^\phi:=(1-D_t^\phi) X_t^\phi$ jumps to zero at default and stays there.
Applying It\^{o}'s formula in the form of Theorem~\ref{th.xD.1}, this
semimartingale has the canonical representation [$Y^\phi_0=(1-D_0^\phi
)X_0^\phi$]
%
%
\begin{eqnarray}
\label{teq.45} Y^\phi_t&=&Y^\phi_0+
\int_0^t \bigl(1-D_{s-}^\phi
\bigr) \bigl(b^{1,\phi
}\bigl(Y_{s-}^\phi
\bigr)-k^\phi\bigl(Y_{s-}^\phi
\bigr)Y_{s-}^\phi \bigr) \,ds
\nonumber\\
&&{}+\int_0^t \int_{{\mathbb R}} y{
\mathbf{1}}_{\{|y|>1\}} \bigl(1-D_{s-}^\phi\bigr) \hat{\mu}(ds
\,dy)
\nonumber
\\[-8pt]
\\[-8pt]
\nonumber
&&{}+\int_0^t \int_{{\mathbb R}} y{
\mathbf{1}}_{\{|y|\leq1\}
}\bigl(1-D_{s-}^\phi\bigr) \bigl(\hat{
\mu}(ds \,dy)-\pi^{1,\phi}\bigl(Y_{s-}^\phi,y\bigr)\,dy\,ds
\bigr)
\\
&&{}+\int_0^t \bigl(1-D_{s-}^\phi
\bigr)\,dX_s^{\phi,c} -\int_0^t
Y_{s-}^\phi \,dM_s^\phi.\nonumber
\end{eqnarray}
This canonical representation decomposes $Y^\phi$ into the ``drift,''
``large jumps'' prior to default,
a purely discontinuous local martingale of ``small jumps'' prior to
default with the compensator measure $(1-D_{s-}^\phi)\pi^{1,\phi
}(X_{s-}^\phi,y)\,dy\,ds$ [observe from equation~\eqref{teq.15} that
$(1-D_{s-}^\phi)\hat{\nu}(ds,dy)=(1-D_{s-}^\phi)\pi^{1,\phi
}(X_{s-}^\phi
,\break y)\,dy\,ds$],
a continuous local martingale component that can be further represented
as $\int_0^t (1-D_{s-}^\phi)\sqrt{\gamma}\sigma(Y_{s-}^\phi
)\,d\tilde
{B}_s$ in terms of a Brownian motion, and a final jump to zero (the
default term $-\int_0^t Y_s^\phi \,dM_s^\phi$). In the credit-equity
context, one identifies the process $Y^\phi_t$ with the defaultable
stock price process; see, for example, \citet{mendoza-carr-linetsky-1}
and \citet{linetsky-mendoza-arriaga-1} for the multi-firm case.
We further remark that \citet{lorig-mendoza-1} apply the canonical
representation~\eqref{teq.45} of the stock price process $Y^\phi_t$ to
the valuation of variance swaps on individual stocks with the risk of
bankruptcy.


So far we have considered recovery payments at maturity. Recovery at
the time of default can also be treated in our framework. Suppose that
if default occurs prior to maturity $T$, the recovery is received by
the investor at the time of default $\tau$ and is equal to some
function of the state variable $X_{\tau}^\phi$ at the time of default,
${\cal R}(X_{\tau}^\phi)$. By the standard calculation in credit risk
modeling [cf. Lemma 7.3.4.3(i) in \citet{jeanblanc-yor-chesney-1},
page~421], the value of such recovery at time $t$ prior to maturity $T$
is then given by
\begin{eqnarray*}
\label{teq.52} {\mathbb E}\bigl[e^{-r(\tau-t)}{\cal R}\bigl(X_{\tau}^\phi
\bigr)|{\cal H}^\phi_t\bigr] &=&\bigl(1-D_{t}^\phi
\bigr)\int_t^T e^{-r(u-t)}{\cal
P}^{1,\phi}_{u-t}\bigl({\cal R}\cdot k^\phi\bigr)
\bigl(X_t^\phi\bigr)\,du
\\
&&{}+e^{r(t-\tau)}{\cal R}\bigl(X_{\tau}^\phi
\bigr)D^\phi_t,
\end{eqnarray*}
where $({\cal R}\cdot k^\phi)(x)={\cal R}(x)k^\phi(x)$.

\begin{remark}[(Risk-free interest rates)]\label{rfir}
We remark that
stochastic risk-free interest rates can be handled in our subordinate
diffusion framework as follows.
The subordinate semigroup $({\cal P}^{1,\phi}_t)_{t\geq0}$ is taken to
be the \emph{pricing semigroup}. Namely, the state variable $Z^\phi$
driving the term structure of interest rates is assumed to be a
Markovian It\^{o} semimartingale with the following dynamics under the
equivalent martingale measure:
\begin{eqnarray*}
\label{teq.45.b} Z^\phi_t&=&Z^\phi_0+
\int_0^t b^{1,\phi}\bigl(Z_{s-}^\phi
\bigr) \,ds+ \int_0^t \int_{{\mathbb R}}
y\mathbf{1}_{\{|y|>1\}} \mu^{Z^\phi
}(ds,dy)+Z_t^{\phi,c}
\\
&&{}+\int_0^t \int_{{\mathbb R}} y{
\mathbf{1}}_{\{|y|\leq1\}} \bigl(\mu ^{Z^\phi}(ds,dy)-\pi^{1,\phi}
\bigl(Z_{s-}^\phi,y\bigr)\,dy\,ds \bigr),
\end{eqnarray*}
where $Z_t^{\phi,c}=\int_0^t \sqrt{\gamma}\sigma(Z_{s}^\phi
)\,d\tilde
{B}_s$ with a standard Brownian motion $\tilde{B}$.
The random measure $\mu^{Z^\phi}$ on ${\mathbb R}_+ \times({\mathbb
R}\backslash\{0\})$ associated to jumps of $Z^\phi$ has a compensator
$\nu^{Z^\phi}(ds,dy)=\pi^{1,\phi}(Z_{s-}^\phi,y)\,dy\,ds$ with $\pi
^{1,\phi
}(x,y)$ given by equation~\eqref{sublevmeas.1} with $\beta=1$. The
function $b^{1,\phi}(x)$ in the drift is given by equation~\eqref
{subdrift.1} with $\beta=1$.
Similarly to Theorem~\ref{spec.sem}, it is easy to show the following.
%
\begin{proposition}[(Condition for
specialness of $Z^\phi$)]\label{cond.spec.prop}
If
\[
\lim_{t\rightarrow\infty}p^0(t,x,y):=\pi(y) \label{limeq.1b}
\]
with the finite first moment $\int_{I}y\pi(y)\,dy<\infty$, then
$Z^\phi$
is a special semimartingale.
\end{proposition}
\begin{pf}
Recall that for the density of the Feynman--Kac semigroup ${\cal P}^1$
we have [cf. \citet{revuz-yor}, page 358]
\[
p^1(t,x,y)= \bbE_x\bigl[e^{-\int_0^tk(X_u)\,du}|X_t=y
\bigr]p^0(t,x,y)\leq p^0(t,x,y)
\]
for each $x,y\in I$ and $t>0$.
Under our assumptions, this implies that $\int_{\mathbb R}(|y|^2\wedge
|y|)\pi^{1,\phi}(x,y)\,dy\leq\int_{\mathbb R}(|y|^2\wedge|y|)\pi
^{0,\phi
}(x,y)\,dy<\infty$, where the second inequality follows from the proof of
Theorem~\ref{spec.sem}.
Thus $Z^\phi$ is special by Proposition 2.29 of \citet{jacod-shiryaev-1}, page 82.
\end{pf}

In the case of special $Z^\phi$, the canonical decomposition of
$Z^\phi
$ reads
\[
Z^\phi_t=x+A_t^{Z^\phi}+Z_t^{\phi,c}+
\int_0^t \int_{{\mathbb R}} y \bigl(
\mu^{Z^\phi}(ds,dy)-\pi^{1,\phi}\bigl(Z^\phi_{t-},y
\bigr)\,dy \,ds \bigr) \label{specialzcd}
\]
with the predictable finite variation part
\[
A_t^{Z^\phi}=\int_0^t
\biggl(\gamma b\bigl(Z_t^\phi\bigr)+\int
_{(0,\infty
)} \biggl(\int_{\mathbb R}yp^{1}
\bigl(u,Z_t^\phi,Z_t^\phi+y\bigr)\,dy
\biggr)\nu (du) \biggr)\,ds \label{limeq.6}
\]
with respect to the truncation function $h^{X^\phi}(x)=x$, the
continuous local martingale part that can be represented as $Z_t^{\phi
,c}=\int_0^t \sqrt{\gamma} \sigma(Z_s^\phi)\,d\tilde{B}_s$ and the purely
discontinuous local martingale with jumps with the compensator $\pi
^{1,\phi}(Z^\phi_{t-},y)\,dy \,ds$.
Most popular short-rate diffusions, such as CIR, $3/2$, etc., have
stationary densities due to mean-reversion. By Proposition~\ref{cond.spec.prop}, the corresponding subordinate short-rate models are
driven by jump-diffusion or pure jump processes $Z^\phi$ that are
special semimartingales.
The short rate process is taken to be
\[
r_t=k^\phi\bigl(Z^\phi_t\bigr),
\label{limeq.7}
\]
where $k^\phi(x)$ is given by equation~\eqref{subkill.1}, and the money
market account is $A_t=e^{\int_0^t r_s \,ds}=e^{\int_0^t k^\phi(Z^\phi_s)\,ds}$.
The \emph{pricing semigroup} is then the semigroup $({\cal P}^{1,\phi
}_t)_{t\geq0}$ with generator ${\cal A}^{1,\phi}$, the subordinate
semigroup of the Feynman--Kac semigroup $({\cal P}^{1}_t)_{t\geq0}$
with generator ${\cal A}^1$ and, in particular, for the risk-free
zero-coupon bond we have
\[
P\bigl(Z_t^\phi,t;T\bigr)={\cal P}_{T-t}^{1,\phi}
1\bigl(Z^\phi_t\bigr)=P^{1,\phi
}_{T-t}
\bigl(Z_t^\phi,I\bigr). \label{teq.55}
\]
%

Now an extension to the combined model that includes both the
subordinate diffusion risk-free interest rate model and the subordinate
default intensity model is immediate, as long as the interest rate
model and the default model are assumed independent. At the expense of
increased complexity, dependence can be further introduced either by
starting with independent factors, each following a subordinate
diffusion, and then combining them in a multi-dimensional model by
taking linear combinations of independent factors, or by means of \emph{multivariate subordination} as in \citet{linetsky-mendoza-arriaga-1}.
\end{remark}

\section{Eigenfunction expansions of subordinate semigroups}\label{sub.spec.dec}

We now show how to explicitly compute the semigroups $({\cal
P}_t^{\beta
,\phi})_{t\geq0}$ by the eigenfunction expansion method.
We start by observing that for any $f\in C^2_c(I)$ the infinitesimal
generator ${\cal A}^\beta$ of $({\cal P}_t^{\beta})_{t\geq0}$ can be
re-written in the \emph{formally} self-adjoint form using the scale and
speed densities~\eqref{eq40}
\[
{\cal A}^\beta f(x)=\frac{1}{m(x)} \biggl(\frac
{f'(x)}{s(x)}
\biggr)'-\beta k(x)f(x).\label{eq41}
\]
Indeed, ${\cal A}^\beta$ can be extended to a self-adjoint operator in the
Hilbert space $L^2(I,m)$ of functions on $I$ square-integrable with the
speed measure $m(dx) = m(x)\,dx$ and endowed with the inner product
\[
(f,g)=\int_I f(x)g(x)m(x)\,dx.\label{eq42}
\]
Furthermore, the restriction of $({\cal P}_t^{\beta})_{t\geq0}$ to
$C([0,\infty])\cap L^2(I,m)$ can then be extended to a
strongly-continuous semigroup of symmetric contractions in the Hilbert
space $L^2(I,m)$.
Thus, the spectral theorem for self-adjoint operators in Hilbert space
can be applied to write down the spectral decomposition of ${\cal
A}^\beta$ and $({\cal P}_t^{\beta})_{t\geq0}$. The spectral
representation for one-dimensional diffusions goes back to the
classical work of \citet{mckean-1} [see also \citet{ito-mckean-1},
Section~4.11]. More generally, one-dimensional diffusions are examples
of \emph{symmetric Markov processes} whose transition semigroups admit
symmetric extensions to the Hilbert space $L^2(E,m)$, where $E$ is the
state space of the Markov process, and $m$ is a positive Radon measure
on $E$ with full support. \citet{fukushima-oshima-takeda-1} and \citet{chen-fukushima-1} are the standard references on the subject.
In the case of one-dimensional diffusions, $E=I$ is the interval on the
real line, and $m$ is the speed measure.
An excellent exposition of the spectral theorem and applications to
subordination can be found in \citet{schilling-song-vondraceck-1},
Chapters~10 and 11. Surveys of applications of the spectral expansion
method to diffusion models in finance can be found in
\citeauthor{li-linetsky-2} (\citeyear{linetsky-2,linetsky-3}), where an extensive bibliography is
given. Recent applications of subordinate diffusion models in finance
can be found in \citet{boyarchenko-levendor-2}, \citet{mendoza-carr-linetsky-1}, \citeauthor{li-linetsky-2} (\citeyear{li-linetsky-2,li-linetsky-1}),
\citet{linetsky-mendoza-arriaga-1}, \citet{lim-li-linetsky-1}. Here we give
a brief account limited to needs of the present paper.

For computational simplicity here we limit ourselves to the special
case when the diffusion $X$ and the function $k$ are such that $({\cal
P}_t^{\beta})_{t\geq0}$ in $L^2(I,m)$ are \emph{trace-class semigroups}
for $\beta\geq0$, that is, the operators ${\cal P}_t^\beta$ are
trace-class for all $t>0$ and $\beta\geq0$.
Recall that for a positive semi-definite operator $A$ on a separable
Hilbert space ${\cal H}$, the trace of $A$ is defined by $\operatorname{tr}
A=\sum_{n=1}^\infty(\varphi_n,A\varphi_n)\in[0,\infty]$, where
$\varphi_n$
is some orthonormal basis in ${\cal H}$. The trace is independent of
the orthonormal basis chosen; cf. \citet{reed-simon-1}, page 206. A
positive semi-definite operator is called \emph{trace-class} if and
only if its trace is finite. The semigroup operators $\mathcal
{P}_t^\beta$ are positive
semi-definite. Under the assumption that $\mathcal{P}_t^\beta$ are
trace-class for all $t>0$, the spectra of each $\mathcal{P}_t^\beta$,
as well as of the generators ${\cal A}^\beta$ of the semigroups
$(\mathcal{P}_t^\beta)_{t\geq0}$ in $L^2(I,m)$, are purely discrete
with eigenvalues $(e^{-\lambda_n^\beta t})_{n\geq1}$ (for $t>0$) and
$(-\lambda^{\beta}_n)_{n\geq1}$
respectively, and
%
%
\begin{equation}
\label{eq:TraceClassCondition} \operatorname{tr} {\cal P}^{\beta}_t=\sum
_{n=1}^\infty e^{-\lambda_n^\beta t}<\infty
\end{equation}
for all $t>0$; cf. Lemma 7.2.1 of \citet{davies-3}.
Here
$0\leq\lambda_1^\beta\leq\lambda_2^\beta\leq\cdots$ are arranged in
increasing
order and repeated according to multiplicity. Then the function ${\cal
P}_t^\beta f(x)$ has
an eigenfunction expansion of the form
%
%
\begin{eqnarray}
\label{eq:PrtEigenfunctionExpansion} \mathcal{P}_t^\beta f(x)=\sum
_{n=1}^\infty f_n^\beta
e^{-\lambda_n^\beta t}\varphi_n^\beta(x), \quad f_n^\beta=
\bigl(f,\varphi _n^\beta \bigr)
\nonumber
\\[-8pt]
\\[-8pt]
\eqntext{\mbox{for any } f\in
L^2(I,m) \mbox{ and all } t\geq0,}
\end{eqnarray}
where $\varphi_n^\beta$ is the $n$th-eigenfunction
%
%
\begin{equation}
{\cal P}_t^\beta\varphi_n^\beta=
e^{-\lambda_n^\beta t}\varphi _n^\beta\quad\mbox{and}\quad {\cal
A}^\beta\varphi_n = -\lambda_n^\beta
\varphi _n^\beta. \label{eigenfuncs}
\end{equation}
The eigenfunctions
$(\varphi_n^\beta)_{n\geq1}$ form a complete orthonormal basis
in $L^2(I,m)$, and $f_n^\beta$ is the $n$th expansion coefficient in this
basis.

For a trace-class semigroup, each $\mathcal{P}^\beta_t$ with $t>0$
admits a symmetric kernel $p^\beta_m(t,x,y)\in
L^2(I\times I,m\times m)$ with respect to the measure $m$ [i.e.,
$p^\beta_m(t,x,y)=p^\beta_m(t,y,x)$,
$\mathcal{P}_t^\beta f(x)=\int_I p^\beta_m(t,x,y)f(y)m(dy)$ for
$f\in
L^2(I,m)$, and $\int_{I\times I}(p^\beta_m(t,x,y))^2
m(dx)m(dy)<\infty$],
which has the following bi-linear expansion:
%
%
\begin{equation}
\label{eq:SymTPDBilinearExpansion} p^\beta_m(t,x,y)=\sum
_{n=1}^\infty e^{-\lambda_n^\beta t}\varphi ^\beta
_n(x)\varphi^\beta_n(y).
\end{equation}
The expansions in
\eqref{eq:PrtEigenfunctionExpansion} and
\eqref{eq:SymTPDBilinearExpansion} in general converge under the
$L^2(I,m)$ and $L^2(I\times I,m\times m)$ norms, respectively.
Moreover, since for one-dimensional diffusions
for each
$t>0$ the kernel $p_m^\beta(t,x,y)$ with respect to the speed measure
is jointly continuous in $x$ and $y$ (and $t$) by the results of \citet{mckean-1},
then
each eigenfunction $\varphi^\beta_n$ is continuous,
and satisfies the estimate
\[
\bigl\llvert \varphi_n^\beta(x)\bigr\rrvert \leq
e^{\lambda^\beta_nt/2}\sqrt {p^\beta _m(t,x,x)}
\label{limeq.8}
\]
for all $n$, $x$ and $t>0$.
Moreover, for any $f\in L^2(I,m)$, expansion
\eqref{eq:PrtEigenfunctionExpansion} converges uniformly in $x$ on
compacts for each $t>0$ to the function $\mathcal{P}^\beta_tf(x)$
continuous in~$x$,
and the bi-linear expansion \eqref{eq:SymTPDBilinearExpansion}
converges uniformly on compacts; cf. Theorem~7.2.5 of \citet{davies-3}.

The spectral representation for the density of a 1D diffusion with
respect to the speed measure was obtained by \citet{mckean-1}; see also
\citet{ito-mckean-1}, Section~4.11. In general, the spectrum contains
some continuous spectrum, and the spectral representation is in terms
of the integral with respect to the spectral measure. Nevertheless,
many diffusions arising in finance applications have purely discrete
spectra with explicitly known eigenfunctions and eigenvalues satisfying
the trace class condition~\eqref{eq:TraceClassCondition} for all $t>0$,
including OU, CIR, CEV and JDCEV diffusions; see surveys \citeauthor{linetsky-2}
(\citeyear{linetsky-2,linetsky-3})
and references therein for finance applications.

We now summarize key results about the eigenfunction expansion of the
subordinate semigroups $({\cal P}_t^{\beta,\phi})_{t\geq0}$ defined in
Section~\ref{sub.sec}.
%
\begin{theorem}\label{traceclass}
Suppose the semigroup $({\cal P}_t^\beta)_{t\geq0}$ defined in
Section~\ref{Biv} is trace-class with eigenvalues and eigenfunctions
$e^{-\lambda_n^\beta t}$ and $\varphi_n^\beta(x)$, respectively.
Further suppose that the eigenfunctions have bounds
%
%
\begin{equation}
\bigl|\varphi_n^\beta(x)\bigr|\leq C^\beta_K
\label{eigenbound}
\end{equation}
on each compact set $K\subset I$ with $C_K^\beta$ independent of $n$
but possibly dependent on~$K$.
Let ${\cal T}$ be a subordinator with the Laplace exponent satisfying
the following condition for all $t>0$:
%
%
\begin{equation}
\sum_{n=1}^\infty e^{-\phi(\lambda^\beta_{n})t}<\infty.
\label{trace.mult.1}
\end{equation}
Then the subordinate semigroup $({\cal P}_t^{\beta,\phi})_{t\geq0}$ is
a strongly continuous semigroup of symmetric contractions on $L^2(I,m)$,
trace-class for all $t>0$ with the eigenvalues $e^{-\phi(\lambda
^\beta
_{n})t}$ and normalized eigenfunctions
$\varphi^\beta_{n}(x)$, and possesses
a continuous in $x, y$ density with respect to the speed measure
$m(dx)$ that is given by the bi-linear expansion
%
%
\begin{equation}
p^{\beta,\phi}_m(t,x,y)= \sum_{n=0}^\infty
e^{-\phi(\lambda^\beta
_{n})t}\varphi^\beta_{n}(x)\varphi^\beta_{n}(y)
\label{transmult.sub.1}
\end{equation}
uniformly convergent in $x,y$ on compacts in $I\times I$ for all $t>0$.
For each $f\in L^2(I,m)$ and $t>0$ the function ${\cal P}^{\beta,\phi
}_t f(x)$ has the eigenfunction expansion
%
%
\begin{equation}
{\cal P}_t^{\beta,\phi} f (x)=\sum_{n=1}^\infty
e^{-\phi(\lambda
^\beta
_{n})t}f_n^\beta\varphi^\beta_{n}(x),\qquad
f_n^\beta= \bigl(f,\varphi _{n}^\beta
\bigr) \label{eigenfsub}
\end{equation}
uniformly convergent in $x$ on compacts in $I$.
\end{theorem}

Without bound~\eqref{eigenbound} on the eigenfunctions $\varphi
_n^\beta
(x)$ and the trace-class condition~\eqref{trace.mult.1} on the Laplace
exponent of the subordinator, the eigenfunction expansions~\eqref
{transmult.sub.1}--\eqref{eigenfsub} generally converge in
$L^2(I\times
I,m\times m)$ and $L^2(I,m)$, respectively, but not necessarily uniformly.
The bound on eigenfunctions and the trace-class condition on the
subordinator are sufficient to ensure uniform convergence.
The bound on eigenfunctions is satisfied for many diffusions important
in finance applications, such as OU, CIR, CEV, JDCEV and models related
to these diffusions. Condition~\eqref{trace.mult.1} also turns out to
be mild and is satisfied in many applications in finance. For example,
it is satisfied for tempered stable subordinators of Example~\ref{TempStab} with $\alpha\in(0,1)$ when eigenvalues grow linearly in the
eigenvalue number, as is the case for OU, CIR, CEV and JDCEV diffusions.
The key observation of practical importance is that, in the context of
the eigenfunction expansion method, subordination simply replaces the
eigenvalues $\lambda_n$ with the new eigenvalues $\phi(\lambda_n)$,
while the original and the subordinated semigroup share the same
eigenfunctions [compare with~\eqref{eigenfuncs}],
%
%
\begin{equation}
{\cal P}_t^{\beta,\phi} \varphi_n^\beta=
e^{-\phi(\lambda_n^\beta)
t}\varphi_n^\beta\quad\mbox{and}\quad {\cal
A}^{\beta,\phi} \varphi_n^\beta= -\phi\bigl(
\lambda_n^\beta\bigr)\varphi_n^\beta.
\label{limeq.9}
\end{equation}
Therefore, \emph{if the eigenfunction expansion is known for the
original semigroup, then it is immediately known for the subordinate
semigroup as well}. This fact was already pointed out in the original
work of \citet{bochner-1}; see equation (11).
This allows us to extend analytical tractability of classical diffusion
models in finance, such OU, CIR, CEV, etc., to their time-changed
(subordinate) counterparts with jumps. This observation has been
applied to subordinate OU processes in \citet{li-linetsky-1}, to
subordinate JDCEV processes in \citet{mendoza-carr-linetsky-1}, \citet{linetsky-mendoza-arriaga-1} and to subordinate CIR default
intensities in Section~\ref{subcirspec} of the present paper.

Applying the eigenfunction expansions of semigroups $({\cal P}_t^{\beta
,\phi})_{t\geq0}$ with $\beta=0,1$, to the pricing of credit-sensitive
securities, assuming the payoffs $f_i(x)\in L^2(I,m)$ in
equation~\eqref
{teq.41}, we immediately obtain the eigenfunction expansion of the
value function~\eqref{teq.42},
%
%
\begin{eqnarray}
\label{payoff.exp.0} f\bigl(t,X^\phi_t,D^\phi_t;T
\bigr)&=&e^{-r(T-t)}\sum_{n=1}^\infty
e^{-\phi
(\lambda
^0_n)(T-t)}f_n^1 \varphi_n^0
\bigl(X^\phi_t\bigr)
\nonumber
\\[-8pt]
\\[-8pt]
\nonumber
&&{}+e^{-r(T-t)}\sum_{n=1}^\infty
e^{-\phi(\lambda^1_n)(T-t)}f_n^{0-1} \bigl(1-D^\phi_t
\bigr)\varphi_n^1\bigl(X^\phi_t
\bigr)
\end{eqnarray}
with the expansion coefficients
%
%
\begin{equation}
f_n^{0-1}=\bigl(f_0-f_1,
\varphi_n^1\bigr) \quad\mbox{and} \quad f_n^1=
\bigl(f_1,\varphi_n^0\bigr). \label{coef.pr.1}
\end{equation}
We note that the eigenfunction expansion has the following
probabilistic interpretation. Due to the eigenfunction property~\eqref
{limeq.9} each process\break  $\{e^{\phi(\lambda^0_n)t}\varphi_n^0(X^\phi
_t),t\geq0\}$ and $\{e^{\phi(\lambda^1_n)t}(1-D^\phi_t)\varphi
_n^1(X^\phi
_t),t\geq0\}$ is an ${\mathbb H}^\phi$-martingale. Thus, the
eigenfunction expansion can be viewed as a \emph{martingale expansion}.

In particular, if $f_0(x)=1\in L^2(I,m)$ and $f_1(x)=0$, we obtain an
eigenfunction expansion of the
survival probability
%
%
\begin{eqnarray}
\calQ\bigl(t,X^\phi_t,D^\phi_t;T
\bigr)=\bigl(1-D^\phi_t\bigr)\sum
_{n=1}^\infty e^{-\phi
(\lambda
^1_n)(T-t)}f_n
\varphi_n^1\bigl(X^\phi_t\bigr),
\nonumber
\\[-8pt]
\\[-8pt]
\eqntext{f_n=\bigl(1,\varphi_n^1\bigr).}
\label{payoff.exp.1}
\end{eqnarray}
We note that, due to the existence of the stationary density, $1\in
L^2(I,m)$ in the SubCIR model, as well as many other default intensity
models, and the survival probability has an eigenfunction expansion.
We also remark that in those cases where the speed measure is an
infinite measure on $I$, constants are not in $L^2(I,m)$. However, it
sometimes happens that, while $1\notin L^2(I,m)$, ${\cal P}_t^{1,\phi
}1\in L^2(I,m)$ for $t>0$ if the semigroup has the property ${\cal
P}_t^{1,\phi} C_b(I)\subset L^2(I,m)$ for $t>0$.

We conclude this section with an observation that the long-maturity
asymptotics of the credit spread of a defaultable zero-coupon bond with
zero recovery is simply equal to the principal eigenvalue of the
negative of the generator ${\cal A}^{1,\phi}$,
\[
{\cal S}_{\infty}:=\lim_{T\rightarrow\infty} {\cal S}
\bigl(t,X^\phi_t;T\bigr) =\phi \bigl(\lambda_0^1
\bigr). \label{spread.1c}
\]
This immediately follows from the definition of the credit
spread~\eqref
{spread.1a} and the structure of the eigenfunction expansion of the
survival probability~\eqref{payoff.exp.1}.

\section{The SubCIR intensity model with two-sided mean-reverting
jumps}\label{subcirspec}

We now come back to the CIR model of Examples~\ref{EG.1} and~\ref{EG.2}.
We start with the bi-variate process $(X,D)$, where $X$ is a CIR
diffusion, and $D$ is a one-point point process with the compensator
$A_t=\int_0^t(1-D_s)X_s \,ds$ and time change it with a subordinator. We
call the resulting process $(X^\phi,D^\phi)$ the \emph{subordinate CIR
\textup{(}SubCIR\textup{)} default intensity model}. The default time $\tau$ in this
model is the first time default indicator $D^\phi$ equals one, and its
default intensity process is $\lambda^\phi_t=(1-D_t^\phi)k^\phi
(X_t^\phi)$.

We recall that the CIR process on $I=(0,\infty)$, if the Feller
condition is satisfied so zero is inaccessible, or on $I=[0,\infty)$,
if the Feller condition is not satisfied so zero is instantaneously
reflecting, has a stationary density~\eqref{eqr.4}. We choose $x_0$ in
the definition of the speed density~\eqref{eq40} so that $m(x)=\pi(x)$
[i.e., $\int_I m(x)\,dx=1$].
Then for all $\beta\geq0$ the semigroup $(\calP_t^\beta)_{t\geq0}$
defined by~\eqref
{fk.1} with the CIR diffusion $X$ and $k(x)=\beta x$ has a symmetric density
$p_m^\beta(t,x,y)$ with respect to the stationary distribution $\pi
(y)\,dy$ given by
%
%
\begin{eqnarray}
\label{trn.1a} p_{m}^\beta(t,x,y)&=&\frac{ \rho\Gamma(b)\sqrt{xy}}{\sigma^2\sinh
( t\rho
/2)}
\biggl(\frac{ e^{\rho t/2}}{a\sqrt{x y}} \biggr)^{b}I_{b-1} \biggl(
\frac{
2\rho\sqrt{x y}}{\sigma^2\sinh( t\rho/2)} \biggr)
\nonumber
\\[-8pt]
\\[-8pt]
\nonumber
&&{}\times\exp \biggl\{(x+y) \biggl(\frac{\kappa\tanh( t\rho/2)- \rho
}{\sigma
^2\tanh( t\rho/2)} \biggr)-\lambda^\beta_0
t \biggr\},
\end{eqnarray}
where $I_\nu(x)$ is the modified Bessel function of the first kind and
\[
\lambda^\beta_1:=\frac{b}{2}(\rho-\kappa)\quad \mbox
{and} \quad \rho:=\rho (\beta)=\sqrt{\kappa^2+2 \beta\sigma^2},\label{trn.1b}
\]
and $a$ and $b$ are defined in equation~\eqref{eqr.4}. This explicit
solution in terms of the Bessel function is due to the fact that the
CIR process can be obtained by a deterministic time change from the
squared Bessel process in a similar way as the OU process can be
obtained from Brownian motion by a deterministic time change [cf.
Proposition 6.3.1.1 and 6.3.2.1 of \citet{jeanblanc-yor-chesney-1},
pages 357--358] combined with the absolute continuity relationships for
Bessel processes; see Section~6.3 in \citet{jeanblanc-yor-chesney-1},
page 340, for more details. For $\beta=1$ this density has appeared in
the seminal work of~\citet{cox-ingersoll-ross-1} on their interest
rate model.

The bi-linear eigenfunction expansion~\eqref
{eq:SymTPDBilinearExpansion} for the density $p_m^\beta(t,x,y)$ can be
obtained from the expression~\eqref{trn.1a} by applying the
Hille--Hardy formula to expand the Bessel function in the bi-linear
expansion of generalized Laguerre polynomials $L_n^{\nu}(x)$ [cf.
\citet{erdelyi-2}, page 189; valid for all $|t|<1$, $\nu>-1$, $a,b>0$]
%
%
\begin{eqnarray}\label{HH.1}
&&\frac{ (abt )^{-\nu/2}}{1-t}\exp \biggl\{ -\frac{
(a+b )t}{1-t} \biggr\}I_{\nu}
\biggl(\frac{2\sqrt {abt}}{1-t} \biggr)
\nonumber
\\[-8pt]
\\[-8pt]
\nonumber
&&\qquad= \sum_{k=0}^{\infty}
\frac{t^{k} k!}{\Gamma (k+\nu+1
)}L_{k}^{\nu
} (a )L_{k}^{\nu}
(b ).
\end{eqnarray}
The application of the Hille--Hardy formula thus yields the
eigenfunctions and eigenvalues of the semigroup $({\cal P}_t^\beta
)_{t\geq0}$ and its generator ${\cal A}^\beta$ in the Hilbert space
$L^2(I,m)$ with $m(dx)=\pi(x)\,dx$ (the CIR stationary distribution).
Due to the appearance of Laguerre polynomials, semigroups of this type
are sometimes called \emph{Laguerre semigroups} in analysis; cf. \citet{nowak-stempak-2}.
The following theorem summarizes the explicit results for eigenvalues
and eigenfunctions.

\begin{theorem}[(CIR eigenfunction
expansion)]\label{prop.cir.2}The semigroup $(\calP_t^\beta)_{t\geq0}$ is a symmetric
trace-class
semigroup in $L^2(I,m)$ with the eigenvalues and continuous eigenfunctions
of the negative of its self-adjoint infinitesimal generator ${\cal
A}^\beta$
given by
%
%
\begin{eqnarray}
\lambda_n^\beta&=&(n-1)\rho+\frac{b}{2} (\rho-
\kappa ), \label{eig.0}
\\
\label{eig.1}
\varphi_n^\beta(x)&=&{\cal N}_n^\beta
e^{((\kappa-\rho)x)/\sigma
^2}L_{n-1}^{b-1} \biggl(\frac{2 x\rho}{\sigma^2}
\biggr),
\nonumber
\\[-8pt]
\\[-8pt]
\nonumber
 {\cal N}_n^\beta &=&\sqrt{\frac{(n-1)!}{(b)_{n-1}}}
\biggl(\frac{\rho}{\kappa} \biggr)^{b/2},\qquad  n=1,2,\ldots,
\end{eqnarray}
where $(a)_n=\Gamma(a+n)/\Gamma(a)=a(a+1)\cdots(a+n-1)$ is the Pochhammer
symbol. Moreover, on each compact interval $K\subset I$ there exists a
constant $C_K$ independent of $n$ such that
\[
\bigl|\varphi_n^\beta(x)\bigr|\leq C_K
n^{-1/4} \label{f}
\]
or all $n\geq1$.
\end{theorem}
\begin{pf} The bi-linear expansion for the density of the form~\eqref
{eq:SymTPDBilinearExpansion} with $\varphi_n^\beta(x)$ and $\lambda
_n^\beta$ given by equations~\eqref{eig.0} and~\eqref{eig.1} is
directly obtained by applying the Hille--Hardy formula~\eqref{HH.1} to
the right-hand side of equation~\eqref{trn.1a}. It is then easy to
directly verify from the properties of Laguerre polynomials that
$\varphi^\beta_n(x)$ are eigenfunctions of the operator
\[
{\cal A}^\beta f=\tfrac{1}{2}\sigma^2 x
f^{\prime\prime}(x)+\kappa (\theta -x)f^\prime(x)-\beta x f(x)
\label{limeq.10}
\]
with eigenvalues $\lambda_n^\beta$ satisfying the boundary condition at
zero $\lim_{x\downarrow0}(\varphi^\beta(x))'/\break  s(x)=0$, where $s(x)$ is
the scale density defined in equation~\eqref{eq40}. The eigenfunctions are
normalized with respect to the inner product with $m(dx)=\pi(x)\,dx$,
$(\varphi^\beta_n,\varphi^\beta_m)=\delta_{n,m}$.
The trace class condition~\eqref{eq:TraceClassCondition} is verified
due to the linear growth of eigenvalues. The bound for the eigenfunctions
is obtained from the estimate in equation (27a) of~\citet{nikiforov-uvarov-1}, page 54.
\end{pf}


This CIR eigenfunction expansion has been applied in finance in~\citet{davydov-linetsky-3} and \citet{gorovoi-linetsky-1}
[we note that our normalization factor ${\cal N}_n$ in the expression
for eigenfunctions differs from~\citet{davydov-linetsky-3}, Proposition
9, due to different normalization of the speed measure; here we
normalize the speed measure so it integrates to one and thus coincides
with the stationary distribution].

For any $f\in L^2(I,m)$ the computation of ${\cal P}^\beta_t f(x)$
reduces to computing the expansion coefficients.
In particular, consider the discounted CIR characteristic function
known in closed form due to the fact that the CIR diffusion is a
CBI/affine process; cf. \citet{cox-ingersoll-ross-1}, \citet{duffie-garleanu-1},
Appendix~A. For any complex $z$ with $\Re z\geq0$,
%
%
\begin{eqnarray} \label{characaffine.2}
\Psi_t(x,\beta,z)&: =& \bbE_x \bigl[e^{-\beta\int
_0^tX_u\,du}e^{-zX_t}
\bigr]
\nonumber
\\[-8pt]
\\[-8pt]
\nonumber
&=&A(t,\beta,z) \exp\bigl\{-B(t,\beta,z) x\bigr\},
\end{eqnarray}
where
\begin{eqnarray*}
\label{ricB} A(t,\beta,z)&:=& \biggl(\frac{2\rho
e^{(\kappa
+\rho)t/2}}{2\rho+(\rho+\kappa+z \sigma^2)(e^{\rho t}-1)} \biggr)^b,
\\
B(t,\beta,z)&:=& \frac{2\beta(e^{\rho t}-1)+z (\rho-\kappa
)e^{\rho t}
+z (\rho+\kappa)}{2\rho+(\rho+\kappa+z \sigma^2)(e^{\rho t}-1)}.
\end{eqnarray*}
We have the following eigenfunction expansion of the characteristic function.

\begin{proposition}\label{CIRBasic}
The characteristic function has the eigenfunction expansion~\eqref
{eigenfsub} with the coefficients given by
%
%
\begin{eqnarray}\label{LtCIR.1}
f_1^\beta(z)=1, \qquad f_n^\beta(z)=
\frac{1}{{\cal N}^\beta_n} \biggl(\frac
{\kappa
-\rho+\sigma^2z}{\kappa+\rho+\sigma^2z} \biggr)^{n-1} \biggl(
\frac
{2\rho
}{\kappa+\rho+\sigma^2z} \biggr)^{b},
\nonumber
\\[-8pt]
\\[-8pt]
  \eqntext{n=2,\ldots.}
\end{eqnarray}
\end{proposition}

\begin{pf}
Obtained immediately from the identity for the generating
function of the generalized Laguerre polynomials
(valid for all complex $|y|<1$ and $a>-1$),
\[
{\sum_{k=0}^\infty y^k
L_k^{a}(x)} ={(1-y)^{-1-a} \exp
\bigl((yx)/(y-1) \bigr)}. \label{lag.pol.1}
\]
Alternatively, the integrals in $f_n^\beta(z)=(e^{-z\cdot},\varphi
_n^\beta)$ can be explicitly calculated due to the integral identity
for Laguerre polynomials in equation (2.19.3.3),
\citet{prudnikov-2}, page 462.
\end{pf}

We note that if one is only interested in the CIR characteristic
function, the affine closed-form expression~\eqref{characaffine.2} is
certainly simpler than the eigenfunction expansion. However, while the
affine expression~\eqref{characaffine.2} does \emph{not} generalize to
the SubCIR model, the eigenfunction expansion generalizes immediately, yielding
\[
\Psi^\phi_t(x,\beta,z): =\bigl({\cal P}_t^{\beta,\phi}e^{-z\cdot}
\bigr) (x) = \sum_{n=1}^\infty
e^{-\phi(\lambda_n^\beta)t} f_n^\beta(z)\varphi
_n^\beta(x) \label{characaffine.ee}
\]
with the same eigenfunctions and expansion coefficients~\eqref{LtCIR.1}
but with new eigenvalues $\phi(\lambda_n^\beta)$, where $\phi$ is the
Laplace exponent of the subordinator.

In particular, the eigenfunction expansion for the
survival probability~\eqref{surv.p.1b} in the SubCIR default intensity
model is then immediately obtained
\begin{eqnarray*}
\label{limeq.11} {\cal Q}\bigl(t,X_t^\phi,D_t^\phi;T
\bigr)&=&\bigl(1-D_t^\phi\bigr) P_{T-t}^{1,\phi}
\bigl(X_t^\phi,I\bigr)
\\
& =&\bigl(1-D_t^\phi\bigr)\Psi^\phi_{T-t}
\bigl(X_t^\phi,1,0\bigr)
\\
&=&\bigl(1-D_t^\phi\bigr)\sum_{n=1}^\infty
e^{-\phi(\lambda_n^1)(T-t)} f_n^1(0)\varphi_n^1
\bigl(X_t^\phi\bigr)
\end{eqnarray*}
by setting $z=0$ in the expansion for the characteristic function.
The pricing of zero-coupon bonds with constant recovery~\eqref{teq.43}
is then immediate. The pricing of other credit-sensitive securities in
the SubCIR default intensity model then reduces to computing the
corresponding expansion coefficients in equations~\eqref
{payoff.exp.0}--\eqref{coef.pr.1}. In particular, the pricing and
calibration of credit default swaptions is considered in~\citet{mendoza-1}.

\begin{figure}[b]
\centering
\begin{tabular}{@{}cc@{}}

\includegraphics{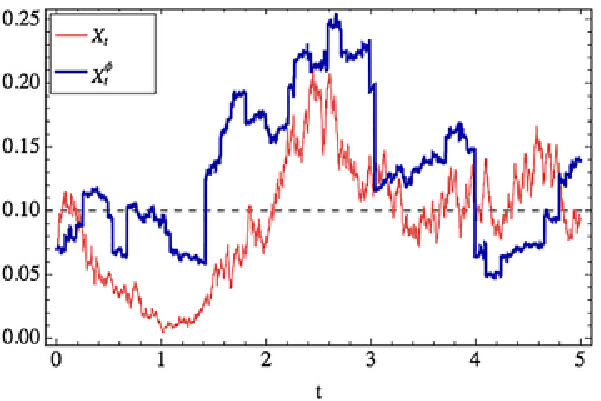}
 & \includegraphics{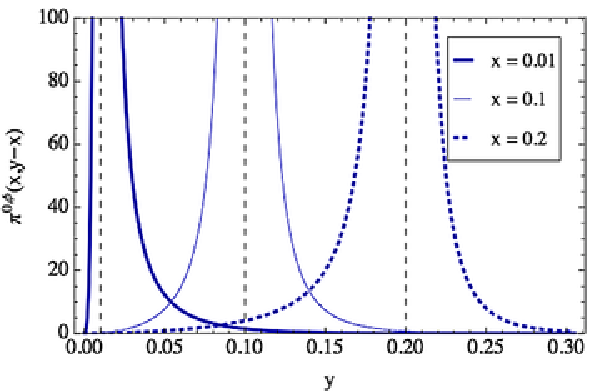}\\
\footnotesize{(a) Sample paths} & \footnotesize{(b) L\'evy densities}
\end{tabular}
\caption{\textup{(a)} Sample paths of a CIR process $(X_t)_{t\geq0}$ and the SubCIR
process $(X_t^\phi)_{t\geq0}$. The horizontal line
(dashed) corresponds to the long run mean level $\theta=0.1$. Figure
\textup{(b)} contains three jump densities $\pi^{0,\phi}(x,y-x)$ corresponding
to the initial states $x=0.01$,
$x=\theta=0.1$ and $x=0.2$, which are indicated by the vertical lines
(dashed).}\label{fig:samp}
\end{figure}

We also remark that the same eigenfunction expansion yields the pricing
of default-free zero-coupon bonds in the SubCIR interest rate model of
Remark~\ref{rfir},
\begin{eqnarray*}
P\bigl(Z_t^\phi,t;T\bigr)&=& P_{T-t}^{1,\phi}
\bigl(Z_t^\phi,I\bigr) =\Psi^\phi_{T-t}
\bigl(X_t^\phi,1,0\bigr)\\
&=&\sum_{n=1}^\infty
e^{-\phi(\lambda_n^1)(T-t)} f_n^1(0)\varphi _n^1
\bigl(Z_t^\phi\bigr). \label{limeq.12}
\end{eqnarray*}

We now present a numerical illustration of the qualitative properties
of the SubCIR default intensity model.
We start with a CIR process $X$ with $\kappa=1$, $\theta=0.1$ and
$\sigma=0.25$. The SubCIR process $X^\phi$ is constructed by subordinating
$X$ with an inverse Gaussian subordinator $({\cal T}_t)_{t\geq0}$ with
the L\'{e}vy
measure~\eqref{Subord.7} with parameters $\alpha=0.5$, $\eta=1$ and
$C=0.5$, and zero drift $\gamma=0$.
Since the subordinator is driftless, $X^\phi$ is a pure jump process in
our example.
Figure~\ref{fig:samp}(a) shows simulation of a typical sample path of the
CIR process $X$ and the SubCIR process $X^\phi$ with these parameters.
While the CIR process diffuses around its long-run level $\theta$ with
volatility $\sigma$, while being pulled back toward it by the
mean-reverting drift at the rate $\kappa$, the SubCIR process is a pure
jump process with state-dependent mean-reverting L\'{e}vy measure. The
mean-reverting nature of jumps is evident in the sample path plot (a),
as well as in the plot (b) of the L\'evy density $\pi^{0,\phi
}(x,y-x)=m(y)\int_{(0,\infty)}p_{m}^0(s,x,y)\nu(ds)$ plotted as a
function of $y$ for three fixed values $x$. This plot shows three L\'
{e}vy densities of jumps from the three initial states $x=0.01$,
$x=\theta=0.1$ and $x=0.2$. Here $x$ is the pre-jump state, and $y$ is
the post-jump state, so that the jump size is $y-x$.
When $x=\theta=0.1$, that is, jumping from the long-run mean, the L\'
{e}vy density looks nearly symmetrical.
In contrast, the L\'{e}vy density of jumps starting from the state
$x=0.01<0.1$ significantly below the long-run mean is highly skewed to
the right, as the process tends to jump back up toward its long run
mean at $0.1$ from this low value of $0.01$.
On the other hand, the L\'{e}vy density of jumps starting from the
state $x=0.2>0.1$ significantly above the long-run mean is highly
skewed to the left, as the process tends to jump back down toward its
long run mean at $0.1$ from this high value of $0.2$.
Either way, the process stays nonnegative. This is in sharp contrast
with the behavior of affine jump-diffusion/CBI-processes that can only
jump up and cannot jump down to ensure that the process stays
nonnegative. In the framework of subordinate diffusions, the
nonnegativity of SubCIR process is immediate, as the subordinate
process and the original process share the same state
space.\looseness=-1

%
\begin{figure}[b]

\includegraphics{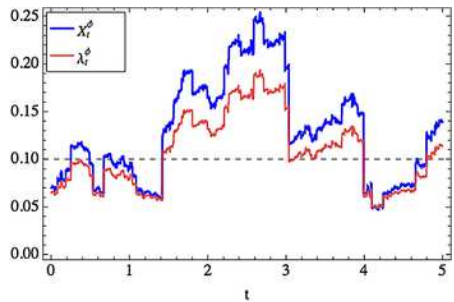}

\caption{Default intensity $\lambda_t^\phi=(1-D^\phi_t)k^\phi
(X^\phi_t)$.
This figure illustrates the sample path
of the default intensity process $(\lambda^\phi_t)_{t\geq0}$ induced
by the SubCIR
process $(X_t^\phi)_{t\geq0}$, which is also depicted. The horizontal
line (dashed)
corresponds to the long run mean level $\theta=0.1$.}\label{fig:SubKill}
\end{figure}

From the expression for $k^\phi(x)$ arising from Theorem~\ref{mkvian.sub.biv},
\[
k^\phi(x)=\gamma\beta x + \int_{(0,\infty)} \bigl(1-A(s,
\beta,0) \exp \bigl\{-B(s,\beta,0) x\bigr\} \bigr)\nu(ds),\label{subkill.1.cir}
\]
where we substituted the closed-form expression for the survival
probability of the CIR process,
it is clear that the default intensity is no longer affine as in the
SubCIR model.
Figure~\ref{fig:SubKill} illustrates a sample path of the default
intensity process $\lambda_t^\phi=(1-D^\phi_t)k^\phi(X^\phi_t)$, along
with a
sample path of the pure jump process $X^\phi$.

\begin{figure}
\centering
\begin{tabular}{@{}cc@{}}

\includegraphics{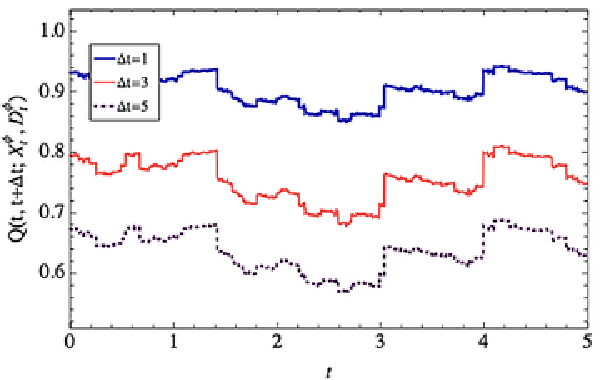}
 & \includegraphics{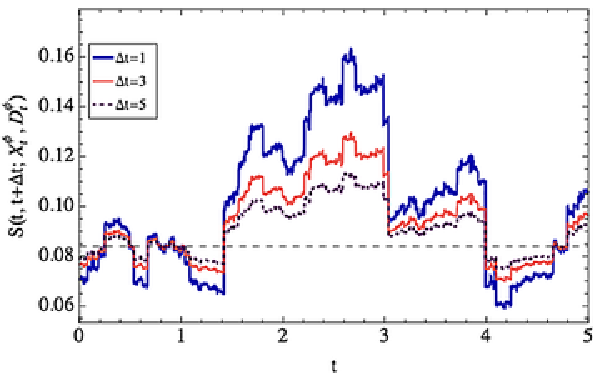}\\
\footnotesize{(a) Survival probability} & \footnotesize{(b) Credit
spreads}\\

\includegraphics{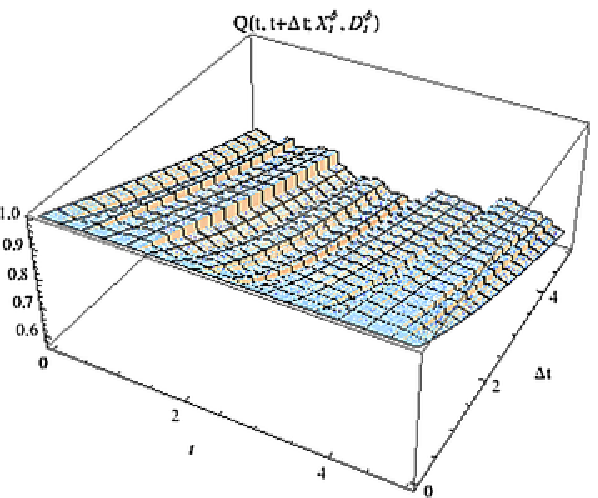}
 & \includegraphics{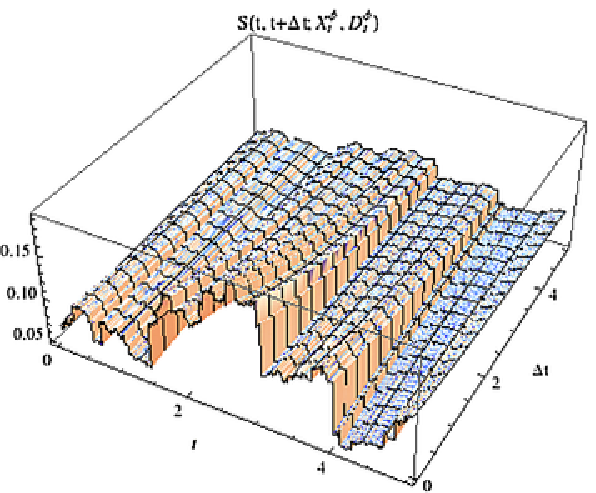}\\
\footnotesize{(c) Survival probability} & \footnotesize{(d) Credit spreads}
\end{tabular}
\caption{Survival Probabilities and Credit Spreads sample
paths.}\label {fig:subfigP1}
\end{figure}

Finally, Figure~\ref{fig:subfigP1} shows sample paths of the survival
probabilities~\eqref{surv.p.1b} and defaultable credit spreads on
zero-coupon bonds~\eqref{spread.1a} over a five-year period simulated
under this SubCIR default intensity specification. Figure~\ref{fig:subfigP1}(a)
and~\ref{fig:subfigP1}(b) show sample paths of the
survival probabilities for one, three and five years, that is, $\calQ
(t, t+\Delta t; X^\phi_t,D^\phi_t)$ with $\Delta t = 1,3,5$ years,
and one-
three- and five-year credit spreads, $\calS(t, X^\phi_t,D^\phi_t,
t+\Delta
t)$, respectively. The dashed horizontal line in (b) corresponds to
the asymptotic credit spread $\calS_\infty= 0.084$ equal to the
principal eigenvalue $\phi(\lambda_1^1)$ of the semigroup ${\cal
P}^{1,\phi}$. Figure~\ref{fig:subfigP1}(c) and~\ref{fig:subfigP1}(d) show
sample paths over five years of the evolution of the term structure of
survival probabilities $\calQ(t,X^\phi_t,D^\phi_t; t+\Delta t)$ and credit
spreads $\calS(t,X^\phi_t,D^\phi_t; t+\Delta t)$, respectively.
Since the SubCIR state variable $X^\phi$ is a jump process, prices of
credit-sensitive securities, such as bond prices, as well as credit
spreads, are also jump processes in this model.


%

\section{Conclusion}\label{conc.s}

The present paper introduces a jump-diffusion extension of the
classical diffusion default intensity model by means of subordination
in the sense of Bochner. We start with the bi-variate process of the
diffusion state variable and default indicator $(X,D)$ in the diffusion
intensity framework and time change it with a L\'evy subordinator
$\calT
$. We characterize the resulting
time changed process $(X^\phi_t,D^\phi_t)=(X(\calT_t),D(\calT_t))$
as a
Markovian It\^{o} semimartingale and, in particular, show from the
Doob--Meyer decomposition of $D^\phi$ that the default time in the
time-changed model has a jump-diffusion or pure jump intensity. When
$X$ is a CIR diffusion with mean-reverting drift, the default intensity
of the subordinate model (SubCIR) is a nonnegative jump-diffusion or
pure jump process with two-sided mean-reverting jumps that stays
nonnegative. The SubCIR default intensity model is fully analytically
tractable by means of the explicitly computed eigenfunction expansion
of the relevant semigroups. This yields explicit closed-form pricing of
credit-sensitive securities.


%
%

%
%
%



\printaddresses


\begin{thebibliography}{79}

\bibitem[\protect\citeauthoryear{Ahn and Gao}{1999}]{ahn-gao-1}
%
\begin{barticle}[author]
\bauthor{\bsnm{Ahn},~\bfnm{Dong-Hyun}\binits{D.-H.}} \AND
\bauthor{\bsnm{Gao},~\bfnm{Bin}\binits{B.}}
(\byear{1999}).
\btitle{{A parametric nonlinear model of term structure dynamics}}.
\bjournal{The Review of Financial Studies}
\bvolume{12}
\bpages{721--762}.
\bptok{imsref}%
\end{barticle}
%
\endbibitem

\bibitem[\protect\citeauthoryear{Andreasen}{2001}]{andreasen-1}
%
\begin{bmisc}[author]
\bauthor{\bsnm{Andreasen},~\bfnm{Jesper}\binits{J.}}
(\byear{2001}).
\bhowpublished{Credit explosives. Bank of America Fixed Income
Research Working
Paper}.
\bptok{imsref}%
\end{bmisc}
%
\endbibitem

\bibitem[\protect\citeauthoryear{Barlow}{2002}]{Barlow-1}
%
\begin{barticle}[author]
\bauthor{\bsnm{Barlow},~\bfnm{Martin~T.}\binits{M.~T.}}
(\byear{2002}).
\btitle{{A diffusion model for electricity prices}}.
\bjournal{Math. Finance}
\bvolume{12}
\bpages{287--298}.
\bptok{imsref}%
\end{barticle}
%
\endbibitem

\bibitem[\protect\citeauthoryear
{Barndorff-Nielsen}{1998}]{barndorff-nielsen-1}
%
\begin{barticle}[mr]
\bauthor{\bsnm{Barndorff-Nielsen},~\bfnm{Ole~E.}\binits{O.~E.}}
(\byear{1998}).
\btitle{Processes of normal inverse {G}aussian type}.
\bjournal{Finance Stoch.}
\bvolume{2}
\bpages{41--68}.
\bid{doi={10.1007/s007800050032}, issn={0949-2984}, mr={1804664}}
\bptok{imsref}%
\end{barticle}
%
\endbibitem

\bibitem[\protect\citeauthoryear{Beaglehole and
Tenney}{1992}]{beaglehole-tenney-1}
%
\begin{barticle}[author]
\bauthor{\bsnm{Beaglehole},~\bfnm{David}\binits{D.}} \AND
\bauthor{\bsnm{Tenney},~\bfnm{Mark}\binits{M.}}
(\byear{1992}).
\btitle{{Corrections and additions to ``A nonlinear equilibrium model
of the
term structure of interest rates.''}}
\bjournal{Math. Finance}
\bvolume{32}
\bpages{345--353}.
\bptok{imsref}%
\end{barticle}
%
\endbibitem

\bibitem[\protect\citeauthoryear{Bertoin}{1996}]{bertoin-1}
%
\begin{bbook}[mr]
\bauthor{\bsnm{Bertoin},~\bfnm{Jean}\binits{J.}}
(\byear{1996}).
\btitle{L\'evy Processes}.
\bseries{Cambridge Tracts in Mathematics}
\bvolume{121}.
\bpublisher{Cambridge Univ. Press}, \blocation{Cambridge}.
\bid{mr={1406564}}
\bptnote{check year}%
\bptok{imsref}%
\end{bbook}
%
\endbibitem

\bibitem[\protect\citeauthoryear{Bertoin}{1999}]{bertoin-2}
%
\begin{bbook}[mr]
\bauthor{\bsnm{Bertoin},~\bfnm{J.}\binits{J.}}
(\byear{1999}).
\btitle{Lectures on Probability Theory and Statistics}.
\bseries{Lecture Notes in Math.}
\bvolume{1717}.
\bpublisher{Springer}, \blocation{Berlin}.
\bid{doi={10.1007/b72002}, mr={1746299}}
\bptok{imsref}%
\end{bbook}
%
\endbibitem

\bibitem[\protect\citeauthoryear{Bielecki, Jeanblanc and
Rutkowski}{2011}]{bielecki-jeanblanc-rutkowski-1}
%
\begin{barticle}[mr]
\bauthor{\bsnm{Bielecki},~\bfnm{Tomasz~R.}\binits{T.~R.}},
\bauthor{\bsnm{Jeanblanc},~\bfnm{Monique}\binits{M.}} \AND
\bauthor{\bsnm{Rutkowski},~\bfnm{Marek}\binits{M.}}
(\byear{2011}).
\btitle{Hedging of a credit default swaption in the {CIR} default intensity
model}.
\bjournal{Finance Stoch.}
\bvolume{15}
\bpages{541--572}.
\bid{doi={10.1007/s00780-010-0143-7}, issn={0949-2984}, mr={2833099}}
\bptok{imsref}%
\end{barticle}
%
\endbibitem

\bibitem[\protect\citeauthoryear{Bielecki and
Rutkowski}{2004}]{bielecki-rutkowski-1}
%
\begin{bbook}[author]
\bauthor{\bsnm{Bielecki},~\bfnm{Tomasz~R.}\binits{T.~R.}} \AND
\bauthor{\bsnm{Rutkowski},~\bfnm{Marek}\binits{M.}}
(\byear{2004}).
\btitle{{Credit Risk: Modeling, Valuation and Hedging}}.
\bpublisher{Springer}, \blocation{Berlin}.
\bptok{imsref}%
\end{bbook}
%
\endbibitem

\bibitem[\protect\citeauthoryear{Bielecki
et~al.}{2008}]{bielecki-crepey-jeanblanc-rutkowski-2}
%
\begin{barticle}[mr]
\bauthor{\bsnm{Bielecki},~\bfnm{Tomasz~R.}\binits{T.~R.}},
\bauthor{\bsnm{Cr{\'e}pey},~\bfnm{St{\'e}phane}\binits{S.}},
\bauthor{\bsnm{Jeanblanc},~\bfnm{Monique}\binits{M.}} \AND
\bauthor{\bsnm{Rutkowski},~\bfnm{Marek}\binits{M.}}
(\byear{2008}).
\btitle{Defaultable options in a {M}arkovian intensity model of credit risk}.
\bjournal{Math. Finance}
\bvolume{18}
\bpages{493--518}.
\bid{doi={10.1111/j.1467-9965.2008.00345.x}, issn={0960-1627}, mr={2454669}}
\bptok{imsref}%
\end{barticle}
%
\endbibitem

\bibitem[\protect\citeauthoryear{Bielecki
et~al.}{2013}]{bielecki-cousin-crepey-herbertsson-1}
%
\begin{bmisc}[author]
\bauthor{\bsnm{Bielecki},~\bfnm{Tomasz~R.}\binits{T.~R.}},
\bauthor{\bsnm{Cousin},~\bfnm{Areski}\binits{A.}},
\bauthor{\bsnm{Cr{\'{e}}pey},~\bfnm{St{\'{e}}phane}\binits{S.}}
\AND
\bauthor{\bsnm{Herbertsson},~\bfnm{Alexander}\binits{A.}}
(\byear{2013}).
\bhowpublished{Dynamic modeling of portfolio credit risk with common shocks.
\textit{J. Optim. Theory Appl.} To appear.}
\bptok{imsref}%
\end{bmisc}
%
\endbibitem

\bibitem[\protect\citeauthoryear{Bielecki
et~al.}{2012}]{bielecki-crepey-jeanblanc-zargari-1}
%
\begin{barticle}[mr]
\bauthor{\bsnm{Bielecki},~\bfnm{T.~R.}\binits{T.~R.}},
\bauthor{\bsnm{Cr{\'e}pey},~\bfnm{S.}\binits{S.}},
\bauthor{\bsnm{Jeanblanc},~\bfnm{M.}\binits{M.}} \AND
\bauthor{\bsnm{Zargari},~\bfnm{B.}\binits{B.}}
(\byear{2012}).
\btitle{Valuation and hedging of {CDS} counterparty exposure in a {M}arkov
copula model}.
\bjournal{Int. J. Theor. Appl. Finance}
\bvolume{15}
\bpages{1250004, 39}.
\bid{doi={10.1142/S0219024911006498}, issn={0219-0249}, mr={2902965}}
\bptok{imsref}%
\end{barticle}
%
\endbibitem

\bibitem[\protect\citeauthoryear{Bochner}{1949}]{bochner-1}
%
\begin{barticle}[mr]
\bauthor{\bsnm{Bochner},~\bfnm{S.}\binits{S.}}
(\byear{1949}).
\btitle{Diffusion equation and stochastic processes}.
\bjournal{Proc. Natl. Acad. Sci. USA}
\bvolume{35}
\bpages{368--370}.
\bid{issn={0027-8424}, mr={0030151}}
\bptok{imsref}%
\end{barticle}
%
\endbibitem

\bibitem[\protect\citeauthoryear{Borodin and
Salminen}{2002}]{borodin-salminen}
%
\begin{bbook}[mr]
\bauthor{\bsnm{Borodin},~\bfnm{Andrei~N.}\binits{A.~N.}} \AND
\bauthor{\bsnm{Salminen},~\bfnm{Paavo}\binits{P.}}
(\byear{2002}).
\btitle{Handbook of {B}rownian Motion---Facts and Formulae},
\bedition{2nd} ed.
\bpublisher{Birkh\"auser}, \blocation{Basel}.
\bid{doi={10.1007/978-3-0348-8163-0}, mr={1912205}}
\bptok{imsref}%
\end{bbook}
%
\endbibitem

\bibitem[\protect\citeauthoryear{Boyarchenko and
Levendorski{\u\i}}{2007}]{boyarchenko-levendor-2}
%
\begin{barticle}[mr]
\bauthor{\bsnm{Boyarchenko},~\bfnm{Nina}\binits{N.}} \AND
\bauthor{\bsnm{Levendorski{\u\i}},~\bfnm{Sergei}\binits{S.}}
(\byear{2007}).
\btitle{The eigenfunction expansion method in multi-factor quadratic term
structure models}.
\bjournal{Math. Finance}
\bvolume{17}
\bpages{503--539}.
\bid{doi={10.1111/j.1467-9965.2007.00314.x}, issn={0960-1627}, mr={2352904}}
\bptok{imsref}%
\end{barticle}
%
\endbibitem

\bibitem[\protect\citeauthoryear{Brigo and Alfonsi}{2005}]{brigo-alfonsi-1}
%
\begin{barticle}[mr]
\bauthor{\bsnm{Brigo},~\bfnm{Damiano}\binits{D.}} \AND
\bauthor{\bsnm{Alfonsi},~\bfnm{Aur{\'e}lien}\binits{A.}}
(\byear{2005}).
\btitle{Credit default swap calibration and derivatives pricing with
the {SSRD}
stochastic intensity model}.
\bjournal{Finance Stoch.}
\bvolume{9}
\bpages{29--42}.
\bid{doi={10.1007/s00780-004-0131-x}, issn={0949-2984}, mr={2210926}}
\bptok{imsref}%
\end{barticle}
%
\endbibitem

\bibitem[\protect\citeauthoryear{Brigo and
El-Bachir}{2006}]{brigo-el-bachir-2}
%
\begin{bmisc}[author]
\bauthor{\bsnm{Brigo},~\bfnm{Damiano}\binits{D.}} \AND
\bauthor{\bsnm{El-Bachir},~\bfnm{Naoufel}\binits{N.}}
(\byear{2006}).
\bhowpublished{Credit derivatives pricing with a smile-extended jump stochastic
intensity model. Working paper}.
\bptok{imsref}%
\end{bmisc}
%
\endbibitem

\bibitem[\protect\citeauthoryear{Brigo and
El-Bachir}{2010}]{brigo-el-bachir-1}
%
\begin{barticle}[mr]
\bauthor{\bsnm{Brigo},~\bfnm{Damiano}\binits{D.}} \AND
\bauthor{\bsnm{El-Bachir},~\bfnm{Naoufel}\binits{N.}}
(\byear{2010}).
\btitle{An exact formula for default swaptions' pricing in the {SSRJD}
stochastic intensity model}.
\bjournal{Math. Finance}
\bvolume{20}
\bpages{365--382}.
\bid{doi={10.1111/j.1467-9965.2010.00401.x}, issn={0960-1627}, mr={2667895}}
\bptok{imsref}%
\end{barticle}
%
\endbibitem

\bibitem[\protect\citeauthoryear{Carr and Linetsky}{2006}]{carr-linetsky-1}
%
\begin{barticle}[mr]
\bauthor{\bsnm{Carr},~\bfnm{Peter}\binits{P.}} \AND
\bauthor{\bsnm{Linetsky},~\bfnm{Vadim}\binits{V.}}
(\byear{2006}).
\btitle{A jump to default extended {CEV} model: An application of {B}essel
processes}.
\bjournal{Finance Stoch.}
\bvolume{10}
\bpages{303--330}.
\bid{doi={10.1007/s00780-006-0012-6}, issn={0949-2984}, mr={2244347}}
\bptok{imsref}%
\end{barticle}
%
\endbibitem

\bibitem[\protect\citeauthoryear{{\c{C}}inlar and
Jacod}{1981a}]{cinlar-jacod-2}
%
\begin{bincollection}[mr]
\bauthor{\bsnm{{\c{C}}inlar},~\bfnm{E.}\binits{E.}} \AND
\bauthor{\bsnm{Jacod},~\bfnm{J.}\binits{J.}}
(\byear{1981}a).
\btitle{Representation of semimartingale {M}arkov processes in terms of
{W}iener processes and {P}oisson random measures}.
In \bbooktitle{Seminar on {S}tochastic {P}rocesses, 1981 ({E}vanston, {I}ll.,
1981)}
(\beditor{\bfnm{Erhan}\binits{E.}~\bsnm{Cinlar}},
\beditor{\bfnm{K.~L.}\binits{K.~L.}~\bsnm{Chung}} \AND
\beditor{\bfnm{R.~K.}\binits{R.~K.}~\bsnm{Getoor}}, eds.).
\bseries{Progr. Prob. Statist.}
\bvolume{1}
\bpages{159--242}.
\bpublisher{Birkh\"auser}, \blocation{Boston, MA}.
\bid{mr={0647786}}
\bptok{imsref}%
\end{bincollection}
%
\endbibitem

\bibitem[\protect\citeauthoryear{{\c{C}}inlar and
Jacod}{1981b}]{cinlar-jacod-1}
%
\begin{bincollection}[mr]
\bauthor{\bsnm{{\c{C}}inlar},~\bfnm{E.}\binits{E.}} \AND
\bauthor{\bsnm{Jacod},~\bfnm{J.}\binits{J.}}
(\byear{1981}b).
\btitle{Semimartingales defined on {M}arkov processes}.
In \bbooktitle{Stochastic Differential Systems ({V}isegr\'ad, 1980)}
(\beditor{\bfnm{M.}\binits{M.}~\bsnm{Arat{\'{o}}}},
\beditor{\bfnm{D.}\binits{D.}~\bsnm{Vermes}} \AND
\beditor{\bfnm{A.~V.}\binits{A.~V.}~\bsnm{Balakrishnan}}, eds.).
\bseries{Lecture Notes in Control and Information Sci.}
\bvolume{36}
\bpages{13--24}.
\bpublisher{Springer}, \blocation{Berlin}.
\bid{mr={0653642}}
\bptok{imsref}%
\end{bincollection}
%
\endbibitem

\bibitem[\protect\citeauthoryear{{\c{C}}inlar
et~al.}{1980}]{cinlar-jacod-protter-sharpe-1}
%
\begin{barticle}[mr]
\bauthor{\bsnm{{\c{C}}inlar},~\bfnm{E.}\binits{E.}},
\bauthor{\bsnm{Jacod},~\bfnm{J.}\binits{J.}},
\bauthor{\bsnm{Protter},~\bfnm{P.}\binits{P.}} \AND
\bauthor{\bsnm{Sharpe},~\bfnm{M.~J.}\binits{M.~J.}}
(\byear{1980}).
\btitle{Semimartingales and {M}arkov processes}.
\bjournal{Z. Wahrsch. Verw. Gebiete}
\bvolume{54}
\bpages{161--219}.
\bid{doi={10.1007/BF00531446}, issn={0044-3719}, mr={0597337}}
\bptok{imsref}%
\end{barticle}
%
\endbibitem

\bibitem[\protect\citeauthoryear{Chen and Fukushima}{2011}]{chen-fukushima-1}
%
\begin{bbook}[mr]
\bauthor{\bsnm{Chen},~\bfnm{Zhen-Qing}\binits{Z.-Q.}} \AND
\bauthor{\bsnm{Fukushima},~\bfnm{Masatoshi}\binits{M.}}
(\byear{2011}).
\btitle{{Symmetric Markov Processes, Time Change, and Boundary Theory}}.
\bpublisher{Princeton Univ. Press}, \blocation{Princeton, NJ}.
\bid{mr={2849840}}
\bptok{imsref}%
\end{bbook}
%
\endbibitem

\bibitem[\protect\citeauthoryear{Cox, Ingersoll and
Ross}{1985}]{cox-ingersoll-ross-1}
%
\begin{barticle}[mr]
\bauthor{\bsnm{Cox},~\bfnm{John~C.}\binits{J.~C.}},
\bauthor{\bsnm{Ingersoll},~\bfnm{Jonathan~E.}\binits{J.~E.}
\bsuffix{Jr.}}
\AND\bauthor{\bsnm{Ross},~\bfnm{Stephen~A.}\binits{S.~A.}}
(\byear{1985}).
\btitle{A theory of the term structure of interest rates}.
\bjournal{Econometrica}
\bvolume{53}
\bpages{385--407}.
\bid{doi={10.2307/1911242}, issn={0012-9682}, mr={0785475}}
\bptok{imsref}%
\end{barticle}
%
\endbibitem

\bibitem[\protect\citeauthoryear{Cuchiero
et~al.}{2011a}]{cuchiero-filipovic-mayerhofer-teichmann-1}
%
\begin{barticle}[mr]
\bauthor{\bsnm{Cuchiero},~\bfnm{Christa}\binits{C.}},
\bauthor{\bsnm{Filipovi{\'c}},~\bfnm{Damir}\binits{D.}},
\bauthor{\bsnm{Mayerhofer},~\bfnm{Eberhard}\binits{E.}} \AND
\bauthor{\bsnm{Teichmann},~\bfnm{Josef}\binits{J.}}
(\byear{2011}a).
\btitle{Affine processes on positive semidefinite matrices}.
\bjournal{Ann. Appl. Probab.}
\bvolume{21}
\bpages{397--463}.
\bid{doi={10.1214/10-AAP710}, issn={1050-5164}, mr={2807963}}
\bptok{imsref}%
\end{barticle}
%
\endbibitem

\bibitem[\protect\citeauthoryear{Cuchiero
et~al.}{2011b}]{Cuchiero-Keller-Ressel-Mayerhofer-Teichmann-1}
%
\begin{bmisc}[author]
\bauthor{\bsnm{Cuchiero},~\bfnm{Christa}\binits{C.}},
\bauthor{\bsnm{Keller-Ressel},~\bfnm{Martin}\binits{M.}},
\bauthor{\bsnm{Mayerhofer},~\bfnm{Eberhard}\binits{E.}} \AND
\bauthor{\bsnm{Teichmann},~\bfnm{Josef}\binits{J.}}
(\byear{2011}b).
\bhowpublished{Affine processes on symmetric cones. Unpublished manuscript.}
\bptok{imsref}%
\end{bmisc}
%
\endbibitem

\bibitem[\protect\citeauthoryear{Davies}{2007}]{davies-3}
%
\begin{bbook}[mr]
\bauthor{\bsnm{Davies},~\bfnm{E.~Brian}\binits{E.~B.}}
(\byear{2007}).
\btitle{Linear Operators and Their Spectra}.
\bseries{Cambridge Studies in Advanced Mathematics}
\bvolume{106}.
\bpublisher{Cambridge Univ. Press}, \blocation{Cambridge}.
\bid{doi={10.1017/CBO9780511618864}, mr={2359869}}
\bptok{imsref}%
\end{bbook}
%
\endbibitem

\bibitem[\protect\citeauthoryear{Davydov and
Linetsky}{2003}]{davydov-linetsky-3}
%
\begin{barticle}[mr]
\bauthor{\bsnm{Davydov},~\bfnm{Dmitry}\binits{D.}} \AND
\bauthor{\bsnm{Linetsky},~\bfnm{Vadim}\binits{V.}}
(\byear{2003}).
\btitle{Pricing options on scalar diffusions: An eigenfunction expansion
approach}.
\bjournal{Oper. Res.}
\bvolume{51}
\bpages{185--209}.
\bid{doi={10.1287/opre.51.2.185.12782}, issn={0030-364X}, mr={1964993}}
\bptok{imsref}%
\end{barticle}
%
\endbibitem\vadjust{\goodbreak}

\bibitem[\protect\citeauthoryear{Duffie, Filipovi{\'c} and
Schachermayer}{2003}]{duffie-filipovic-schachermayer-1}
%
\begin{barticle}[mr]
\bauthor{\bsnm{Duffie},~\bfnm{D.}\binits{D.}},
\bauthor{\bsnm{Filipovi{\'c}},~\bfnm{D.}\binits{D.}} \AND
\bauthor{\bsnm{Schachermayer},~\bfnm{W.}\binits{W.}}
(\byear{2003}).
\btitle{Affine processes and applications in finance}.
\bjournal{Ann. Appl. Probab.}
\bvolume{13}
\bpages{984--1053}.
\bid{doi={10.1214/aoap/1060202833}, issn={1050-5164}, mr={1994043}}
\bptok{imsref}%
\end{barticle}
%
\endbibitem

\bibitem[\protect\citeauthoryear{Duffie and
Garleanu}{2001}]{duffie-garleanu-1}
%
\begin{barticle}[author]
\bauthor{\bsnm{Duffie},~\bfnm{Darrell}\binits{D.}} \AND
\bauthor{\bsnm{Garleanu},~\bfnm{Nicolae}\binits{N.}}
(\byear{2001}).
\btitle{{Risk and valuation of collateralized debt obligations}}.
\bjournal{Financial Analysts Journal}
\bvolume{57}
\bpages{41--59}.
\bptok{imsref}%
\end{barticle}
%
\endbibitem

\bibitem[\protect\citeauthoryear{Duffie and Kan}{1996}]{duffie-kan-2}
%
\begin{barticle}[author]
\bauthor{\bsnm{Duffie},~\bfnm{Darrell}\binits{D.}} \AND
\bauthor{\bsnm{Kan},~\bfnm{Rui}\binits{R.}}
(\byear{1996}).
\btitle{{A yield-factor model of interest rates}}.
\bjournal{Math. Finance}
\bvolume{6}
\bpages{379--406}.
\bptok{imsref}%
\end{barticle}
%
\endbibitem

\bibitem[\protect\citeauthoryear{Duffie, Pan and
Singleton}{2000}]{duffie-pan-singleton-1}
%
\begin{barticle}[mr]
\bauthor{\bsnm{Duffie},~\bfnm{Darrell}\binits{D.}},
\bauthor{\bsnm{Pan},~\bfnm{Jun}\binits{J.}} \AND
\bauthor{\bsnm{Singleton},~\bfnm{Kenneth}\binits{K.}}
(\byear{2000}).
\btitle{Transform analysis and asset pricing for affine jump-diffusions}.
\bjournal{Econometrica}
\bvolume{68}
\bpages{1343--1376}.
\bid{doi={10.1111/1468-0262.00164}, issn={0012-9682}, mr={1793362}}
\bptok{imsref}%
\end{barticle}
%
\endbibitem

\bibitem[\protect\citeauthoryear{Duffie and
Singleton}{1999}]{duffie-singleton-1}
%
\begin{barticle}[author]
\bauthor{\bsnm{Duffie},~\bfnm{Darrell}\binits{D.}} \AND
\bauthor{\bsnm{Singleton},~\bfnm{Kenneth~J.}\binits{K.~J.}}
(\byear{1999}).
\btitle{{Modeling term structures of defaultable bonds}}.
\bjournal{The Review of Financial Studies}
\bvolume{12}
\bpages{687--720}.
\bptok{imsref}%
\end{barticle}
%
\endbibitem

\bibitem[\protect\citeauthoryear{Duffie and
Singleton}{2003}]{duffie-singleton-2}
%
\begin{bbook}[author]
\bauthor{\bsnm{Duffie},~\bfnm{Darrell}\binits{D.}} \AND
\bauthor{\bsnm{Singleton},~\bfnm{Kenneth~J.}\binits{K.~J.}}
(\byear{2003}).
\btitle{{Credit Risk: Pricing, Measurement, and Management}}.
\bpublisher{Princeton Univ. Press}, \blocation{Princeton, NJ}.
\bptok{imsref}%
\end{bbook}
%
\endbibitem

\bibitem[\protect\citeauthoryear{Elkamhi
et~al.}{2012}]{Elkamhi-Jacobs-Langlois-Ornthanalai-1}
%
\begin{bmisc}[author]
\bauthor{\bsnm{Elkamhi},~\bfnm{Redouane}\binits{R.}},
\bauthor{\bsnm{Jacobs},~\bfnm{Kris}\binits{K.}},
\bauthor{\bsnm{Langlois},~\bfnm{Hugues}\binits{H.}} \AND
\bauthor{\bsnm{Ornthanalai},~\bfnm{Chayawat}\binits{C.}}
(\byear{2012}).
\bhowpublished{Accounting information releases and CDS spreads.
Working paper.}
\bptok{imsref}%
\end{bmisc}
%
\endbibitem

\bibitem[\protect\citeauthoryear{Erdelyi}{1953}]{erdelyi-2}
%
\begin{bbook}[author]
\bauthor{\bsnm{Erdelyi},~\bfnm{A.}\binits{A.}}
(\byear{1953}).
\btitle{{Higher Transcendental Functions II}}.
\bpublisher{McGraw-Hill}, \blocation{New York}.
\bptok{imsref}%
\end{bbook}
%
\endbibitem

\bibitem[\protect\citeauthoryear{Ethier and Kurtz}{1986}]{ethier-kurtz-1}
%
\begin{bbook}[mr]
\bauthor{\bsnm{Ethier},~\bfnm{Stewart~N.}\binits{S.~N.}} \AND
\bauthor{\bsnm{Kurtz},~\bfnm{Thomas~G.}\binits{T.~G.}}
(\byear{1986}).
\btitle{Markov Processes: Characterization and Convergence}.
\bpublisher{Wiley}, \blocation{New York}.
\bid{doi={10.1002/9780470316658}, mr={0838085}}
\bptok{imsref}%
\end{bbook}
%
\endbibitem

\bibitem[\protect\citeauthoryear{Feller}{1951}]{feller-1}
%
\begin{barticle}[mr]
\bauthor{\bsnm{Feller},~\bfnm{William}\binits{W.}}
(\byear{1951}).
\btitle{Two singular diffusion problems}.
\bjournal{Ann. of Math. (2)}
\bvolume{54}
\bpages{173--182}.
\bid{issn={0003-486X}, mr={0054814}}
\bptok{imsref}%
\end{barticle}
%
\endbibitem

\bibitem[\protect\citeauthoryear{Filipovi{\'c}}{2001}]{filipovic-1}
%
\begin{barticle}[mr]
\bauthor{\bsnm{Filipovi{\'c}},~\bfnm{Damir}\binits{D.}}
(\byear{2001}).
\btitle{A general characterization of one factor affine term structure models}.
\bjournal{Finance Stoch.}
\bvolume{5}
\bpages{389--412}.
\bid{doi={10.1007/PL00013540}, issn={0949-2984}, mr={1850789}}
\bptok{imsref}%
\end{barticle}
%
\endbibitem

\bibitem[\protect\citeauthoryear{Fukushima, Oshima and
Takeda}{2011}]{fukushima-oshima-takeda-1}
%
\begin{bbook}[mr]
\bauthor{\bsnm{Fukushima},~\bfnm{Masatoshi}\binits{M.}},
\bauthor{\bsnm{Oshima},~\bfnm{Yoichi}\binits{Y.}} \AND
\bauthor{\bsnm{Takeda},~\bfnm{Masayoshi}\binits{M.}}
(\byear{2011}).
\btitle{Dirichlet Forms and Symmetric {M}arkov Processes},
\bedition{extended} ed.
\bseries{de Gruyter Studies in Mathematics}
\bvolume{19}.
\bpublisher{de Gruyter}, \blocation{Berlin}.
\bid{mr={2778606}}
\bptok{imsref}%
\end{bbook}
%
\endbibitem

\bibitem[\protect\citeauthoryear{Geman and
Roncoroni}{2006}]{Geman-Roncoroni-1}
%
\begin{barticle}[author]
\bauthor{\bsnm{Geman},~\bfnm{H{\'{e}}lyette}\binits{H.}} \AND
\bauthor{\bsnm{Roncoroni},~\bfnm{Andrea}\binits{A.}}
(\byear{2006}).
\btitle{{Understanding the fine structure of electricity prices}}.
\bjournal{Journal of Business}
\bvolume{79}
\bpages{1225--1261}.
\bptok{imsref}%
\end{barticle}
%
\endbibitem

\bibitem[\protect\citeauthoryear{G{\"o}ing-Jaeschke and
Yor}{2003}]{going-jaeschke-yor-1}
%
\begin{barticle}[mr]
\bauthor{\bsnm{G{\"o}ing-Jaeschke},~\bfnm{Anja}\binits{A.}} \AND
\bauthor{\bsnm{Yor},~\bfnm{Marc}\binits{M.}}
(\byear{2003}).
\btitle{A survey and some generalizations of {B}essel processes}.
\bjournal{Bernoulli}
\bvolume{9}
\bpages{313--349}.
\bid{doi={10.3150/bj/1068128980}, issn={1350-7265}, mr={1997032}}
\bptok{imsref}%
\end{barticle}
%
\endbibitem

\bibitem[\protect\citeauthoryear{Gorovoi and
Linetsky}{2004}]{gorovoi-linetsky-1}
%
\begin{barticle}[mr]
\bauthor{\bsnm{Gorovoi},~\bfnm{Viatcheslav}\binits{V.}} \AND
\bauthor{\bsnm{Linetsky},~\bfnm{Vadim}\binits{V.}}
(\byear{2004}).
\btitle{Black's model of interest rates as options, eigenfunction expansions
and {J}apanese interest rates}.
\bjournal{Math. Finance}
\bvolume{14}
\bpages{49--78}.
\bid{doi={10.1111/j.0960-1627.2004.00181.x}, issn={0960-1627}, mr={2030835}}
\bptok{imsref}%
\end{barticle}
%
\endbibitem

\bibitem[\protect\citeauthoryear{It{\^o} and McKean}{1974}]{ito-mckean-1}
%
\begin{bbook}[mr]
\bauthor{\bsnm{It{\^o}},~\bfnm{Kiyosi}\binits{K.}} \AND
\bauthor{\bsnm{McKean},~\bfnm{Henry~P.}\binits{H.~P.}}
(\byear{1974}).
\btitle{Diffusion Processes and Their Sample Paths},
\bedition{corrected 2nd} ed.
\bpublisher{Springer}, \blocation{Berlin}.
\bid{mr={0345224}}
\bptok{imsref}%
\end{bbook}
%
\endbibitem

\bibitem[\protect\citeauthoryear{Jacod}{1979}]{jacod-1}
%
\begin{bbook}[mr]
\bauthor{\bsnm{Jacod},~\bfnm{Jean}\binits{J.}}
(\byear{1979}).
\btitle{Calcul Stochastique et Probl\`emes de Martingales}.
\bseries{Lecture Notes in Math.}
\bvolume{714}.
\bpublisher{Springer}, \blocation{Berlin}.
\bid{mr={0542115}}
\bptok{imsref}%
\end{bbook}
%
\endbibitem

\bibitem[\protect\citeauthoryear{Jacod and Protter}{2011}]{jacod-protter-1}
%
\begin{bbook}[author]
\bauthor{\bsnm{Jacod},~\bfnm{Jean}\binits{J.}} \AND
\bauthor{\bsnm{Protter},~\bfnm{Philip}\binits{P.}}
(\byear{2011}).
\btitle{{Discretization of Processes}}.
\bpublisher{Springer}, \blocation{Berlin}.
\bptok{imsref}%
\end{bbook}
%
\endbibitem

\bibitem[\protect\citeauthoryear{Jacod and Shiryaev}{2002}]{jacod-shiryaev-1}
%
\begin{bbook}[author]
\bauthor{\bsnm{Jacod},~\bfnm{Jean}\binits{J.}} \AND
\bauthor{\bsnm{Shiryaev},~\bfnm{Albert~N.}\binits{A.~N.}}
(\byear{2002}).
\btitle{{Limit Theorems for Stochastic Processes}},
\bedition{2nd.} ed.
\bseries{Comprenhensive Studies in Mathematics}
\bvolume{288}.
\bpublisher{Springer}, \blocation{Berlin}.
\bptok{imsref}%
\end{bbook}
%
\endbibitem

\bibitem[\protect\citeauthoryear{Jamshidian}{1996}]{jamshidian-3}
%
\begin{barticle}[author]
\bauthor{\bsnm{Jamshidian},~\bfnm{Farshid}\binits{F.}}
(\byear{1996}).
\btitle{{Bond, futures and option evaluation in the quadratic interest rate
model}}.
\bjournal{Appl. Math. Finance}
\bvolume{3}
\bpages{93--115}.
\bptok{imsref}%
\end{barticle}
%
\endbibitem

\bibitem[\protect\citeauthoryear{Janson, M'Baye and
Protter}{2011}]{janson-mbay-protter-1}
%
\begin{barticle}[mr]
\bauthor{\bsnm{Janson},~\bfnm{Svante}\binits{S.}},
\bauthor{\bsnm{M'Baye},~\bfnm{Sokhna}\binits{S.}} \AND
\bauthor{\bsnm{Protter},~\bfnm{Philip}\binits{P.}}
(\byear{2011}).
\btitle{Absolutely continuous compensators}.
\bjournal{Int. J. Theor. Appl. Finance}
\bvolume{14}
\bpages{335--351}.
\bid{doi={10.1142/S0219024911006565}, issn={0219-0249}, mr={2804101}}
\bptok{imsref}%
\end{barticle}
%
\endbibitem

\bibitem[\protect\citeauthoryear{Jarrow, Lando and
Turnbull}{1997}]{jarrow-lando-turnbull-1}
%
\begin{barticle}[author]
\bauthor{\bsnm{Jarrow},~\bfnm{Robert~A.}\binits{R.~A.}},
\bauthor{\bsnm{Lando},~\bfnm{David}\binits{D.}} \AND
\bauthor{\bsnm{Turnbull},~\bfnm{Stuart~M.}\binits{S.~M.}}
(\byear{1997}).
\btitle{{A Markov model for the term structure of credit risk spreads}}.
\bjournal{The Review of Financial Studies}
\bvolume{10}
\bpages{481--523}.
\bptok{imsref}%
\end{barticle}
%
\endbibitem

\bibitem[\protect\citeauthoryear{Jeanblanc, Yor and
Chesney}{2009}]{jeanblanc-yor-chesney-1}
%
\begin{bbook}[mr]
\bauthor{\bsnm{Jeanblanc},~\bfnm{Monique}\binits{M.}},
\bauthor{\bsnm{Yor},~\bfnm{Marc}\binits{M.}} \AND
\bauthor{\bsnm{Chesney},~\bfnm{Marc}\binits{M.}}
(\byear{2009}).
\btitle{Mathematical Methods for Financial Markets}.
\bpublisher{Springer}, \blocation{London}.
\bid{doi={10.1007/978-1-84628-737-4}, mr={2568861}}
\bptok{imsref}%
\end{bbook}
%
\endbibitem

\bibitem[\protect\citeauthoryear{Kawazu and
Watanabe}{1971}]{kawazu-watanabe-1}
%
\begin{barticle}[mr]
\bauthor{\bsnm{Kawazu},~\bfnm{Kiyoshi}\binits{K.}} \AND
\bauthor{\bsnm{Watanabe},~\bfnm{Shinzo}\binits{S.}}
(\byear{1971}).
\btitle{Branching processes with immigration and related limit theorems}.
\bjournal{Theory Probab. Appl.}
\bvolume{16}
\bpages{36--54}.
\bptok{imsref}%
\end{barticle}
%
\endbibitem

\bibitem[\protect\citeauthoryear{Keller-Ressel, Schachermayer and
Teichmann}{2011}]{keller-ressel-schachermayer-teichmann-1}
%
\begin{barticle}[mr]
\bauthor{\bsnm{Keller-Ressel},~\bfnm{Martin}\binits{M.}},
\bauthor{\bsnm{Schachermayer},~\bfnm{Walter}\binits{W.}} \AND
\bauthor{\bsnm{Teichmann},~\bfnm{Josef}\binits{J.}}
(\byear{2011}).
\btitle{Affine processes are regular}.
\bjournal{Probab. Theory Related Fields}
\bvolume{151}
\bpages{591--611}.
\bid{doi={10.1007/s00440-010-0309-4}, issn={0178-8051}, mr={2851694}}
\bptok{imsref}%
\end{barticle}
%
\endbibitem

\bibitem[\protect\citeauthoryear{Kita}{2012}]{Kita-1}
%
\begin{bmisc}[author]
\bauthor{\bsnm{Kita},~\bfnm{Arben}\binits{A.}}
(\byear{2012}).
\bhowpublished{CDS spreads explained with credit spread volatility and jump
risk of individual firms. Working paper}.
\bptok{imsref}%
\end{bmisc}
%
\endbibitem\vadjust{\goodbreak}

\bibitem[\protect\citeauthoryear{Lewis}{1994}]{lewis-4}
%
\begin{bmisc}[author]
\bauthor{\bsnm{Lewis},~\bfnm{Alan~L.}\binits{A.~L.}}
(\byear{1994}).
\bhowpublished{Three expansion regimes for interest rate term
structure models.
\textit{Analytic Investment Management}. Available at
\url{http://optioncity.net/}.}
\bptok{imsref}%
\end{bmisc}
%
\endbibitem

\bibitem[\protect\citeauthoryear{Li and Linetsky}{2013a}]{li-linetsky-2}
%
\begin{barticle}[author]
\bauthor{\bsnm{Li},~\bfnm{Lingfei}\binits{L.}} \AND
\bauthor{\bsnm{Linetsky},~\bfnm{Vadim}\binits{V.}}
(\byear{2013}a).
\btitle{Optimal stopping and early exercise: An eigenfunction expansion
approach}.
\bjournal{Oper. Res.}
\bvolume{61}
\bpages{625--644}.
\bptok{imsref}%
\end{barticle}
%
\endbibitem

\bibitem[\protect\citeauthoryear{Li and Linetsky}{2013b}]{li-linetsky-1}
%
\begin{bmisc}[author]
\bauthor{\bsnm{Li},~\bfnm{Lingfei}\binits{L.}} \AND
\bauthor{\bsnm{Linetsky},~\bfnm{Vadim}\binits{V.}}
(\byear{2013}b).
\bhowpublished{Time-changed Ornstein--Uhlenbeck processes and their applications
in commodity derivative models. \textit{Math. Finance}. To appear.}
\bptok{imsref}%
\end{bmisc}
%
\endbibitem

\bibitem[\protect\citeauthoryear{Lim, Li and
Linetsky}{2012}]{lim-li-linetsky-1}
%
\begin{barticle}[mr]
\bauthor{\bsnm{Lim},~\bfnm{Dongjae}\binits{D.}},
\bauthor{\bsnm{Li},~\bfnm{Lingfei}\binits{L.}} \AND
\bauthor{\bsnm{Linetsky},~\bfnm{Vadim}\binits{V.}}
(\byear{2012}).
\btitle{Evaluating callable and putable bonds: An eigenfunction expansion
approach}.
\bjournal{J. Econom. Dynam. Control}
\bvolume{36}
\bpages{1888--1908}.
\bid{doi={10.1016/j.jedc.2012.06.002}, issn={0165-1889}, mr={2982959}}
\bptok{imsref}%
\end{barticle}
%
\endbibitem

\bibitem[\protect\citeauthoryear{Linetsky}{2004}]{linetsky-2}
%
\begin{barticle}[mr]
\bauthor{\bsnm{Linetsky},~\bfnm{Vadim}\binits{V.}}
(\byear{2004}).
\btitle{The spectral decomposition of the option value}.
\bjournal{Int. J. Theor. Appl. Finance}
\bvolume{7}
\bpages{337--384}.
\bid{doi={10.1142/S0219024904002451}, issn={0219-0249}, mr={2064020}}
\bptok{imsref}%
\end{barticle}
%
\endbibitem

\bibitem[\protect\citeauthoryear{Linetsky}{2006}]{linetsky-1}
%
\begin{barticle}[mr]
\bauthor{\bsnm{Linetsky},~\bfnm{Vadim}\binits{V.}}
(\byear{2006}).
\btitle{Pricing equity derivatives subject to bankruptcy}.
\bjournal{Math. Finance}
\bvolume{16}
\bpages{255--282}.
\bid{doi={10.1111/j.1467-9965.2006.00271.x}, issn={0960-1627}, mr={2212266}}
\bptok{imsref}%
\end{barticle}
%
\endbibitem

\bibitem[\protect\citeauthoryear{Linetsky}{2008}]{linetsky-3}
%
\begin{bincollection}[mr]
\bauthor{\bsnm{Linetsky},~\bfnm{Vadim}\binits{V.}}
(\byear{2008}).
\btitle{Spectral methods in derivatives pricing}.
In \bbooktitle{Handbooks in Operations Research and Management Science: Financial Engineering}
\bvolume{15}
\bpages{223--300}.
\bpublisher{Elsevier/North-Holland}, \blocation{Amsterdam}.
\bptok{imsref}%
\end{bincollection}
%
\endbibitem

\bibitem[\protect\citeauthoryear{Lorig, Lozano-Carbass\'{e} and
Mendoza-Arriaga}{2013}]{lorig-mendoza-1}
%
\begin{bmisc}[author]
\bauthor{\bsnm{Lorig},~\bfnm{M.}\binits{M.}},
\bauthor{\bsnm{Lozano-Carbass\'{e}},~\bfnm{O.}\binits{O.}}
\AND
\bauthor{\bsnm{Mendoza-Arriaga},~\bfnm{R.}\binits{R.}}
(\byear{2013}).
\bhowpublished{Variance swaps on defaultable assets and market
implied time-changes. Unpublished manuscript.}
\bptok{imsref}%
\end{bmisc}
%
\endbibitem

\bibitem[\protect\citeauthoryear{Madan, Carr and
Chang}{1998}]{madan-carr-chang-1}
%
\begin{barticle}[author]
\bauthor{\bsnm{Madan},~\bfnm{D.~B.}\binits{D.~B.}},
\bauthor{\bsnm{Carr},~\bfnm{P.}\binits{P.}} \AND
\bauthor{\bsnm{Chang},~\bfnm{E.~C.}\binits{E.~C.}}
(\byear{1998}).
\btitle{{The variance gamma process and option pricing}}.
\bjournal{European Finance Review}
\bvolume{2}
\bpages{79--105}.
\bptok{imsref}%
\end{barticle}
%
\endbibitem

\bibitem[\protect\citeauthoryear{McKean}{1956}]{mckean-1}
%
\begin{barticle}[mr]
\bauthor{\bsnm{McKean},~\bfnm{Henry~P.}\binits{H.~P.} \bsuffix{Jr.}}
(\byear{1956}).
\btitle{Elementary solutions for certain parabolic partial differential
equations}.
\bjournal{Trans. Amer. Math. Soc.}
\bvolume{82}
\bpages{519--548}.
\bid{issn={0002-9947}, mr={0087012}}
\bptok{imsref}%
\end{barticle}
%
\endbibitem

\bibitem[\protect\citeauthoryear{Mendoza-Arriaga}{2012}]{mendoza-1}
%
\begin{bmisc}[author]
\bauthor{\bsnm{Mendoza-Arriaga},~\bfnm{Rafael}\binits{R.}}
(\byear{2012}).
\btitle{Credit default swap options under the subordinate diffusion framework.
Working paper.}
\bptok{imsref}%
\end{bmisc}
%
\endbibitem

\bibitem[\protect\citeauthoryear{Mendoza-Arriaga, Carr and
Linetsky}{2010}]{mendoza-carr-linetsky-1}
%
\begin{barticle}[mr]
\bauthor{\bsnm{Mendoza-Arriaga},~\bfnm{Rafael}\binits{R.}},
\bauthor{\bsnm{Carr},~\bfnm{Peter}\binits{P.}} \AND
\bauthor{\bsnm{Linetsky},~\bfnm{Vadim}\binits{V.}}
(\byear{2010}).
\btitle{Time-changed {M}arkov processes in unified credit-equity modeling}.
\bjournal{Math. Finance}
\bvolume{20}
\bpages{527--569}.
\bid{doi={10.1111/j.1467-9965.2010.00411.x}, issn={0960-1627}, mr={2731407}}
\bptok{imsref}%
\end{barticle}
%
\endbibitem

\bibitem[\protect\citeauthoryear{Mendoza-Arriaga and
Linetsky}{2013}]{linetsky-mendoza-arriaga-1}
%
\begin{bmisc}[author]
\bauthor{\bsnm{Mendoza-Arriaga},~\bfnm{Rafael}\binits{R.}} \AND
\bauthor{\bsnm{Linetsky},~\bfnm{Vadim}\binits{V.}}
(\byear{2013}).
\bhowpublished{Multivariate subordination of Markov processes with financial
applications. \textit{Math. Finance}. To appear.}
\bptok{imsref}%
\end{bmisc}
%
\endbibitem

\bibitem[\protect\citeauthoryear{Meyer-Brandis and
Tankov}{2008}]{Meyer-Brandis-Tankov-1}
%
\begin{barticle}[author]
\bauthor{\bsnm{Meyer-Brandis},~\bfnm{Thilo}\binits{T.}} \AND
\bauthor{\bsnm{Tankov},~\bfnm{Peter}\binits{P.}}
(\byear{2008}).
\btitle{{Multi-factor jump-diffusion models of electricity prices}}.
\bjournal{Int. J. Theor. Appl. Finance}
\bvolume{11}
\bpages{503--528}.
\bptok{imsref}%
\end{barticle}
%
\endbibitem

\bibitem[\protect\citeauthoryear{Nikiforov and
Uvarov}{1988}]{nikiforov-uvarov-1}
%
\begin{bbook}[mr]
\bauthor{\bsnm{Nikiforov},~\bfnm{Arnold~F.}\binits{A.~F.}} \AND
\bauthor{\bsnm{Uvarov},~\bfnm{Vasilii~B.}\binits{V.~B.}}
(\byear{1988}).
\btitle{Special Functions of Mathematical Physics}.
\bpublisher{Birkh\"auser}, \blocation{Basel}.
\bid{mr={0922041}}
\bptok{imsref}%
\end{bbook}
%
\endbibitem

\bibitem[\protect\citeauthoryear{Nowak and Stempak}{2010}]{nowak-stempak-2}
%
\begin{bmisc}[author]
\bauthor{\bsnm{Nowak},~\bfnm{Adam}\binits{A.}} \AND
\bauthor{\bsnm{Stempak},~\bfnm{Krzysztof}\binits{K.}}
(\byear{2010}).
\bhowpublished{On $L^p$-contractivity of Laguerre semigroups. Working paper}.
\bptok{imsref}%
\end{bmisc}
%
\endbibitem

\bibitem[\protect\citeauthoryear{Phillips}{1952}]{phillips-1}
%
\begin{barticle}[mr]
\bauthor{\bsnm{Phillips},~\bfnm{R.~S.}\binits{R.~S.}}
(\byear{1952}).
\btitle{On the generation of semigroups of linear operators}.
\bjournal{Pacific J. Math.}
\bvolume{2}
\bpages{343--369}.
\bid{issn={0030-8730}, mr={0050797}}
\bptok{imsref}%
\end{barticle}
%
\endbibitem

\bibitem[\protect\citeauthoryear{Pitman and Yor}{1982}]{pitman-yor-1}
%
\begin{barticle}[mr]
\bauthor{\bsnm{Pitman},~\bfnm{Jim}\binits{J.}} \AND
\bauthor{\bsnm{Yor},~\bfnm{Marc}\binits{M.}}
(\byear{1982}).
\btitle{A decomposition of {B}essel bridges}.
\bjournal{Z. Wahrsch. Verw. Gebiete}
\bvolume{59}
\bpages{425--457}.
\bid{doi={10.1007/BF00532802}, issn={0044-3719}, mr={0656509}}
\bptok{imsref}%
\end{barticle}
%
\endbibitem

\bibitem[\protect\citeauthoryear{Prudnikov, Brychkov and
Marichev}{1986}]{prudnikov-2}
%
\begin{bbook}[author]
\bauthor{\bsnm{Prudnikov},~\bfnm{A.~P.}\binits{A.~P.}},
\bauthor{\bsnm{Brychkov},~\bfnm{Yu.~A.}\binits{Y.~A.}} \AND
\bauthor{\bsnm{Marichev},~\bfnm{O.~I.}\binits{O.~I.}}
(\byear{1986}).
\btitle{{Integrals Series: Special Functions. Integrals and Series~II}}.
\bpublisher{Gordon \& Breach}, \blocation{New York}.
\bptok{imsref}%
\end{bbook}
%
\endbibitem

\bibitem[\protect\citeauthoryear{Reed and Simon}{1980}]{reed-simon-1}
%
\begin{bbook}[mr]
\bauthor{\bsnm{Reed},~\bfnm{Michael}\binits{M.}} \AND
\bauthor{\bsnm{Simon},~\bfnm{Barry}\binits{B.}}
(\byear{1980}).
\btitle{Methods of Modern Mathematical Physics {I}: Functional Analysis},
\bedition{revised} ed.
\bpublisher{Academic Press}, \blocation{San Diego, CA}.
\bid{mr={0751959}}
\bptok{imsref}%
\end{bbook}
%
\endbibitem

\bibitem[\protect\citeauthoryear{Revuz and Yor}{1999}]{revuz-yor}
%
\begin{bbook}[mr]
\bauthor{\bsnm{Revuz},~\bfnm{Daniel}\binits{D.}} \AND
\bauthor{\bsnm{Yor},~\bfnm{Marc}\binits{M.}}
(\byear{1999}).
\btitle{Continuous Martingales and {B}rownian Motion},
\bedition{3rd} ed.
\bseries{Grundlehren der Mathematischen Wissenschaften}
\bvolume{293}.
\bpublisher{Springer}, \blocation{Berlin}.
\bid{mr={1725357}}
\bptok{imsref}%
\end{bbook}
%
\endbibitem

\bibitem[\protect\citeauthoryear{Sato}{1999}]{sato}
%
\begin{bbook}[mr]
\bauthor{\bsnm{Sato},~\bfnm{Ken-iti}\binits{K.-i.}}
(\byear{1999}).
\btitle{L\'evy Processes and Infinitely Divisible Distributions}.
\bseries{Cambridge Studies in Advanced Mathematics}
\bvolume{68}.
\bpublisher{Cambridge Univ. Press}, \blocation{Cambridge}.
\bid{mr={1739520}}
\bptok{imsref}%
\end{bbook}
%
\endbibitem

\bibitem[\protect\citeauthoryear{Schilling, Song and
Vondra{\v{c}}ek}{2010}]{schilling-song-vondraceck-1}
%
\begin{bbook}[mr]
\bauthor{\bsnm{Schilling},~\bfnm{Ren{\'e}~L.}\binits{R.~L.}},
\bauthor{\bsnm{Song},~\bfnm{Renming}\binits{R.}} \AND
\bauthor{\bsnm{Vondra{\v{c}}ek},~\bfnm{Zoran}\binits{Z.}}
(\byear{2010}).
\btitle{Bernstein Functions: Theory and Applications}.
\bseries{de Gruyter Studies in Mathematics}
\bvolume{37}.
\bpublisher{de Gruyter}, \blocation{Berlin}.
\bid{mr={2598208}}
\bptok{imsref}%
\end{bbook}
%
\endbibitem\

\bibitem[\protect\citeauthoryear{Zhang, Zhou and
Zhu}{2009}]{zhang-zhou-zhu-1}
%
\begin{barticle}[author]
\bauthor{\bsnm{Zhang},~\bfnm{Benjamin~Yibin}\binits{B.~Y.}},
\bauthor{\bsnm{Zhou},~\bfnm{Hao}\binits{H.}} \AND
\bauthor{\bsnm{Zhu},~\bfnm{Haibin}\binits{H.}}
(\byear{2009}).
\btitle{{Explaining credit default swap spreads with the equity
volatility and
jump risks of individual firms}}.
\bjournal{Review of Financial Studies}
\bvolume{22}
\bpages{5099--5131}.
\bptok{imsref}%
\end{barticle}
%
\endbibitem

\end{thebibliography}
\end{document}